\documentclass[11pt,oneside,paper=A4,DIV=3,BCOR=0mm]{scrreprt}

\usepackage[utf8]{inputenc}
\usepackage{graphicx}
\usepackage{epstopdf}
\usepackage{bm}
\usepackage{kotex}
\usepackage{stackengine}
\usepackage{feynmp}
\usepackage{feynmp-auto}
\usepackage{slashed}
\usepackage{amsmath,amsfonts,amssymb,amscd,amsxtra,amsthm}
\usepackage[linktocpage=true]{hyperref}
\usepackage{color,ulem}
\usepackage{tocbibind}
\usepackage{apptools}
\usepackage{geometry}
\usepackage{setspace}
\usepackage[format=plain]{caption}
\usepackage[title,toc,titletoc,page]{appendix}
\newgeometry{   left = 38mm,
                right = 38mm,
                top = 40mm,
                bottom = 40mm}

\begin{document}

\newpage
\renewcommand{\baselinestretch}{1.3}%
\thispagestyle{empty}
    \begin{center}
        \fontsize{14pt}{14pt}\selectfont{Thesis for the Degree of Master of Science}
    \end{center}
    \begin{center}
        \vspace{5mm}
        \fontsize{22pt}{22pt}\selectfont{A study of medium effects in elastic $\pi N$ and $\pi A$ scatterings}
    \end{center}
    \begin{center}
        \vspace{90mm}
        \fontsize{14pt}{14pt}\selectfont{by}
        \\[5mm]
        \fontsize{14pt}{14pt}\selectfont{Hyeon-dong Han}
        \\[10mm]
        \fontsize{14pt}{14pt}\selectfont{Department of Physics}
        \\[10mm]
        \fontsize{14pt}{14pt}\selectfont{The Graduate School}
        \\[10mm]
        \fontsize{14pt}{14pt}\selectfont{Pukyong National University}
        \\[10mm]
        \fontsize{12pt}{12pt}\selectfont{February 2023}
    \end{center}

\newpage
\thispagestyle{empty}
    \begin{center}
        \fontsize{22pt}{22pt}\selectfont{A study of medium effects in elastic $\pi N$ and $\pi A$ scatterings\\밀도 영향을 고려한 파이온-핵자 및 파이온-핵 탄성 산란에 관한 연구}
    \end{center}
    \begin{center}
        \vspace{10mm}
        \fontsize{12pt}{12pt}\selectfont{Advisor: Prof. Seung-il Nam}
        \\[15mm]
        \fontsize{14pt}{14pt}\selectfont{by}
        \\[5mm]
        \fontsize{14pt}{14pt}\selectfont{Hyeon-dong Han}
        \\[55mm]
        \fontsize{12pt}{12pt}\selectfont{A thesis submitted in partial fulfillment of the requirements\\}
        \vspace{2mm}
        \fontsize{12pt}
        {12pt}\selectfont{for the degree of}
        \\[10mm]
        \fontsize{12pt}{12pt}\selectfont{Master of Science}
        \\[10mm]
        \fontsize{12pt}{12pt}\selectfont{in Department of Physics, The Graduate School,\\}
        \vspace{2mm}
        \fontsize{12pt}{12pt}\selectfont{Pukyong National University}
        \\[10mm]
        \fontsize{12pt}{12pt}\selectfont{February 2023}
    \end{center}

\newpage
\thispagestyle{empty}
    \begin{center}
        \fontsize{14pt}{14pt}\selectfont{A study of medium effects in elastic $\pi N$ and $\pi A$ scatterings}
        \\[30mm]
        \fontsize{12pt}{12pt}\selectfont{A dissertation}
        \\[5mm]
        \fontsize{12pt}{12pt}\selectfont{by}
        \\[5mm]
        \fontsize{12pt}{12pt}\selectfont{Hyeon-dong Han}
        \\[30mm]
        \fontsize{10pt}{10pt}\selectfont{Approved by:}
        \\[15mm]
        \rule{80mm}{0.05pt}\\
        \fontsize{10pt}{10pt}\selectfont{(Chairman) Chang-ho Hyun}
        \\[15mm]
        \rule{80mm}{0.05pt}\\
        \fontsize{10pt}{10pt}\selectfont{(Member) Seung-il Nam}
        \\[15mm]
        \rule{80mm}{0.05pt}\\
        \fontsize{10pt}{10pt}\selectfont{(Member) Parada T. P. Hutauruk}
        \\[35mm]
        \fontsize{12pt}{12pt}\selectfont{February 17, 2023}
    \end{center}

\newgeometry{   left = 38mm,
                right = 38mm,
                top = 52mm,
                bottom = 50mm}

\newpage
\pagenumbering{roman}
\setcounter{tocdepth}{1}
\tableofcontents

\newpage
{%
\let\oldnumberline\numberline%
\renewcommand{\numberline}{\figurename~\oldnumberline}%
\listoffigures%
}
{%
\let\oldnumberline\numberline%
\renewcommand{\numberline}{\tablename~\oldnumberline}%
\listoftables%
}

\newpage

    \begin{center}
        \fontsize{16pt}{16pt}\selectfont{밀도 영향을 고려한 파이온-핵자 및 파이온-핵 탄성 산란에 관한 연구}
        \\[10mm]
        \fontsize{12pt}{12pt}\selectfont{한 현 동}
        \\[10mm]
        \fontsize{12pt}{12pt}\selectfont{부경대학교 대학원 물리학과}
        \\
    \end{center}

    \phantomsection
    \addcontentsline{toc}{chapter}{Abstract}
    \begin{center}요약\end{center}\par
    
    본 연구는 유한 중입자 밀도 하에 존재하는 $\Delta(1232)$ 공명 에너지 영역에서 아이소스핀 $I=3/2$ 채널을 고려한 파이온-핵자 탄성 산란을 다루고 있으며 계산을 위해 나뭇가지 준위(tree level)에서의 보른 근사를 통한 유효 라그랑지언 방법이 사용되었다. 매질 내에서의 중입자 특성을 고려해주기 위해 쿼크 단계에서 기술되어지는 쿼크-중간자 결합(QMC) 모형을 사용하였으며 이를 이용하여 매질 내에서의 유효 핵자 및 델타입자 질량과 델타입자의 붕괴 폭 등이 계산에 포함되었다. 먼저 진공에서의 총 단면적 결과가 실험값을 잘 설명하는지를 확인한 다음 매질 내에서의 총 단면적과 각 분포, 양성자-스핀 비대칭성을 포함한 여러 물리량들을 분석하였다. 또한 파이온-핵자 탄성 산란의 결과를 바탕으로 가벼운 원자핵(헬륨-4, 탄소-12)에 대해 아이코널-글러우버(Eikonal Glauber) 모형을 이용하여 유한 핵을 고려한 파이온-핵 탄성 산란을 연구하였다. 이때 유한 핵 내에서의 현상을 고려해주기 위하여 우드-삭슨(Wood-Saxon) 형태 및 가우스 함수들의 합으로 표현되는 형태, 총 두 종류의 핵 밀도 분포가 사용되었다. 유한 핵 내에서의 핵 밀도 분포, 유효 중입자 질량 및 매질 내에서의 붕괴 폭, 결합 상수와 같은 물리량들을 분석하였으며 이들을 고려한 저에너지 영역에서의 파이온-핵 탄성 산란의 총 단면적을 실험값과 비교하여 조사하였다. 결과적으로 델타입자 붕괴 폭을 제외한 유효 중입자 질량 및 매질 내에서의 산란단면적 등의 물리량들이 밀도가 증가함에 따라 감소함을 확인할 수 있었고 나뭇가지 준위에서의 보른 근사를 통한 파이온-핵 탄성 산란 계산은 헬륨에서는 실험 결과와 잘 맞았으나 탄소의 경우에는 실험값보다 크게 계산되었다. 매질 내에서의 델타입자의 특성을 포함한 본 연구 결과의 분석은 상대론적 중이온 충돌 실험 분야에서 유의미할 것으로 기대된다.
    \par

\newpage
    \begin{center}
        \fontsize{16pt}{16pt}\selectfont{A study of medium effects in elastic $\pi N$ and $\pi A$ scatterings}
        \\[10mm]
        \fontsize{12pt}{12pt}\selectfont{Hyeon-dong Han}
        \\[10mm]
        \fontsize{12pt}{12pt}\selectfont{Department of Physics, The Graduate School,\\Pukyong National University}
        \\
    \end{center}

\begin{center}\textbf{Abstract}\end{center}\par
The elastic $\pi N$ scattering is investigated for the $I=3/2$ channel dominated by the $\Delta(1232)$ resonance at finite baryon density, employing the effective Lagrangian approach at the tree-level Born approximation. The quark-meson coupling (QMC) model is employed to describe the in-medium baryon properties that are constructed at the quark level, such as the nucleon and $\Delta$ masses, and $\Delta$ full decay width. I reproduce the experimental data of the cross-section in a vacuum as a justification of our approach and then analyze the in-medium total and differential cross-sections as well as proton-spin asymmetry. Following the results of the in-medium elastic $\pi N$ scattering calculation, the elastic $\pi A$ scattering is investigated at finite baryon density in the framework of the Eikonal Glauber model for the light nuclei, $^4\mathrm{He}$ and $^{12}\mathrm{C}$. For the description of the finite nuclei, the Wood-Saxon density profile, and an expansion of the charge distribution as a sum of Gaussians are employed in this study. The nuclear density distribution $\rho_A$, effective baryon mass $m^*_B$, in-medium decay width $\Gamma^*_\Delta$, and in-medium coupling constants $f^*_{\pi NN}$ and $f^*_{\pi N \Delta}$ are analyzed as well as the total cross-section. The results show that the effective baryon mass and cross-sections in the medium decrease as density increases except for the $\Delta$ decay width, which increases as the density increases. The elastic $\pi A$ scattering at the tree-level Born approximation reproduces well the experimental data for $^4\mathrm{He}$ but overestimates for $^{12}\mathrm{C}$. Results for the in-medium $\Delta$ resonance and other findings in this work will be relevant for the relativistic heavy-ion collision experiments.
    \par


\cleardoubleemptypage
\pagenumbering{arabic}

\renewcommand{\thechapter}{\Roman{chapter}}
\renewcommand{\thesection}{\thechapter.\arabic{section}}
\renewcommand{\thesubsection}{\thesection.\arabic{subsection}}

\chapter{Introduction}

It is widely known that hadrons composed of the confined quarks that interact with the gluons in terms of quantum chromodynamics (QCD) are the generic degrees of freedom in strong interactions. Hadron properties such as the mass and decay width are expected to be modified in the medium, due to the partial restoration of chiral symmetry, which is one of the most important ingredients to understand the low-energy dynamics of the non-perturbative QCD~\cite{Brown:1995qt,Hatsuda:1998vb,Rapp:1999ej}. For instance, such in-medium modifications effects have been considered in studying the properties of neutron star (NS) and relativistic heavy-ion collision (RHIC) experiments at finite baryon density $\rho_B$, which attracts the attention of physics communities. Among the various types of hadrons in medium, the appearance of the $\Delta(1232,3/2^+)$ resonance becomes crucial inside NS~\cite{Raduta:2021xiz,Dexheimer:2021sxs,Schurhoff:2010ph,Sen:2018tms,Sahoo:2018xeu,Malfatti:2020onm,Sen:2019kxt,Zhu:2016mtc}, since the energies in the core of NS is far more sufficient to create the resonance, which is heavier than the neutron mass. However, the mass and the coupling constants of the $\Delta$(1232) resonance that couples to scalar and vector meson fields, and the density at which the $\Delta$ resonance appears in a nuclear medium still remain uncertain due to the lack of experimental data. Also, it was expected that the $\Delta N$ potential provoked the instability of $\Delta$-rich matter at some certain ranges of baryon densities, using the covariant density functional theory~\cite{Raduta:2021xiz}. The effects of the $\Delta$ resonance on the NS properties, such as the mass-radius (M-R) relations were investigated in Refs.~\cite{Schurhoff:2010ph,Sahoo:2018xeu,Sen:2019kxt}, where the hadron-quark phase transition was also taken into account~\cite{Sen:2018tms,Malfatti:2020onm}. Moreover, the external magnetic field induced to NS with the $\Delta$ resonance was investigated in Ref.~\cite{Dexheimer:2021sxs}. In Ref.~\cite{Zhu:2016mtc}, the authors studied the $\Delta$-resonance effects in density-dependent relativistic Hartree-Fock theory (DDRHFT) for NS. It is also worth noting that, as in Refs.~\cite{vanHees:2004sv,Rodriguez-Sanchez:2020hfh}, the medium modifications in the RHIC experiments were scrutinized.

Besides those interesting and challenging astrophysical and RHIC phenomena, recent theoretical studies on the elastic $\pi$-$N$ scattering via the Dyson-Schwinger equations (DSEs) method, the $\Delta$-resonance self-energy and $\pi N$ cross-section in the medium was studied at finite density and temperature in Ref.~\cite{Ghosh:2016hln}. In addition, in Ref.~\cite{Cui:2020fhr}, the authors studied the $\Delta$-resonance decays in isospin asymmetric nuclear matter (ANM) by considering the one-boson-exchange (OBE) model. They found that the in-medium cross-section of $N\pi \rightarrow \Delta$ is enhanced in the nuclear medium at around center mass energy $\sqrt{s} \simeq$ 1.11 GeV. Several studies on $\pi$N scattering were performed in the symmetric nuclear matter (SNM)~\cite{Cui:2020fhr,Mao:1998pr} and they mostly describe the nuclear medium in the relativistic mean-field (RMF) model and Walecka or quantum hadrodynamics (QHD) model in terms of the hadron degree of freedom. Motivated by those works, in the present work, I study the elastic $\pi N$ scattering in the SNM of the quark-meson coupling (QMC) model, which is built in terms of the quark degree of freedom~\cite{Guichon:1987jp,Saito:2005rv,Guichon:2018uew,Hutauruk:2018qku,Hutauruk:2019ipp,Saito:1996sf,Guichon:1995ue,Stone:2016qmi,Whittenbury:2013wma}.

Taking into account those impressive progresses for the in-medium modifications of the $\Delta$-resonance properties, in the present work, I focus to study the $\Delta$-resonance production through the elastic $\pi N$ scattering for the $I=3/2$ channel in free space as well as at finite baryon densities ($\rho_B\ne 0$). Here, in my first attempt and study, in the SNM, which is relevant for RHIC, as the target nucleon is set to be surrounded by other nucleons uniformly (homogeneous matter) and the incident and final pion in the nuclear medium is assumed to be the same as those in a vacuum as indicated in the pionic atom experiment. Note that the $I=3/2$ channel $(\pi^+p\to\pi^+ p)$ can eliminate the effects of the nucleon-resonance contributions in the $s-$channel, resulting in clear signals of the $\Delta$ resonance by reducing theoretical background uncertainties. In this work, the scattering observable is computed using the effective Lagrangian method at the tree-level Born approximation in a fully relativistic manner. Phenomenological form factors are taken into account for the spatial extensions of the hadrons involved. In order to consider the medium modifications of the hadrons, I make use of the QMC model for the SNM~\cite{Saito:2005rv,Guichon:2018uew,Hutauruk:2018qku,Hutauruk:2019ipp}. This QMC model has been widely and successfully applied in the nuclear matter~\cite{Saito:2005rv,Guichon:2018uew}, hadron structure~\cite{Hutauruk:2018qku,Hutauruk:2019ipp}, finite nuclei~\cite{Saito:1996sf,Guichon:1995ue,Stone:2016qmi}, and properties of neutron star~\cite{Whittenbury:2013wma,Hutauruk:2020mhl}. In the QMC model, the wave function of the quark inside the nucleon (bag), as well as that of the nucleon, is solved self-consistently via scalar ($\sigma$) and vector ($\omega$) fields in nuclear matter. The coupling constants and other related quantities are then determined by reproducing the binding energy $E_B =$  $-$15.7 MeV at a normal density to guarantee the model stability. The linear-density approximation is employed for the full decay width for the $\Delta$-resonance in medium $\Gamma^*_\Delta$~\cite{Larionov:2003av,Hirata:1978wp,Oset:1987re} which also depends on energies. This approach has been used in many calculations of the in-medium scattering~\cite{Jido:2008bk} as well as in the determination of the decay width and chiral condensate in pionic atom~\cite{Kienle:2004hq}. In addition, in Ref.~\cite{Ericson:1988gk}, they argued that the higher order density (nonlinear) approximation, which was for the first time derived in the SNM, could be included in the elastic $\pi^+ p$ scattering that is parameterized via the scattering length. However, the experimental data confirmations are still required to establish the usefulness and urgencies of this nonlinear terms in the medium. Therefore, this remains for future works.

As a result, it is found that the experimental data of the total and differential cross-sections (TCS and DCS) for the elastic $\pi^+p$ scattering in vacuum are qualitatively well reproduced by fitting the model parameters appropriately, showing the $\Delta$-resonance domination. This obviously justifies that our theory prediction result on the elastic $\pi N$ scattering in free space is quite reasonable. As for the baryon mass modifications with respect to the baryon density via the QMC model, it turns out that the in-medium masses $M^*_{N,\Delta}$ decrease by about $10\%$ at $\rho_B\approx\rho_0$ in comparison to their vacuum values, where $\rho_0$ denotes the normal nuclear density. The in-medium decay width $\Gamma^*_\Delta$ increases as a function of $\rho_B$, indicating the imaginary potential of $\Delta$ resonance modifies at finite density. Using these in-medium masses and decay width, it is respectively shown that TCS becomes wider (broadening), and the $\Delta$-resonance peak gets diminished obviously with respect to $\rho_B$. The peak position moves gradually to the higher-energy region as $\rho_B$ increases. In contrast, it is interesting to note that the background (BKG) contributions, except for the $\Delta$ resonance in the $s-$channel, are insensitive to the density and hardly increase with respect to $\rho_B$. The angular dependence from DCS is almost dominated by the $\Delta$ resonance and the strength of DCS decreases as a function of $\rho_B$ as expected from TCS. However, as the energy increases beyond $\sqrt{s}\approx1.5$ GeV, the $\Delta$-resonance contribution is reduced and DCS almost remains the same as that for vacuum. The proton-spin asymmetry ($P$) is also computed between the target and recoil proton spin states. Hence, as for the $\Delta$-resonance region, $P$ does not change much due to the baryon density, and vice versa for the higher-energy region. It is verified that the forward-scattering differential cross-section $d\sigma/dt$ shows similar tendencies to those of DCS.

Next, extending the method adopted in the $\pi N$ scattering, I investigate the elastic $\pi A$ scattering by considering medium effects using the Eikonal Glauber model for the light nuclei, in particular for the $^4\mathrm{He}$ and $^{12}\mathrm{C}$. The analysis of the elastic $\pi A$ scattering process consists of the $\pi^+ n$ and the $\pi^+ p$ channels. The total cross-section for the elastic $\pi^+ n$ scattering is found to be smaller about nine times than that for the $\pi^+p$ channel, because of the isospin factor difference of the $s-$channel contribution. This result is consistent with other model calculations. The concept of the Glauber model begins by describing the hadronic interaction and integrating the cross-section for the reaction over the entire interaction space that is represented in the nuclear density distribution $\rho_A$. The parameters in the $\rho_A$ are determined by fitting $\rho_A$ with the experimental data~\cite{DEJAGER:1974479, osti:6477756}. With these parameters of $\rho_A$, it must satisfy the normalization to the nuclear mass number $A$, by integrating $\rho_A$ over the volume of the nucleus. The Glauber model has been successfully used in nuclear reactions for several decades even until now~\cite{Miller:2007}. In this study, I choose two kinds of density profiles namely, the Wood-Saxon density profile and the density distribution given by the Sum of Gaussian (SOG). In the common calculation, the Glauber model is used for the $NN$ interaction to describe the $A$-$A$ scattering. However, in this study, the Glauber model is used for the elastic $\pi^+ N$ scattering to reproduce the total cross-section of the elastic $\pi^+ A$ scattering.

Unlike $\pi^+ N$ scattering in the nuclear matter case, at finite nuclei, the density distribution depends on the radial position of the nucleons ($r$) in the nucleus as well as the radius ($R$) of the finite nuclei. Using this density distribution of finite nuclei, the expression for the effective baryon masses can be reconstructed as follows: $M^*_B=M_B+C_1\cdot\rho_A+C_2\cdot\rho^2_A$, where $C_1$ and $C_2$ are the constant coefficients determined by adjusting to the effective mass calculated from the QMC model. Thus, in the present study, other physical variables are also calculated in finite nuclei such as the $\Delta$ decay width $\Gamma^*_\Delta$, the coupling constants of $f^*_{\pi N \Delta}$ and $f^*_{\pi NN}$, and the pion decay constant of $f^*_\pi$, where the similar behavior of the pion decay constant in the nuclear medium found in Refs.~\cite{Kienle:2004hq,Lu:2001mf} are also expected.

As a final goal of this analysis, using a similar procedure as the calculation of the density-dependent TCS of the elastic $\pi N$ scattering $\sigma_{\pi^+ N}(\sqrt{s^*},\rho_A)$ where the density effect is calculated in the QMC model, I calculate the total cross-section of elastic $\pi^+$--$^4\mathrm{He}$ and $\pi^+$--$^{12}\mathrm{C}$ scattering in the Glauber model approach with two phenomenological forms of $\rho_A$ to analyze the cross-section from a more diverse perspective. In this calculation, I consider the processes of two channels: $\pi^+p$ and $\pi^+n$ by taking the ratio of the proton or neutron number divided by the nuclear mass number $A$ into account. In the Glauber approach, the parameters like $\rho^{(n,p)}_A$ and $T^{(n,p)}_A$ are considered, where the superscripts of $(n,p)$ indicate the neutron and proton constituents, respectively. From this study, it can be concluded that the results for the  TCS of $^4\mathrm{He}$ are quite good for describing the data, whereas, for the $^{12}\mathrm{C}$ case, the TCS result rather overestimates the data~\cite{Shcherbakov:1976,Ashery:1981}.

The content of this thesis is organized as follows. In Sec.~II, I briefly introduce the elastic $\pi N$ scattering with medium effects and the theoretical framework starting with the effective Lagrangian approach, the formula for the scattering amplitudes of the corresponding channels, and the description of the QMC model for SNM. In Sec.~III, I explain the elastic $\pi A$ scattering by considering the medium effects and the related physical observables relating to the Eikonal Glauber model in finite nuclei. The final section is devoted to the summary and future perspectives.

\chapter{The elastic $\pi N$ scattering with medium effects}
\section{Effective Lagrangians for the elastic scattering process}
\begin{figure*}[t]
  \centering
  \includegraphics[width=1.05\textwidth]{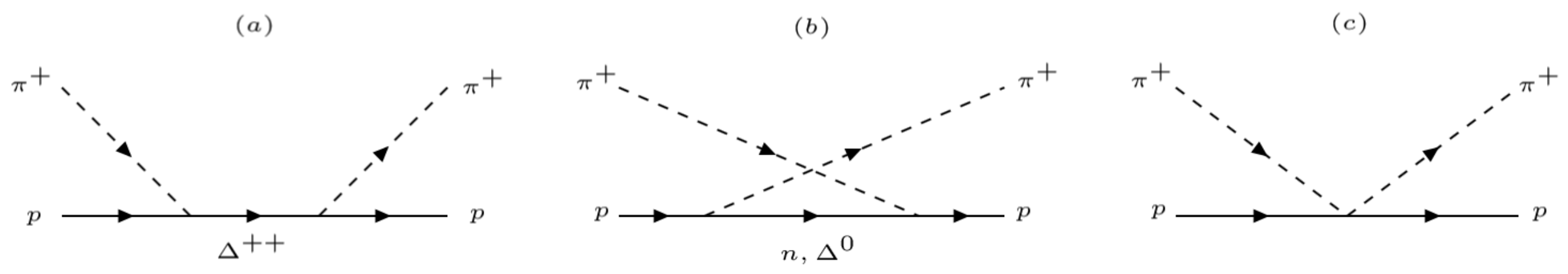}
  \caption{ \label{fig1a} Relevant Feynman diagrams contribute to the $\pi^+ p$ elastic scattering for (a) the $\Delta^{++}$ pole diagram in the $s$-channel, (b) the neutron and $\Delta^0$ intermediate diagrams in the $u$-channel, and (c) Weinberg-Tomozawa (WT) contact interaction. The solid and dashed lines represent the baryon (nucleon and $\Delta$) and pion, respectively.}
\end{figure*}
In this section, I briefly present the effective Lagrangian approach for the elastic scattering process. At the tree-level Born approximation, the relevant Feynman diagrams are depicted in Fig.~\ref{fig1a}: (a) the $\Delta^{++}$ pole diagram in the $s$-channel, (b) the neutron and $\Delta^0$ intermediate-state diagrams in the $u$-channel, and (c) Weinberg-Tomozawa (WT) contact interaction. The solid and dashed lines represent the baryon (nucleon and $\Delta$) and pion, respectively. For the interaction vertices, I introduce the effective Lagrangians for the corresponding channels. The effective Lagrangian for $\pi N \Delta$ vertex for the diagrams (a) and (b) in Fig.~\ref{fig1a} can be given in terms of the Rarita-Schwinger field formalism~\cite{Gasparyan:2003fp}
\begin{eqnarray}
  \label{eq1}
  \mathcal{L}_{\pi N \Delta}=\left(\frac{f_{\pi N\Delta}}{M_\pi}\right)\bar{\Delta}^{\mu}\partial_\mu (\bm{S}\cdot\bm{\pi})N + \mathrm{h.c},
\end{eqnarray}
where $f_{\pi N\Delta}$, and $M_\pi$ are respectively the coupling constant of the $\pi N\Delta$, and the pion mass. $\Delta_\mu $ is the $\Delta$ baryon field and $\bm{S}$ is the isospin transition operator between the isospin 1/2 and 3/2 fields. Thus, the corresponding scattering amplitudes are calculated as follows:
\begin{eqnarray}
\begin{split}
  \label{eq2} 
  i\mathcal{M}_s^{\Delta^{++}} = -\frac{f_{\pi N\Delta}^2}{M_\pi ^2}\bar{u}(p')k'_\mu G^{\mu\nu}(p+k)k_\nu u(p),\\
  i\mathcal{M}_u^{\Delta^{0}}= -\frac{f_{\pi N\Delta}^2}{3M_\pi ^2}\bar{u}(p')k_\mu G^{\mu\nu}(p-k')k'_\nu u(p),
\end{split}
\end{eqnarray}
where the factor of $\frac{1}{3}$ in the $u$-channel amplitude in Eq.~(\ref{eq2}) comes from the isospin factor. The $\Delta$ baryon propagator $G^{\mu\nu}(q)$ in Eq.~(\ref{eq2}) can be defined by
\begin{eqnarray}
  \label{eq3}
  G^{\mu\nu}(q) = i\frac{(\rlap{/}{q}+M_\Delta)}{q^2-M^2_\Delta + iM_\Delta\Gamma_\Delta}
 \left[-g^{\mu\nu}+\frac{1}{3}\gamma^\mu\gamma^\nu+\frac{2q^\mu q^\nu}{3M^2_\Delta}-\frac{q^\mu\gamma^\nu-q^\nu\gamma^\mu}{3M_\Delta}\right],
\end{eqnarray}
where $M_\Delta$, and $\Gamma_\Delta$ indicate the $\Delta$ baryon mass and the decay width for the $\Delta$-resonance, individually.

The effective Lagrangian for the $\pi NN$ interaction is then given by~\cite{Gasparyan:2003fp}
\begin{eqnarray}
  \label{eq4}
  \mathcal{L}_{\pi NN} = -\left(\frac{f_{\pi NN}}{M_\pi} \right)\bar{N}\gamma_5 \rlap{/}{\partial}( \bm{\tau}\cdot\bm{\pi}) N+\mathrm{h.c.},
\end{eqnarray}
where $f_{\pi NN}$ is the pseudovector $\pi NN$ coupling constant (dimensionless), whereas the $N$ and $\bm{\pi}$ denote the nucleon and pion fields, respectively. $\bm{\tau} = \sigma /2$ is the isospin operator of the nucleon, where $\sigma$ stands for Pauli isospin matrices. Using the effective Lagrangian in Eq.~(\ref{eq4}), I straightforwardly calculate the scattering amplitude for the Feynman diagram (b) in Fig.~\ref{fig1a}, resulting in
\begin{eqnarray}
  \label{eq5}
  i\mathcal{M}_u^n = -i \mathcal{I}_{\pi N} \left( \frac{f^2 _{\pi NN}}{M_\pi ^2} \right)\bar{u}(p')\gamma_5\rlap{/}{k}\frac{\rlap{/}{p}-\rlap{/}{k}'+M_N}{(p-k')^2 -M_N^2}\gamma_5\rlap{/}{k}'u(p),
\end{eqnarray}
where $\mathcal{I}_{\pi N}=2/3$. $M_N$, $p$, and $k$ are the nucleon mass in a vacuum, the initial four-momenta for the nucleon and pion, respectively, while $p'$ and $k'$ stand for the final four-momenta for them.

The effective Lagrangian for the Weinberg-Tomozawa (WT) contact interaction for the diagram (c) in Fig.~\ref{fig1a} reads~\cite{Weinberg:1966kf,Tomozawa:1966jm,Hyodo:2011ur}
\begin{eqnarray}
  \label{eq6}
  \mathcal{L}_{\mathrm{WT}} &=& i\frac{C_{\pi N}}{4f^2_\pi}\bar{N}\left[\pi(\rlap{/}{\partial}\pi^\dagger)-(\rlap{/}{\partial}\pi)\pi^\dagger\right]N.
\end{eqnarray}
Similarly, the scattering amplitude for the contact term is calculated, using the effective Lagrangian in Eq.~(\ref{eq6}) as follows:
\begin{eqnarray}
  \label{eq7}
  i\mathcal{M}_{\mathrm{WT}} &=& -i \frac{C_{\pi N}}{4f^2_\pi}\bar{u}(p')(\rlap{/}{k}+\rlap{/}{k}')u(p).
\end{eqnarray}
The pion decay constant $f_\pi = 93.2$ MeV is used in this work and $C_{\pi N}=-1$ is for the isospin state of $I=\frac{3}{2}$~\cite{Sun:2019nyo}.

Then, I obtain the spin-averaged differential cross-section summing all the obtained scattering amplitudes in Eqs.~(\ref{eq2}), (\ref{eq5}), and (\ref{eq7}):
\begin{eqnarray}
  \label{eq8}
  \left(\frac{\partial\sigma}{\partial\Omega}\right)_{\pi^+p} &=& \frac{1}{64\pi^{2}s}\frac{|\textbf{p}|}{|\textbf{k}|}
  \frac{1}{2}\underset{\mathrm{spin}}{\sum}\left|\mathcal{M}^\mathrm{total}_{\pi^+p}\right|^2,
\end{eqnarray}
where $\textbf{k}$ and $\textbf{p}$ are the initial three-momenta for the $\pi^+$ and $p$, respectively. The total amplitude with the phenomenological form factors is written as 
\begin{eqnarray}
  \label{eq9}
  \mathcal{M}^\mathrm{total}_{\pi^+p}= i \mathcal{M}_s^{\Delta^{++}}F_s^{\Delta^{++}}
  +i \mathcal{M}_u^{\Delta^{0}}F_u^{\Delta^{0}}
  +i \mathcal{M}_u^nF_u^n
  +i \mathcal{M}_{\mathrm{WT}}F_{\mathrm{WT}}, 
\end{eqnarray}
and the form factors for each channel are defined by
\begin{eqnarray}
  \label{eq10}
  F^h_{x}(x) &=& \frac{\Lambda^{4}}{\Lambda^{4}+(x-M^{2}_h)^{2}},
\end{eqnarray}
where $x$ and $h$ denote the Mandelstam variable and corresponding hadrons, respectively. As for the WT contact amplitude, it is assumed that $F_\mathrm{WT}=F^\rho_t$ for brevity, since the WT reproduces the exchange of the vector meson $\sim\rho(770)$ in the low-energy limit~\cite{Inoue:2001ip}. In this work, I set $\Lambda$ to be 800 MeV except for the WT channel, which has $\Lambda=450$ MeV, to well reproduce the data as will be shown in Section III.

\section{Nuclear medium effects from the QMC model}
In this section, I present the calculation of the effective $M_N^*$ and $M_\Delta^*$ masses at finite density via the QMC model for calculating in-medium cross-sections. Besides the change of baryon masses at finite density, the momenta of the nucleon and delta baryon are expected to be changed in the nuclear medium, due to the effect of the vector potential. In the QMC model, the nuclear medium effects come from the self-consistent exchange of the scalar $\sigma$ and vector $\omega$ meson fields which are directly coupled to the confined valence quark in baryon~\cite{Guichon:1987jp,Guichon:2018uew,Hutauruk:2018qku,Hutauruk:2019ipp}. The QMC effective Lagrangian in the symmetric nuclear matter (SNM) is expressed as
\begin{eqnarray}
  \label{eq11}
  \mathcal{L}_\mathrm{QMC} &=& \sum_{B= N,\Delta_\nu} \bar{B} [i\gamma \cdot M_B^*(\sigma) - g_\omega \omega^\mu \gamma_\mu]B+\mathcal{L}_{M},
\end{eqnarray}
where the $g_\omega$ is the $\omega$-$N$ coupling constant, $\psi_B$ is the baryon field and $M_B^* (\sigma)$ is the effective baryon mass. Thus, the free meson Lagrangian density in Eq.~(\ref{eq11}) is given by
\begin{eqnarray}
  \label{eq12}
  \mathcal{L}_M = \frac{1}{2} (\partial_\mu\sigma \partial^\mu \sigma - m_\sigma \sigma^2) - \frac{1}{2} \partial_\mu \omega_\nu(\partial^\mu \omega^\nu - \partial^\nu \omega^\mu)
 +\frac{1}{2} m_\omega^2 \omega^\mu \omega_\mu.
\end{eqnarray}
The nucleon Fermi momentum $k_F$, baryon density $\rho_B$, and scalar density $\rho_s$ in SNM at the mean-field approximation are given by
\begin{eqnarray}
\begin{split}
  \rho_B &= \frac{4}{(2\pi)^3} \int d\mathbf{k} \theta(k_F - |\mathbf{k}|) = \frac{2k_F^3}{3\pi^2},\\
  \rho_s &= \frac{4}{(2\pi)^3} \int d\mathbf{k} \theta (k_F - |\mathbf{k}|) \frac{M_N^* (\sigma)}{\sqrt{M_N^{*2}(\sigma) + \mathbf{k}^2}}.
\end{split}
\end{eqnarray}

In the QMC model, the Dirac equations for the light quarks ($q = u,d$) in the assumption of the non-overlapping MIT bags with the collection of nucleons treated as nuclear matter are defined by
\begin{eqnarray}
\begin{split} \label{eqintro5}
   \left[ i \gamma \cdot \partial_{x} - \left( m_q - V_{\sigma}^{q} \right) \mp \gamma^{0} \left( 
    V_{\omega}^{q} + \frac{1}{2} V_{\rho}^{q} \right) \right] \left( \begin{array}{c} \psi_u(x)  \\ 
    \psi_{\bar{u}}(x) \\ \end{array} \right) = 0, \\
   \left[ i \gamma \cdot \partial_{x} - \left( m_q - V_{\sigma}^{q} \right) \mp \gamma^{0} \left( 
    V_{\omega}^{q} - \frac{1}{2} V_{\rho}^{q} \right) \right] \left( \begin{array}{c} \psi_d(x)  \\ 
    \psi_{\bar{d}}(x) \\ \end{array} \right) = 0,
\end{split}
\end{eqnarray}
where the effective quark mass $m_q^{*}$ is defined by
\begin{align}
  \label{eqintro5a}
  m_q^{*} & \equiv m_q - V_{\sigma}^{q}.
\end{align}
Here, $m_q$ is the light-quark current mass ($q=u,d$), and $V_\sigma^{q}$ the scalar potential. In SNM with the Hartree approximation, the isospin dependent $\rho$-meson mean field yields $V_\rho^{q} =$ 0 in Eq.~(\ref{eqintro5}). Thus, the scalar- and vector-mean field potentials in the nuclear matter are defined as
\begin{equation}
  \label{eq:potqqmc}
  V_{\sigma}^{q}  \equiv g_{\sigma}^{q} \sigma = g_{\sigma}^{q} \langle\sigma\rangle, \qquad
  V_{\omega}^{q} \equiv g_{\omega}^{q} \omega = g_{\omega}^{q} \delta^{\mu 0} \langle\omega^{\mu}\rangle.
\end{equation}

The in-medium bag radius of the hadron $R_h^{*}$ is calculated by considering the hadron mass stability condition against the variation of the bag radius, and the eigen-energies in units of $1/R_h^{*}$ are
\begin{equation}
  \label{eq:pionmed9}
  \left( \begin{array}{c}
    \epsilon_u \\
    \epsilon_{\bar{u}}
  \end{array} \right)
  = \Omega_q^* \pm R_h^* \left(  V^q_\omega + \frac{1}{2} V^q_\rho \right),\,\,\,
  \left( \begin{array}{c} \epsilon_d \\
    \epsilon_{\bar{d}}
  \end{array} \right)
 = \Omega_q^* \pm R_h^* \left(
  V^q_\omega
  - \frac{1}{2} V^q_\rho \right).
  \end{equation}
The effective hadron mass in nuclear medium $M_B^{*}$ is calculated by 
\begin{equation}
M_B^{*} = \sum_{j = q, \bar{q}} \frac{n_j \Omega_j^{*} -z_B^{}}{R^{*}_B} + \frac{4}{3} \pi  R_B^{* 3} B,\,\,\,
\Omega^{*}_q = \Omega^{*}_{\bar{q}} = \sqrt{x_q^2 + \left(R_B^{*} m_q^{*} \right)^2},
\end{equation}
while the in-medium bag radius is determined by the condition $\partial M_B^{*}/\partial R_B\vert_{R_B = R_B^{*}} = 0$. $z_B = 3.295$ is related to the bag-model quantity determined by the hadron mass in a vacuum, the bag pressure $B = \mathrm{ (170\, MeV)}^4$ is fixed using the standard input of the QMC model for the nucleon in a vacuum, $R_N = 0.8$ fm, and $m_q = 5$ MeV. For the quarks inside the bag of the hadron $h$, the lowest positive eigenfunctions of the bag satisfy the boundary condition at the bag surface, $j_0 (x_{q}) =  \beta_{q}\, j_1 (x_{q})$, where $\beta_{q} = \sqrt{\Omega^*_{q} -(m^*_{q} R^*_B)/\Omega^*_{q} + (m^*_{q} R^*_B)}$ with $j_0$ and $j_1$ being the spherical Bessel functions. The scalar $\sigma$ and vector $\omega$ meson mean fields at the nucleon level can be related as
\begin{eqnarray}
  \label{eq:pionmed11}
  \omega = \frac{g_\omega}{m_\omega^2}\rho_B,\,\,\,\quad
  \label{eq:sigma}
  \sigma = \frac{g_\sigma C_N (\sigma)}{m_\sigma^2}\rho_s, 
\end{eqnarray}
where $C_N (\sigma)$ is defined as
\begin{eqnarray}
  C_N (\sigma) = \frac{-1}{g_\sigma (\sigma =0)} \left[ \frac{\partial M_N^{*} (\sigma )}{\partial 
      \sigma } \right].
\end{eqnarray}
Note that the value of $C_N (\sigma)$ is unity for the point-like nucleon. Both $C_N (\sigma)$ and $g_\sigma (\sigma)$ originate from the novel properties of the QMC model and contain the dynamics of quark structures of the nucleon. By solving the scalar $\sigma$ mean field in Eq.~(\ref{eq:sigma}) self-consistently, the energy per particle is given by
\begin{eqnarray}
  \label{eq:pionmed12}
  \frac{E^\mathrm{tot}}{A} =\frac{4}{(2\pi)^3 \rho_B} \int d\bm{k} \, \theta (k_F - | \bm{k} |) 
  \sqrt{M_N^{*2} (\sigma) + \bm{k}^2}
  + \frac{m_\sigma^2 \sigma^2}{2\rho_B} + \frac{g_\omega^2 \rho_B}{2 m_\omega^2}.
\end{eqnarray}
All the coupling constants in Eq.~(\ref{eq:pionmed12}) are determined by fitting to the binding energy of $-15.7~\textrm{MeV}$ at the saturation density $\rho_0 =$ 0.15 fm$^{-3}$ for SNM. Then it is obtained that $g_\sigma^2 /4\pi =$ 5.393, $g_\omega^2/4\pi =$ 5.304, $M_N^* =$ 754.6, and $K=$ 279.3.
\begin{figure}[t]
  \centering
  \includegraphics[width=0.5\textwidth]{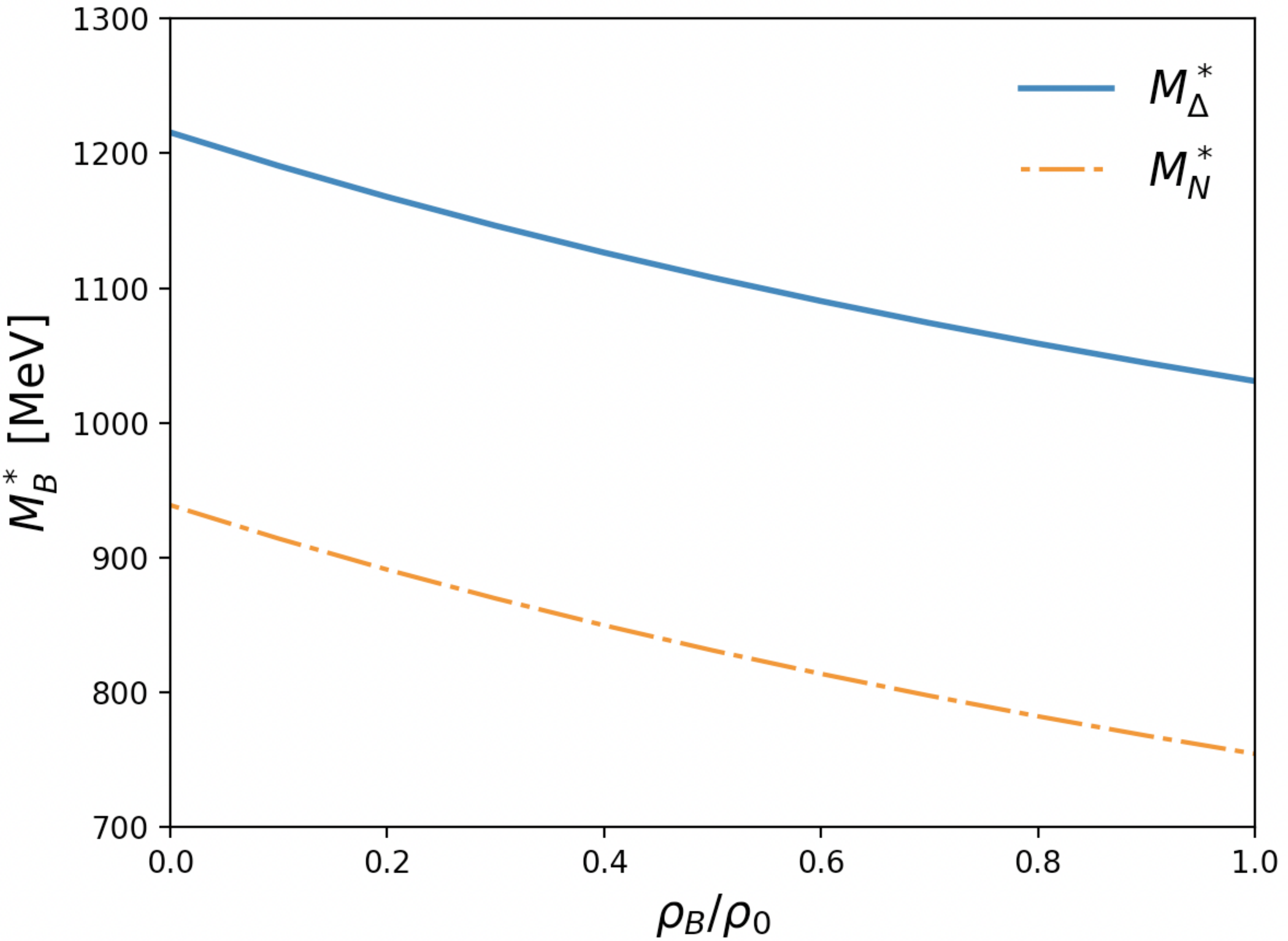}
  \caption{ \label{fig1b} Effective baryon masses for the nucleon (orange dot-dashed line) and $\Delta$ resonance (blue solid line) as functions of the baryon density $\rho_B / \rho_0$.}
\end{figure}

Using the obtained coupling constants, I obtain the effective baryon masses for $N$ and $\Delta$ resonance as shown in Fig.~\ref{fig1b}. The numerical results show that the effective nucleon mass decreases as the baryon density increases as expected. This behavior is followed by the effective delta mass which is consistent with the model calculation in Ref.~\cite{Motta:2019ywl}. These effective baryon masses will be used as inputs to the $\pi^{+}p$ cross-section in the medium. Note that, as already explained before, for the pion, I do not consider its medium modifications, since the incident and final pion in the nuclear medium are considered the same as those in a vacuum. Next, I will calculate the TCS and DCS for elastic $\pi^+p$ scattering in free space and nuclear medium.

\section{Numerical result: elastic $\pi^{+}p$ scattering at $\rho_B=0$}
Here, I present the numerical results for the various scattering observables for the $\pi^{+}p$ elastic process in a vacuum $(\rho_B=0)$~\cite{Han:2021zrh}. In my calculation, I employ $f_{\pi NN} = 0.989$ obtained from the Nijmegen potential~\cite{Gasparyan:2003fp} and $f_{\pi N \Delta} = 2.127$ given by the experiment~\cite{Janssen:1996kx}. In general, the vacuum full decay width of the $\Delta$ resonance, i.e., $\Gamma_\Delta$ depends on energies as found in Ref.~\cite{Larionov:2003av}. Also, a similar expression for the vacuum decay width can be found in Refs.~\cite{Chiang:2001as,Jain:1996bt}. Therefore, I adopt this parameterized vacuum decay width formula in this work, which is defined as
\begin{eqnarray}
  \label{eqN11}
  \Gamma_\Delta \left(\sqrt{s},M_{\pi,N,\Delta}\right)=\Gamma^0_\Delta \Big[ \frac{q \left(M_N,M_\pi,\sqrt{s}\right)}{q \left(M_N,M_\pi,M_\Delta \right)}\Big]^3 \frac{M_\Delta}{\sqrt{s}}\frac{\beta^2_0+q^2 \left(M_N,M_\pi,M_\Delta \right)}{\beta^2_0+q^2 \left(M_N,M_\pi,\sqrt{s}\right)},
\end{eqnarray}
where $\sqrt{s}$ and $\Gamma_{\Delta}^0$ are respectively the center-of-mass (cm) energy and Breit-Wigner width, and the $\beta_0 = 200$ MeV in Eq.~(\ref{eqN11}) is a cut-off parameter~\cite{Larionov:2003av}. The $q$ denotes the three-momenta of the intermediate particle, i.e., the $\Delta$ resonance:
\begin{eqnarray}
  \label{eqN13}
  q \left(M_b,M_c,M_a\right) &=& \frac{\sqrt{\left(M_a^2 + M^2_b - M^2_c\right)^2-4M_a^2 M_b^2}}{2M_a},
\end{eqnarray}
where $M_b$, $M_c$ denote respectively the mass of outgoing particles, and $M_a$ indicates the mass of the intermediate particle.

First, using the vacuum decay width $\Gamma_\Delta$ in Eq.~(\ref{eqN11}), I calculate the vacuum total cross-section (TCS) as a function of $\textbf{p}_{\textrm{Lab}}$ and the angular-dependent differential cross-section (DCS$_\theta$) for the elastic $\pi^{+}p$ scattering as a function of $\cos \theta$ in a vacuum. I obtain $M_\Delta = 1215.5$ MeV and $\Gamma^0_\Delta = 94.0$ MeV, which are slightly different from their PDG values, to reproduce the experimental data. However it is consistent with other calculations~\cite{LopezCastro:2000ep,Mariano:2012zz}.
\begin{figure}[t] 
  \stackinset{r}{4.7cm}{t}{0.4cm}{(a)}{\includegraphics[width=0.51\textwidth]{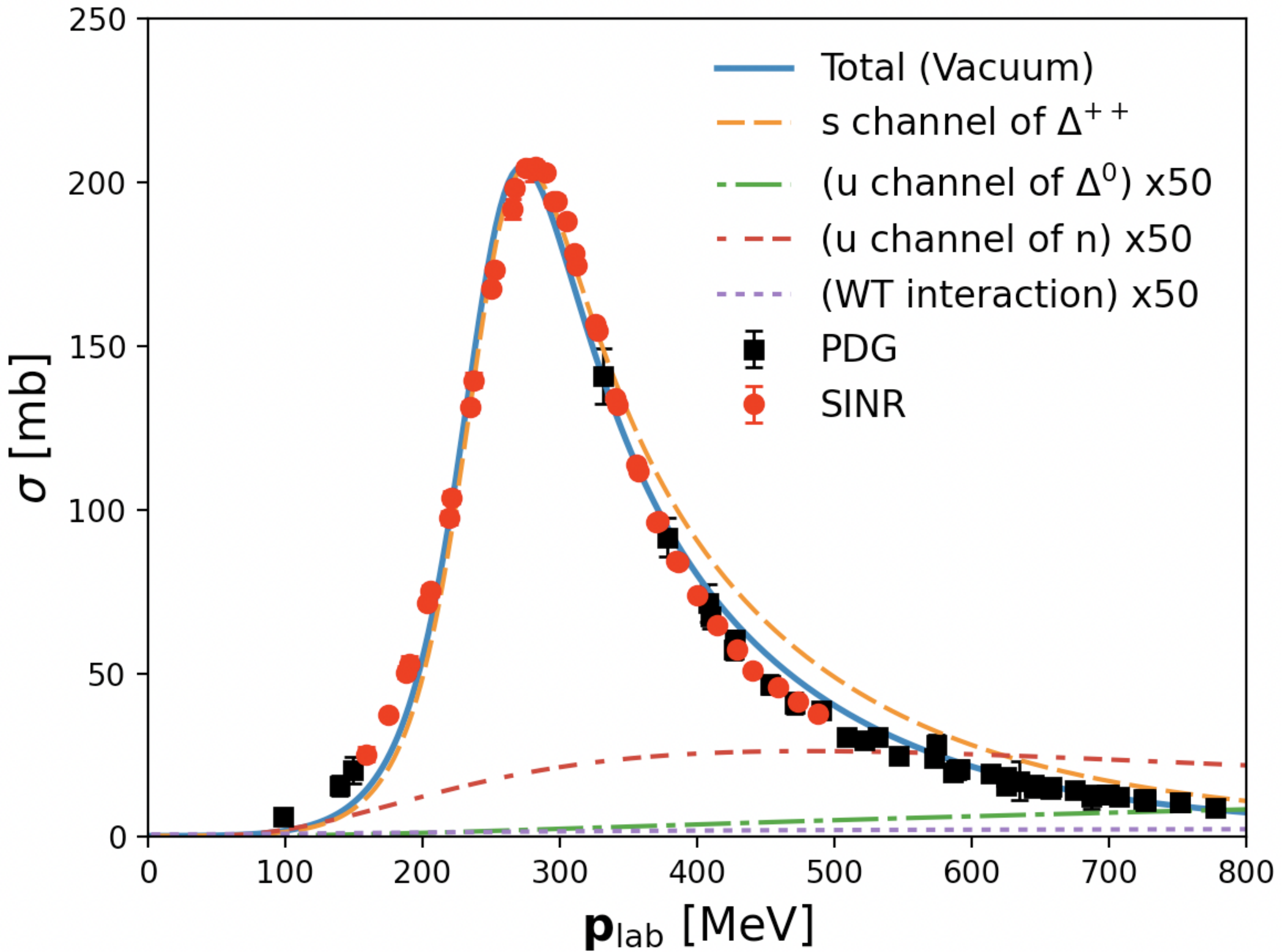}}
  \stackinset{r}{1cm}{t}{0.4cm}{(b)}{\includegraphics[width=0.51\textwidth]{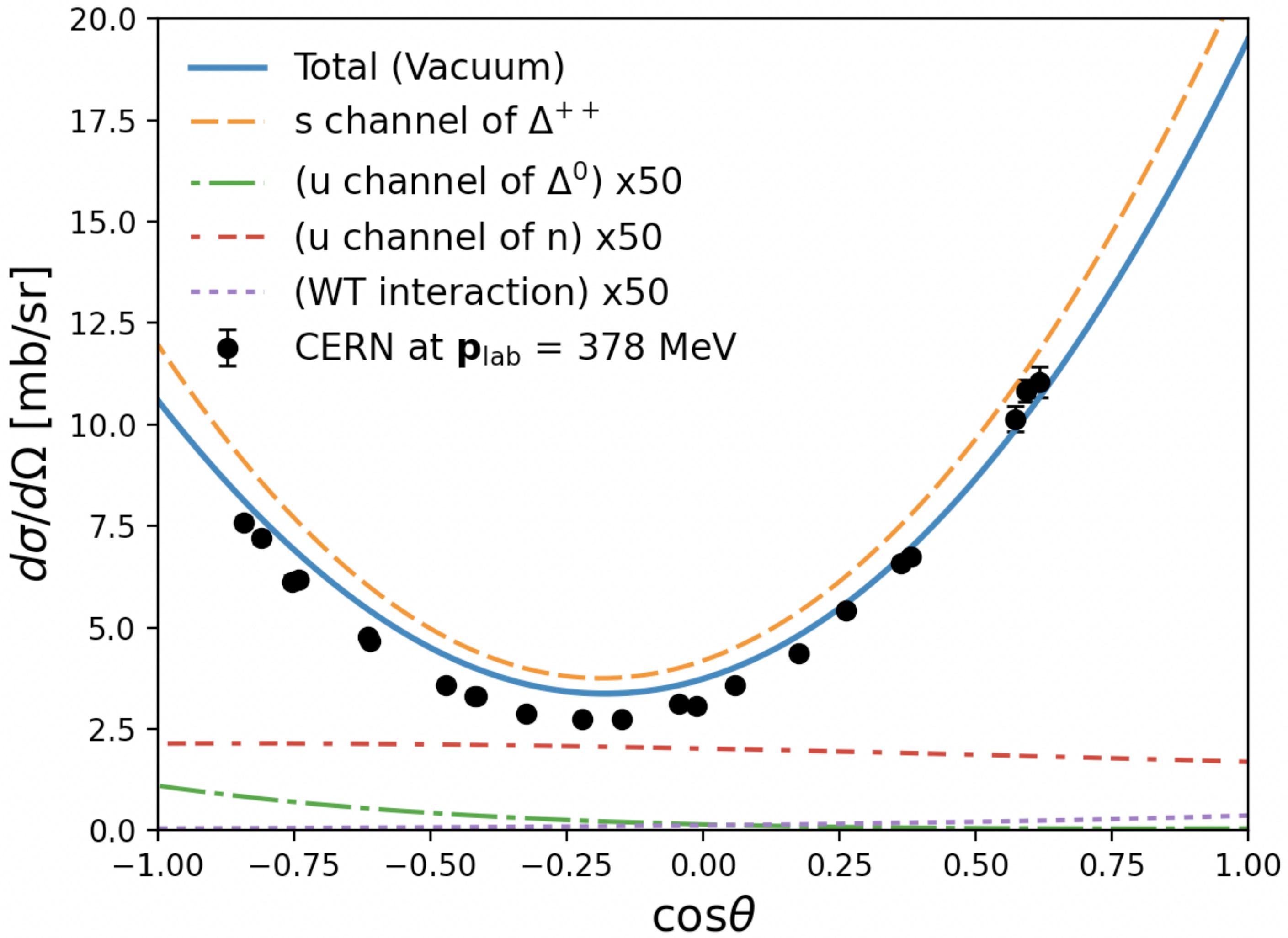}}
  \caption{ \label{fig1} (a) Total cross-section (TCS) of the elastic $\pi^{+}p$ scattering as a function of $\textbf{p}_\mathrm{lab}$ and (b) angular-dependent differential cross-section (DCS$_{\theta}$) as a function of $\cos\theta$ for $\rho_B=0$ (vacuum). Here, I employ $M_\Delta =$ 1215.5 MeV and  $\Gamma^0_\Delta =$ 94.0 MeV. The experimental data for TCS are taken from the Particle Data Group (PDG) compilation~\cite{PDG2020} (square) and Swiss Institute of Nuclear Research (SINR)~\cite{Pedroni:1978it} (circle). Those for DCS$_{\theta}$ are taken from Ref.~\cite{Bussey:1973gz} (circle). Each contribution is separately shown as well.}
\end{figure}

Panel (a) of Fig.~\ref{fig1} shows the comparison between TCS (blue solid line) and the experimental data, which are taken from the Particle data group (PDG) compilation~\cite{PDG2020} (square) and Swiss Institute of Nuclear Research (SINR)~\cite{Pedroni:1978it} (circle). It turns out that the numerical results for TCS are in good agreement with the experimental data. Also, each channel contribution for TCS is given separately. It can be clearly seen that the dominant contribution of $\Delta$(1232) in the $s-$channel is crucial to produce the peak in the cross-section in the vicinity of the threshold.

Panel (b) of Fig.~\ref{fig1} shows the numerical results for the differential cross-section as a function of $\cos\theta$ (DCS$_{\theta}$), in which $\theta$ denotes the cm-frame angle for the scattered pion. This result fits relatively well with the existing data from CERN at $\textbf{p}_{\textrm{lab}} =$ 378 MeV (circle)~\cite{Bussey:1973gz}, showing in a typical $p$-wave scattering, $\propto\textbf{k}'\cdot\textbf{k}$, due to the $\Delta$ resonance but the curve shape seems like a $d$-wave. In the backward-scattering region, however, the numerical result slightly overestimates the data. It is expected that other baryon resonances for the $u$-channel neglected in the present work for brevity are necessary to improve the result, and those improvements will be taken into account in future works.

\section{Numerical result: elastic $\pi^{+}p$ scattering at $\rho_B\ne0$}
\begin{figure}[t]
\centering
\includegraphics[width=0.5\textwidth]{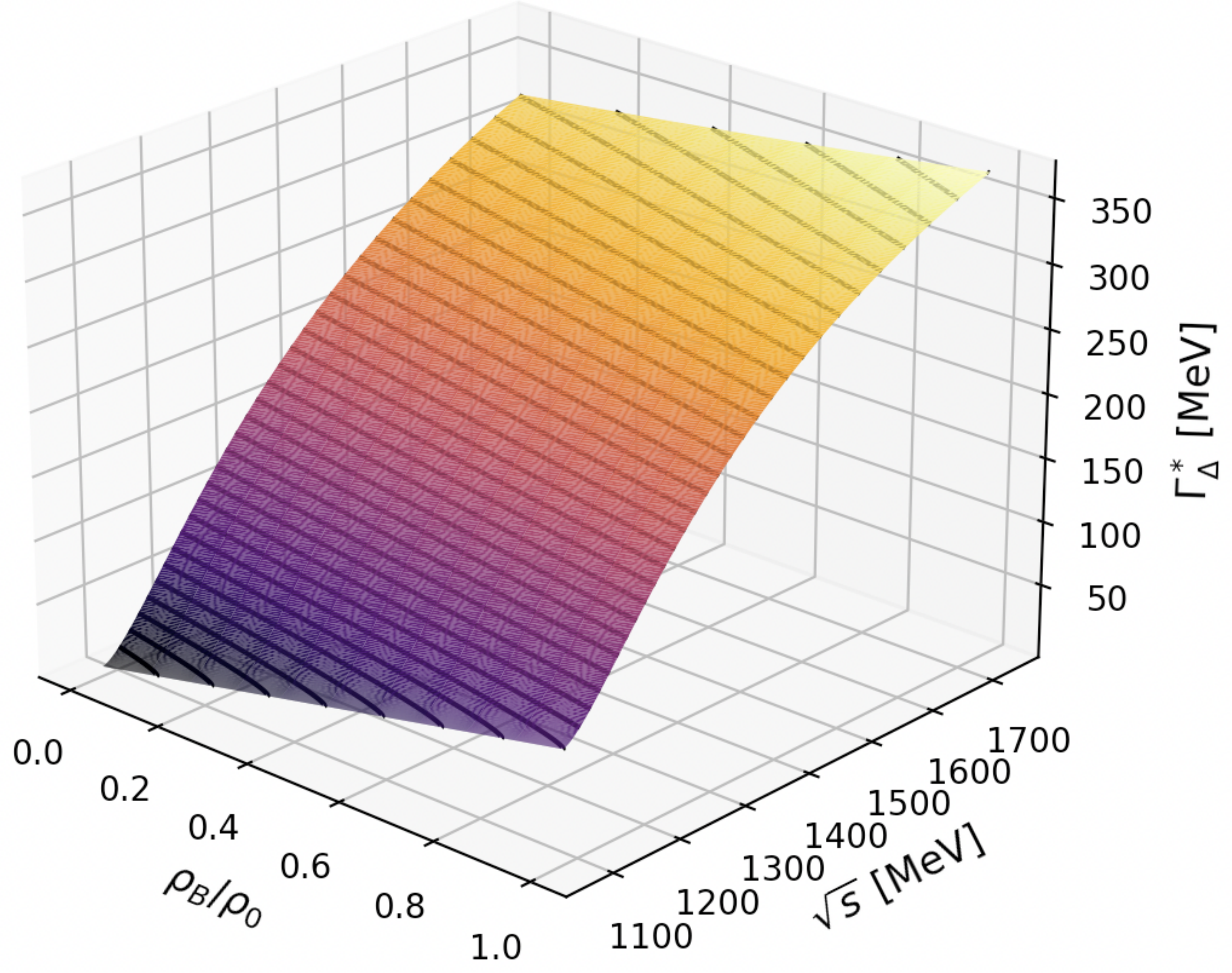}
\caption{ In-medium $\Delta$-resonance full decay width $\Gamma^*_\Delta$ as a function of $E_{\textrm{cm}}$ and $\rho_B/\rho_0$ from Eq.~(\protect\ref{eqN12}).}
\label{fig2}
\end{figure}

Now I am in a position to consider the physical observable at finite baryon density. Note that, in addition to the hadron mass modifications as discussed in the previous Section, the decay width for the $\Delta$ resonance is modified as well in the medium as follows~\cite{Han:2021zrh}:
\begin{eqnarray}
  \label{eqN12}
  \Gamma^*_\Delta \left(\sqrt{s^*} \right) =  \Gamma_\mathrm{sp} \left( \frac{\rho_B}{\rho_0} \right)
  +   \Gamma^*_\Delta \left(\sqrt{s^*},M^*_{\pi,N,\Delta}\right),
\end{eqnarray}
where $\rho_B$ and $\Gamma_\mathrm{sp}$ are the baryon density and spreading width of the resonance in the medium, respectively. In this calculation, I take the values of the $\Gamma_\mathrm{sp}$ = 80 MeV, which are determined by fitting to the medium quantity at finite density~\cite{Larionov:2003av,Hirata:1978wp,Oset:1987re}. The numerical result for the in-medium $\Delta$-resonance full decay width $\Gamma^*_\Delta$ in Eq.~(\ref{eqN12}) as a function of $E_{\textrm{cm}}$ and $\rho_B/\rho_0$ is depicted in Fig.~\ref{fig2}. It shows that $\Gamma^*_\Delta$ increases with respect to the energy as well as the density due to the decreasing of the baryon masses in the dense medium~\cite{Ghosh:2016hln,Cui:2020fhr}.

\begin{figure}[t]
  \centering
  \includegraphics[width=0.5\textwidth]{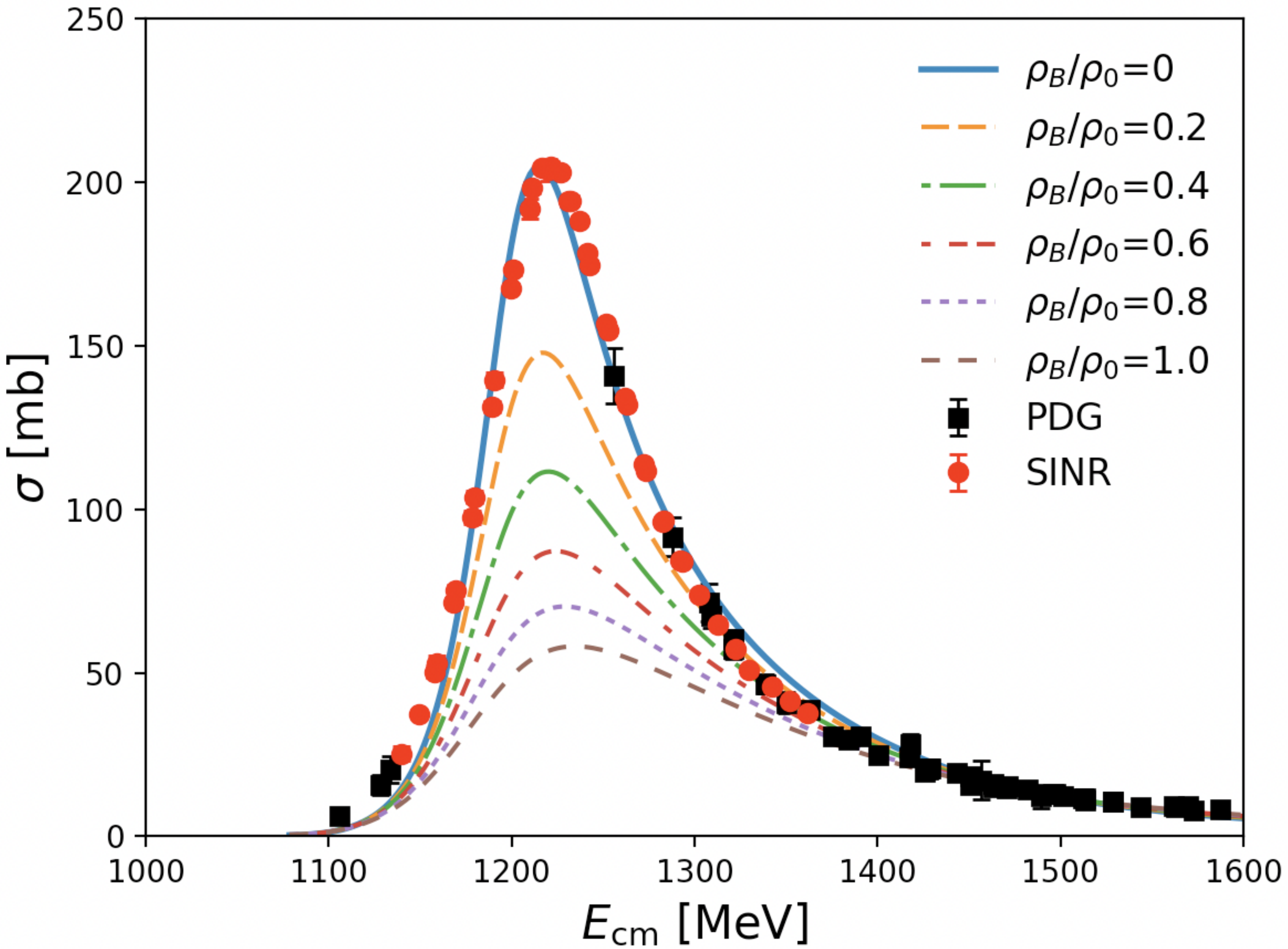}
  \caption{ \label{fig3} Total cross-section for the $\pi^{+}p$  elastic scattering as a function of $E_{\textrm{cm}}$ for various baryon densities. The circle and square denote the experimental data from Refs.~\cite{PDG2020,Pedroni:1978it} for vacuum.}
\end{figure}

Using the density-dependent width and mass, i.e., $\Gamma^*_\Delta$ and  $M^*_B$, I calculate TCS as a function of $E_{\textrm{cm}}$ for various baryon densities of $\rho_B/\rho_0=(0-1)$ in Fig.~\ref{fig3}. The experimental data for the vacuum case are taken again from Refs.~\cite{PDG2020,Pedroni:1978it}. Unfortunately, there is no available data for the density-dependent TCS. Figure~\ref{fig3} shows, as the density increase, the peak of the $\Delta$ resonance in the TCS gets diminished as well as broadened, as expected from the numerical results shown in Fig.~\ref{fig2}. Moreover, it is worth noting that the peak position of the resonance smoothly moves to the higher cm energy. Similar tendencies are reported in other theoretical model calculations~\cite{Ghosh:2016hln,Cui:2020fhr}.

\begin{figure*}[t]
  \centering
  \stackinset{r}{4.5cm}{t}{0.6cm}{(a)}{\includegraphics[width=0.43\textwidth]{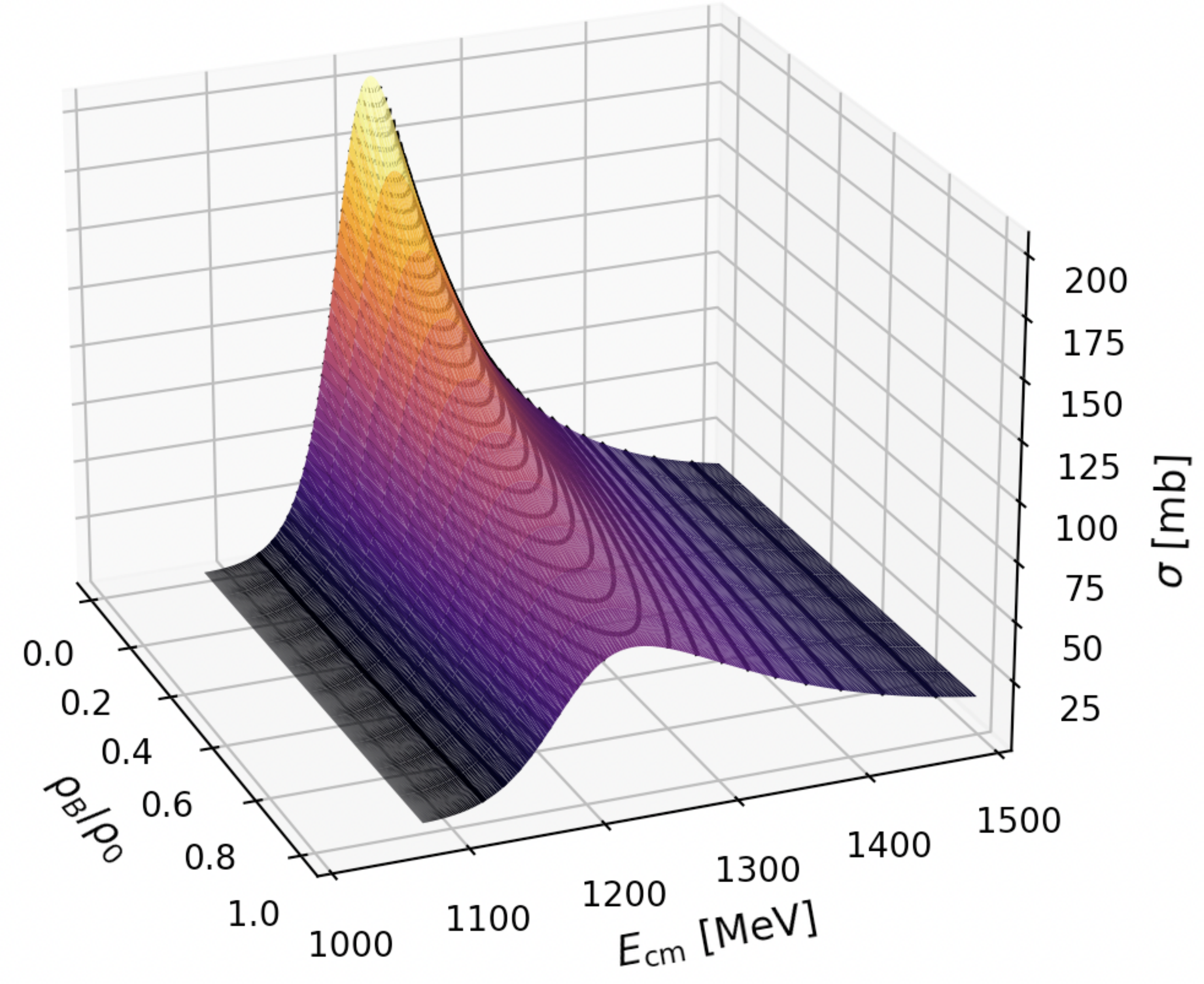}}
  \stackinset{r}{4.5cm}{t}{0.6cm}{(b)}{\includegraphics[width=0.43\textwidth]{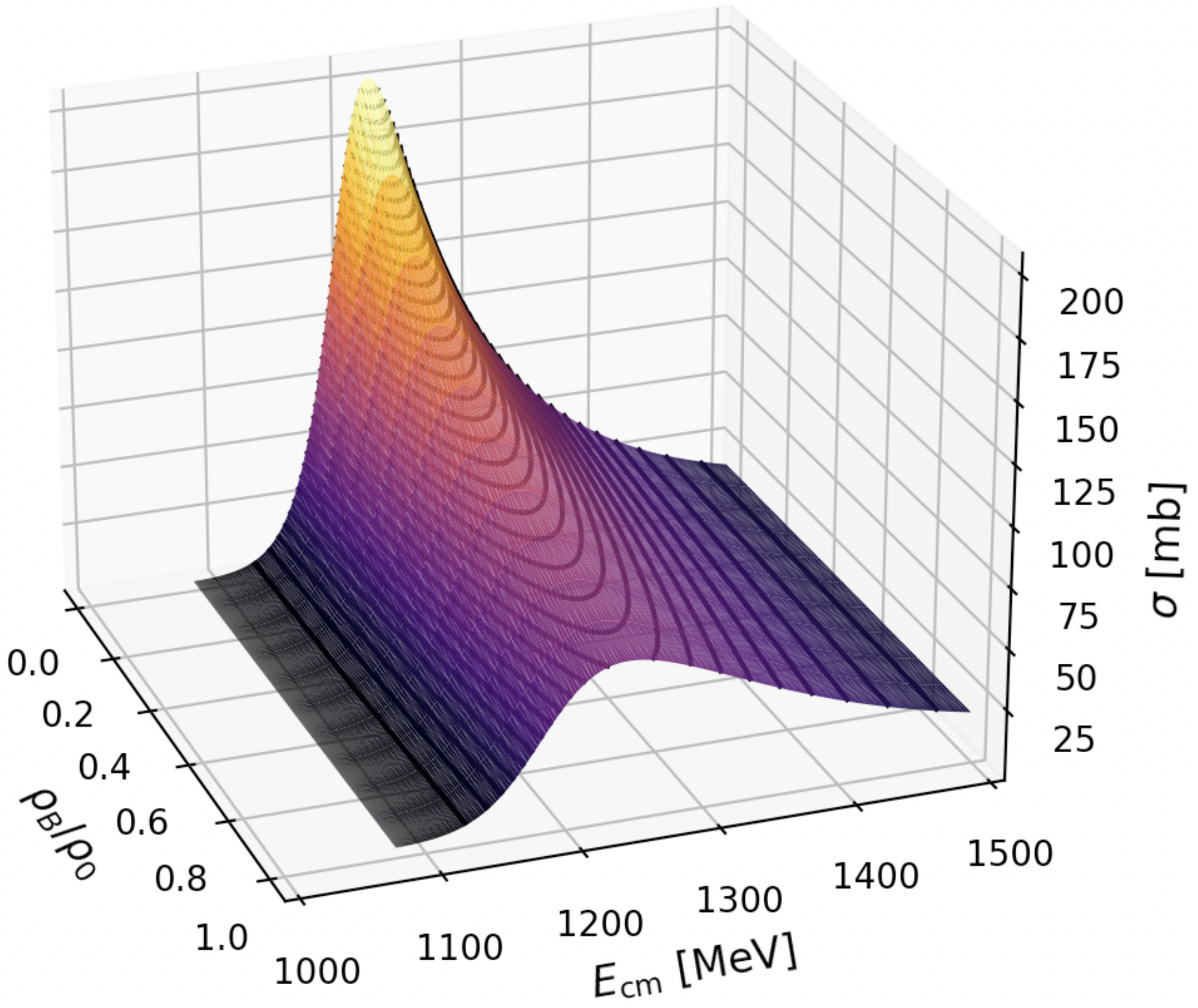}}
  \stackinset{r}{4cm}{t}{0.5cm}{(c)}{\includegraphics[width=0.43\textwidth]{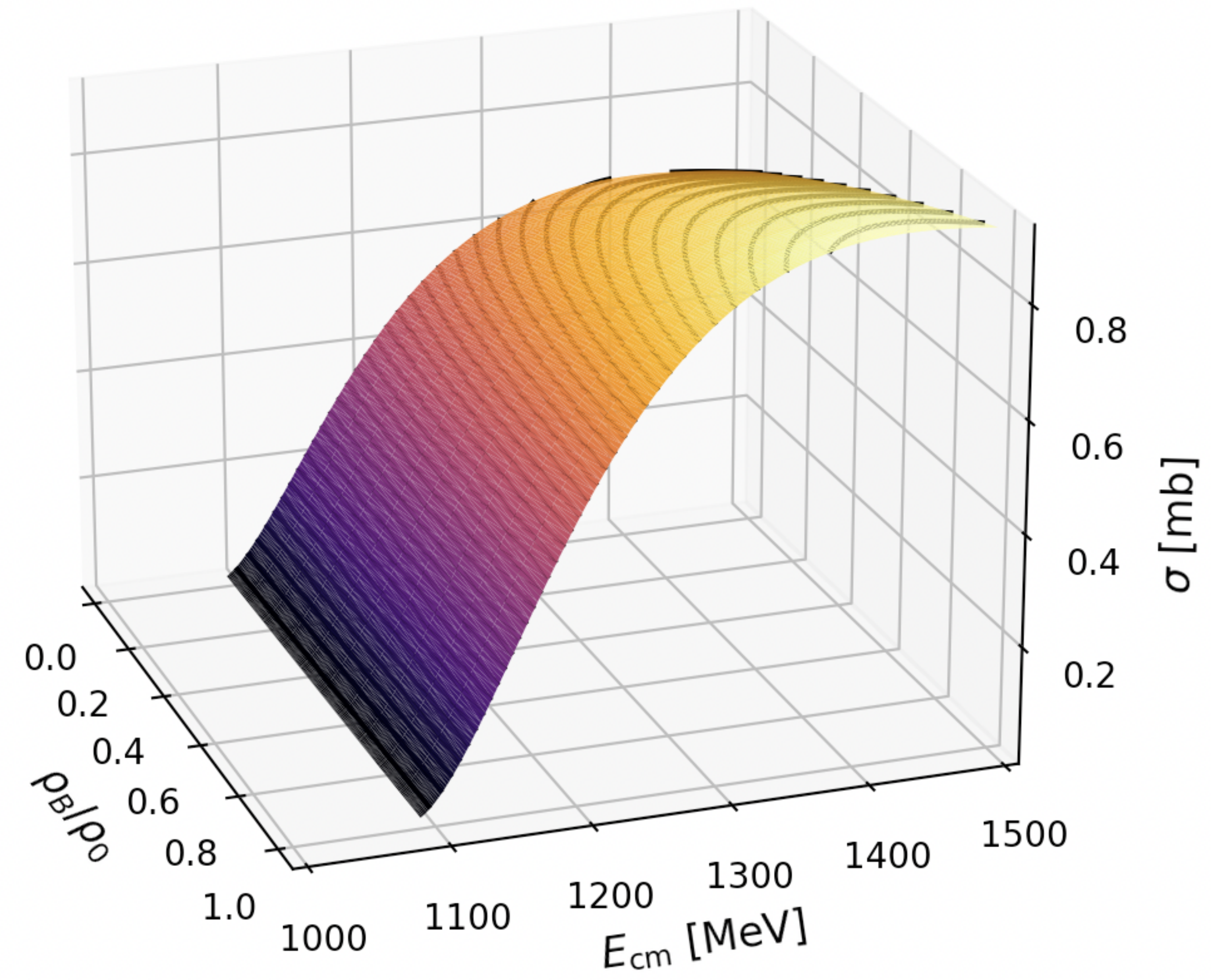}}
  \caption{ \label{fig4} Total cross-section for the elastic $\pi^+p$ scattering as a function of $E_{\textrm{cm}}$ and $\rho_B/\rho_0$ for (a) the total, (b) $s$-channel $\Delta$(1232), and (c) background contributions.}
\end{figure*}

In Fig.~(\ref{fig4}), I depict the total (a), $s$-channel $\Delta$(1232) (b), and background (c) contributions separately for TCS as a function of $E_{\textrm{cm}}$ and $\rho_B/\rho_0$. As shown there, the $\Delta$-resonance contribution dominates the scattering process and decreases rapidly with respect to the density. On the contrary, the BKG contribution slightly increases as a function of the density, but it is almost negligible. 

\begin{figure}[t]
\stackinset{r}{1.5cm}{t}{0.5cm}{(a)}{  \includegraphics[width=0.51\textwidth]{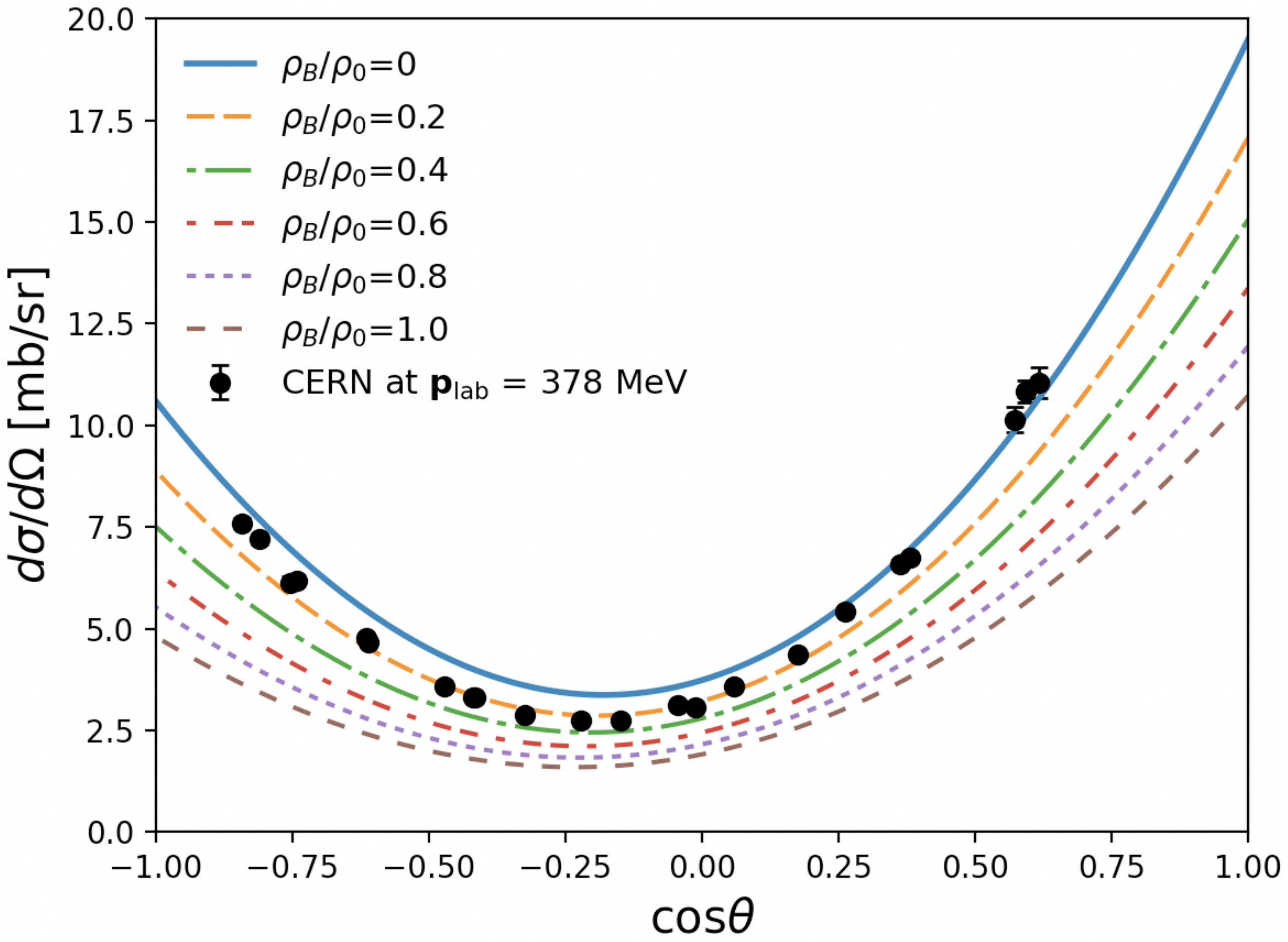}}
 \stackinset{r}{1.5cm}{t}{0.5cm}{(b)}{ \includegraphics[width=0.5\textwidth]{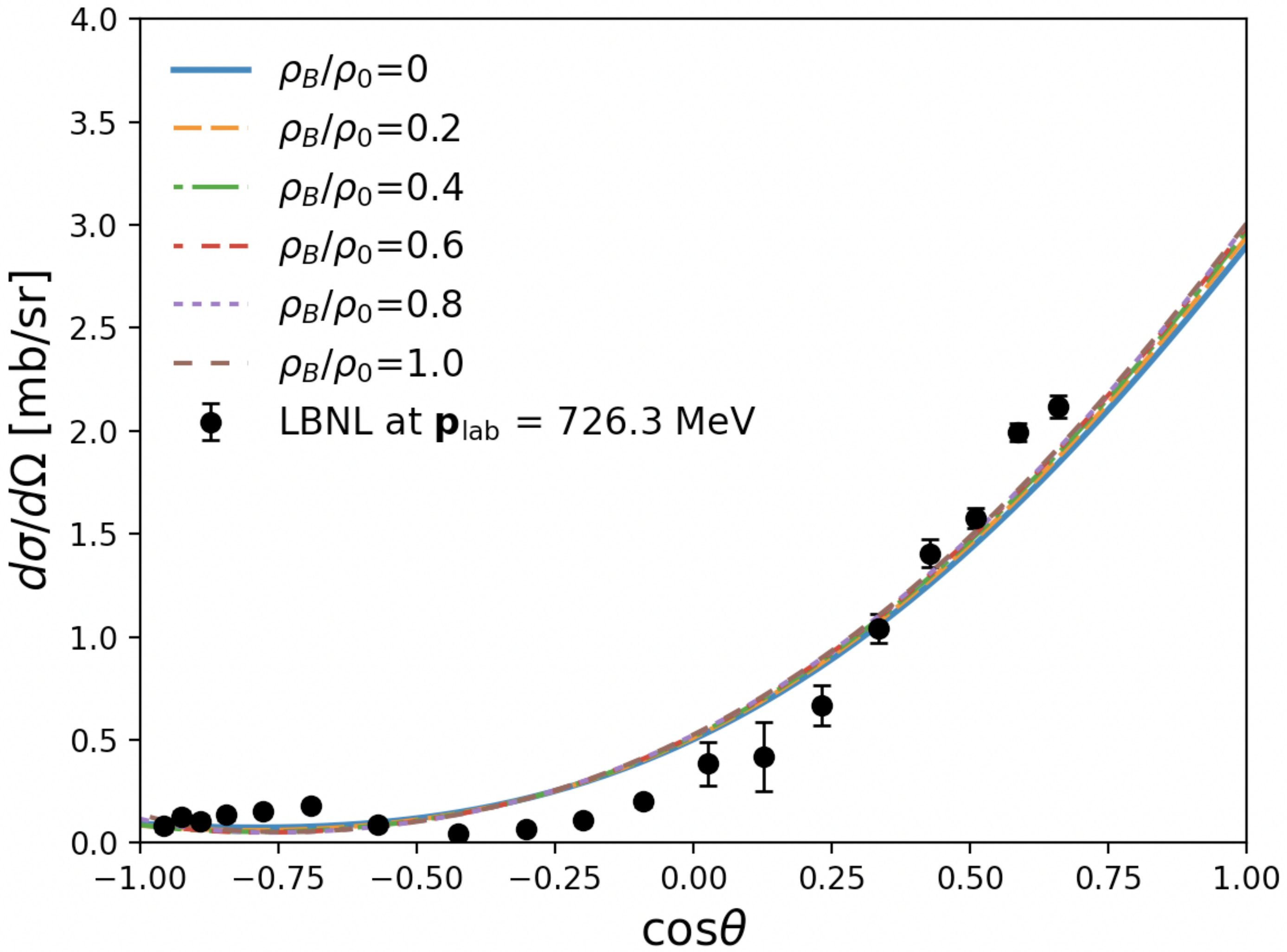}}
  \caption{ \label{fig5} Differential cross-section DCS$_{\theta}$ for the elastic $\pi^+p$ scattering as a function of $\cos\theta$ for $\rho_B/\rho_0=(0-1)$ at (a) $\textbf{p}_{\textrm{lab}}$ = 378 MeV and (b) $726.3$ MeV. The experimental data are taken from Ref.~\cite{Bussey:1973gz,Ogden:1965zz}.}
\end{figure}

The numerical results for the angular dependent DCS$_{\theta}$ are shown in the panel (a) and (b) of Fig.~\ref{fig5} for $\rho_B/\rho_0=(0-1)$ at $\textbf{p}_{\textrm{lab}} = 378$ MeV and $726.3$ MeV, respectively. The experimental data are taken from Ref.~\cite{Bussey:1973gz,Ogden:1965zz} for $\rho_B=0$. For $\textbf{p}_{\textrm{lab}} = 378$ MeV at which the $\Delta$ resonance dominates the process ($E_\mathrm{cm}\approx1290$ MeV), the angular dependence shows a typical $p$-wave curve but the shape seems like a $d$-wave, due to the dominant $\Delta$ resonance, showing the strong forward and mild backward scattering enhancements and the strength gets diminished with respect to the density as expected from Fig.~\ref{fig4}. As the momentum increases up to $\textbf{p}_{\textrm{lab}} = 726.3$ MeV ($E_\mathrm{cm}\approx1510$ MeV), where the $\Delta$-resonance contribution becomes compatible with the BKGs, and interferes with the $u$-channel BKG contributions destructively, resulting in the forward scattering enhancements as shown in panel (b) of Fig.~\ref{fig5}.

Interestingly, at $\textbf{p}_{\textrm{lab}} = 726.3$ MeV, the curves remain almost the same even with the density changes for $\rho_B/\rho_0=(0-1)$, since the density dependence becomes weak and less visible in the relatively higher energy regions as shown in the TCS results as indicated in Figs.~\ref{fig3} and \ref{fig4}. Note that the experimental data for $\textbf{p}_{\textrm{lab}}$ = 726.3 MeV~\cite{Ogden:1965zz} is well reproduced by the theory, indicating the validity of the present model calculation. It is verified that the backward scattering is enhanced as the momentum goes higher than $800$ MeV since the $\Delta$-resonance contribution almost disappears and the $u$-channel BKG contributions start to be important. 

\begin{figure}[t]
\stackinset{r}{5cm}{t}{0.5cm}{(a)}{\includegraphics[width=0.53\textwidth]{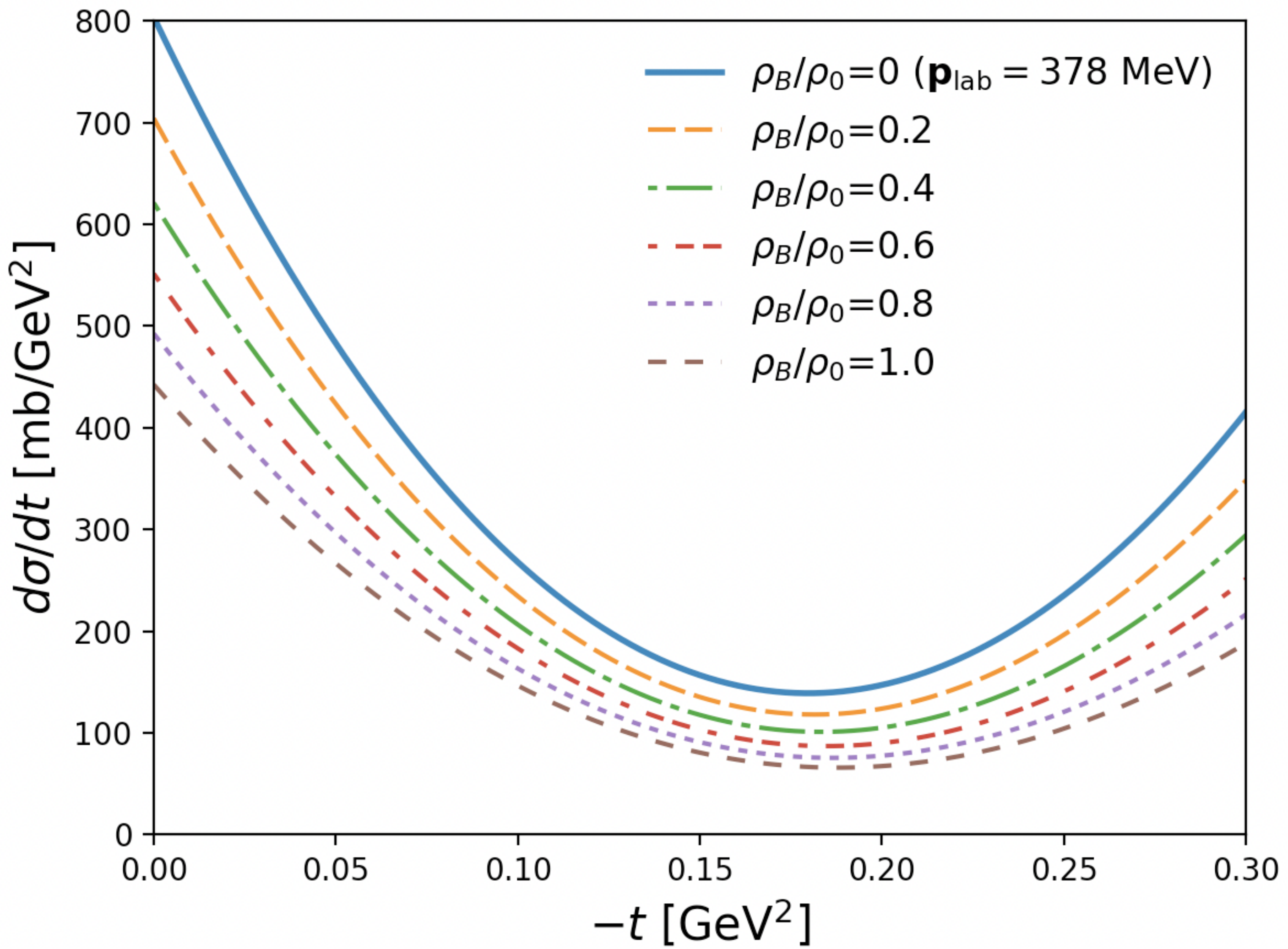}}
\stackinset{r}{5cm}{t}{0.5cm}{(b)}{\includegraphics[width=0.51\textwidth]{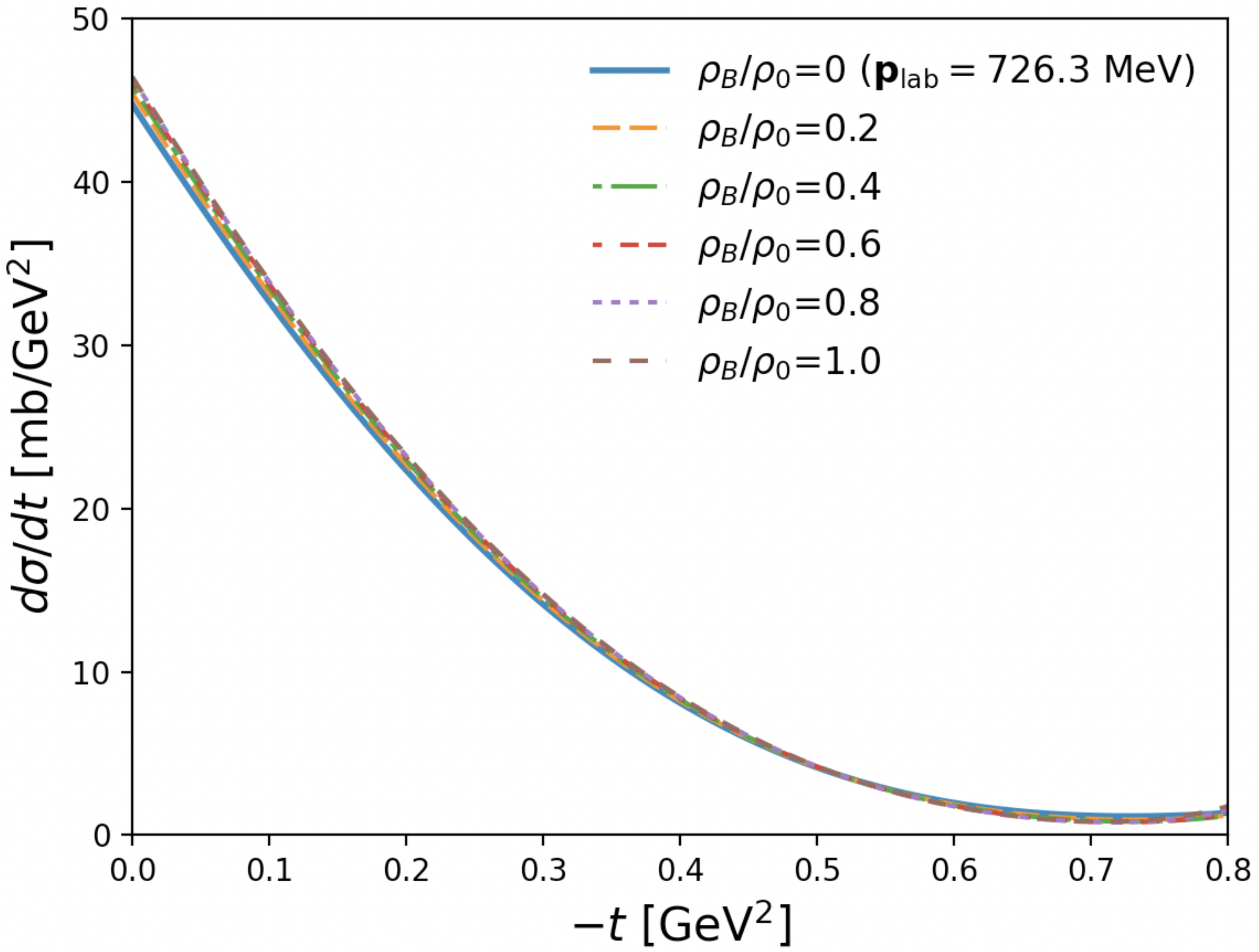}}
\caption{ $t$-dependent differential cross-section (DCS$_{t}$) for the elastic $\pi^{+}p$ scattering as a function of $-t$ for $\rho_B/\rho_0=(0-1)$ at (a) $\textbf{p}_{\textrm{lab}}$ = 378 MeV and (b) $726.3$ MeV.}
\label{fig7}
\end{figure}

The numerical results for the $t$-dependent differential cross-section (DCS$_{t}$) as a function of $-t$ and the densities at $\textbf{p}_{\textrm{lab}}$ = 378 MeV and  $726.3$ MeV are given in the panel (a) and (b) of Fig.~\ref{fig7}, respectively. At $\textbf{p}_{\textrm{lab}}$ = 378 MeV, the strong forward- and backward-scattering enhancements are observed as already shown in DCS$_{t}$, and similar tendencies are found for the higher momentum. The density dependence can be understood in the same way.  

\begin{figure}[t]
 \stackinset{r}{0.5cm}{t}{1.3cm}{(a)}{ \includegraphics[width=0.5\textwidth]{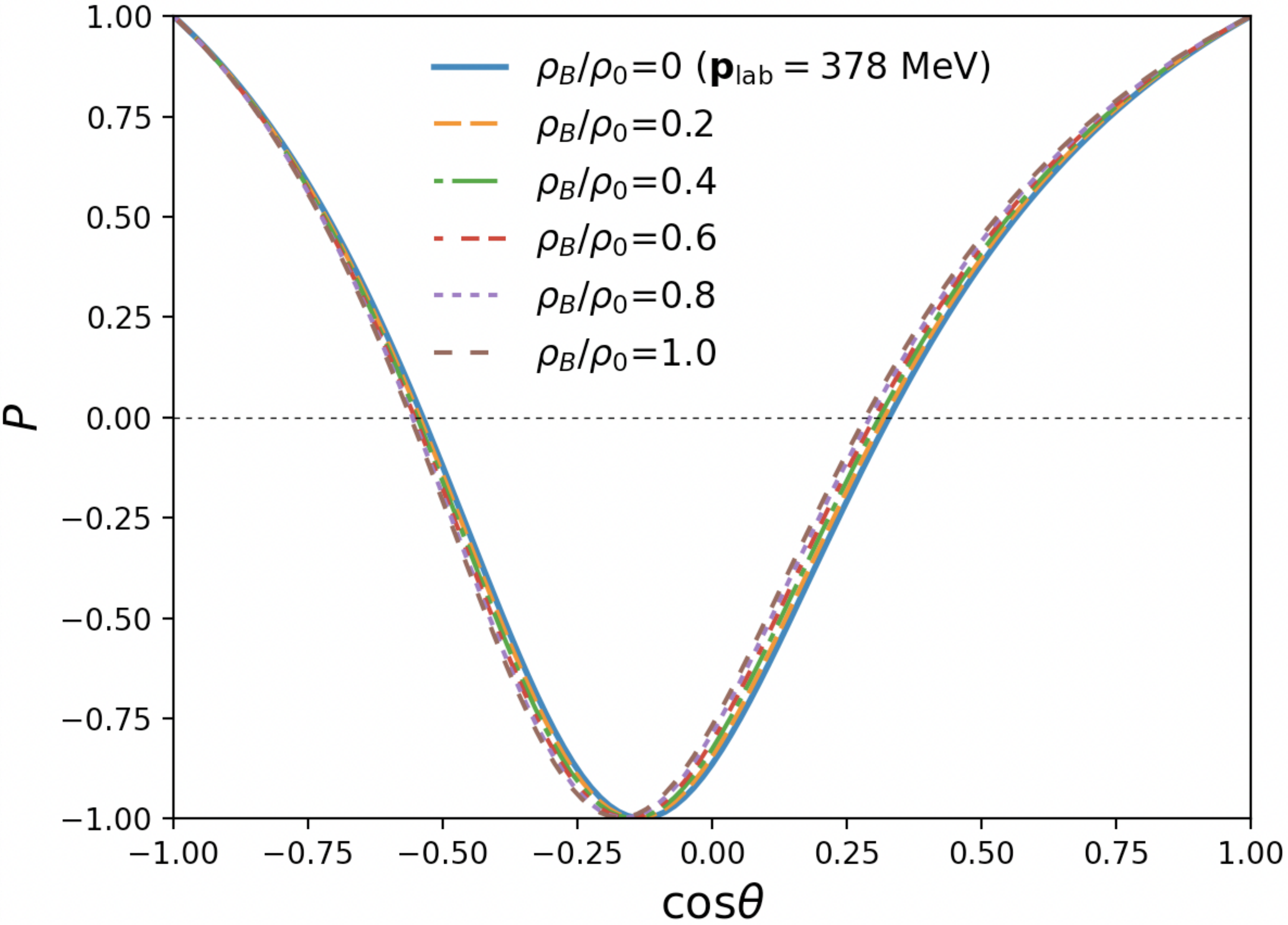}}
 \stackinset{r}{0.5cm}{t}{1.3cm}{(b)}{ \includegraphics[width=0.5\textwidth]{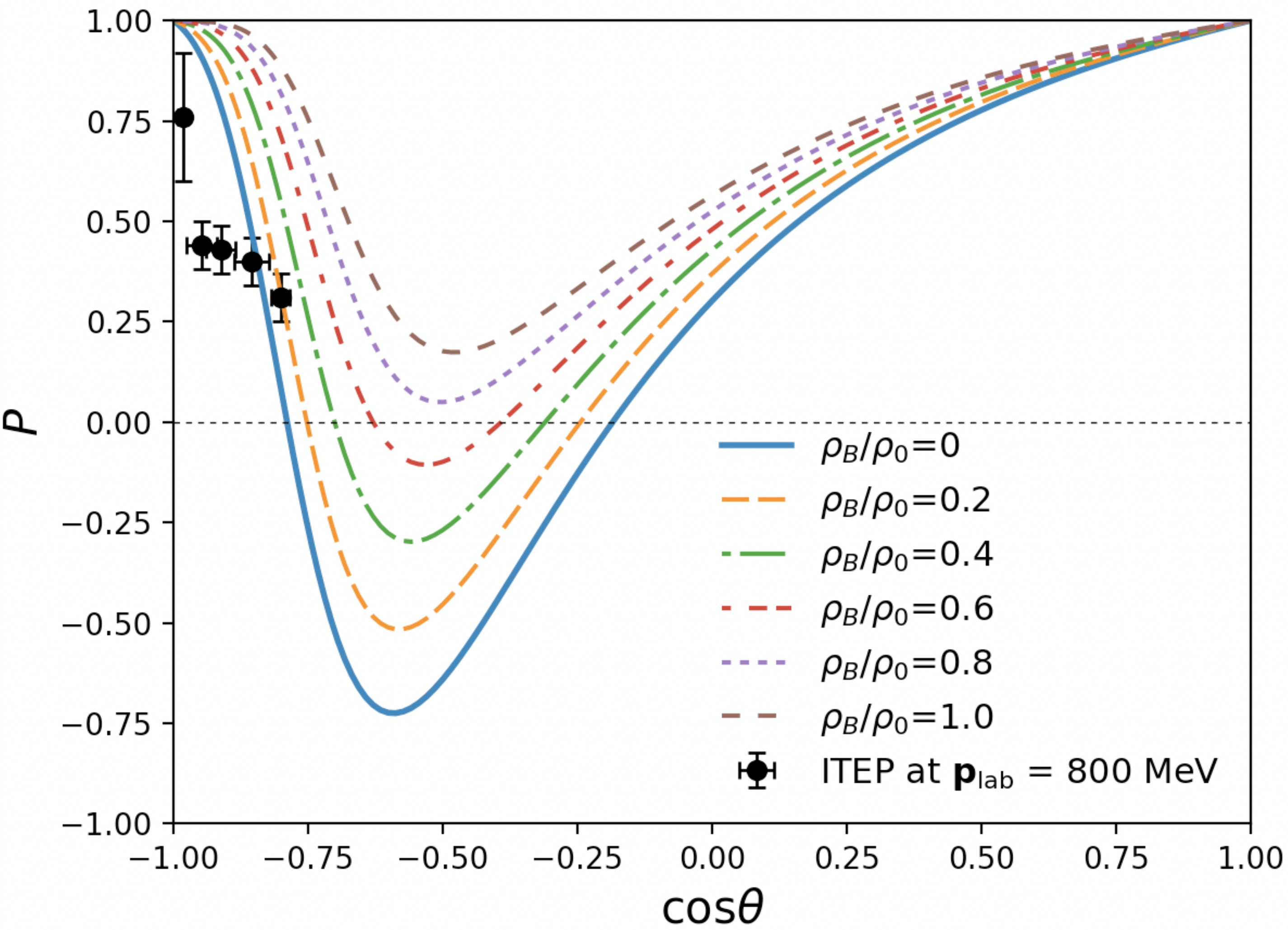}}
  \caption{ \label{fig6} Target- and recoil-proton spin asymmetry $P$ in Eq.~(\protect\ref{eqN14}) for the elastic $\pi^{+}p$ scattering as a function of $\cos \theta$ for $\rho_B/\rho_0=(0-1)$ at (a) $\textbf{p}_{\textrm{lab}}$ = 378 MeV and (b) $800$ MeV. The experimental data from the ITEP-PNPI collaboration at $\textbf{p}_\mathrm{lab}=800$ MeV~\cite{ITEP-PNPI:2008cmv}.}
\end{figure}

Finally, I compute the target- and recoil-proton spin asymmetry $P$ defined as follows:
\begin{eqnarray}
  \label{eqN14}
P&= & \frac{\sum_\mathrm{spin}\left[\left( \frac{\partial \sigma}{\partial \Omega}\right)_\mathrm{S}-\left(\frac{\partial\sigma}{\partial\Omega}\right)_\mathrm{O}\right]}{\sum_\mathrm{spin}\left[\left( \frac{\partial \sigma}{\partial \Omega}\right)_\mathrm{S}+\left(\frac{\partial\sigma}{\partial\Omega}\right)_\mathrm{O}\right]},
\end{eqnarray}
where the subscripts S and O stand for that the target- and recoil-proton spins are aligned in the same and opposite quantization axes, respectively. This physical quantity represents the short-range spin correlation inside the medium, i.e., a nuclei target for instance. In panel (a) and (b) of Fig.~\ref{fig6}, the numerical results of $P$ as a function of $\cos\theta$ are shown for different densities at $\textbf{p}_{\textrm{lab}}$ = 378 MeV and $800$ MeV, respectively. The curve shape can be easily understood by spin statistics. When the spins of the target and recoil proton are in the same direction, the intermediate $\Delta$ resonance is in its $S=1/2$ spin state, and its angular dependence $(d\sigma/d\Omega)_\mathrm{S}$ is given by $\sim(\mathrm{const.}+\cos^2\theta)$. As for the opposite direction, $(d\sigma/d\Omega)_\mathrm{O}$ is described by $\sim\sin^2\theta$. Hence, the combination of these two angular dependencies results in that curve shape commonly for the two momenta. 

Interestingly, however, the density dependence is quite different for the momenta as seen in Fig.~\ref{fig6}. The reason for this finding can be explained as follows: At the lower momentum, the scattering process is dominated by the $\Delta$-resonance contribution. Then, as the density increases, $(d\sigma/d\Omega)_\mathrm{O}$ and $(d\sigma/d\Omega)_\mathrm{S}$ decrease at the same rate, resulting in the stable $P$ with respect to $\rho_B$ as shown in the panel (a) of Fig.~\ref{fig6}. On the contrary, as for the higher momentum, although the $\Delta$-resonance contribution is still larger than other ones, it starts to compete with the BKG contributions, resulting in the obvious density dependence as shown in panel (b) of the figure. It is worth mentioning that the experimental data from the ITEP-PNPI collaboration for $P$ at $\textbf{p}_\mathrm{lab}=800$ MeV~\cite{ITEP-PNPI:2008cmv}  is qualitatively well reproduced by the theory for $\rho_B=0$. 

\chapter{The elastic $\pi A$ scattering with medium effects}


In this section, the elastic $\pi A$ scattering with medium effects is presented. Before explaining the main concept of $\pi A$ scattering in this following study, it is required to describe the elastic $\pi^+ n$ scattering process, which is one possible reaction at finite nuclei as described below. 

\section{The elastic $\pi^+ n$ scattering}
In this section, I briefly present the elastic $\pi^+ n$ scattering to complete the elastic $\pi^+ N$ scattering process. The nucleon $N$ has isospin $I=\frac{1}{2}$ and can be divided into two states, proton, and neutron, except in the case of anti-particles. So between proton and neutron, most of their properties are similar in spite of having different $I_3$ components. Consequently, the expression of the  elastic $\pi^+ n$ scattering is rather similar to that for the $\pi^+ p$ channel.

\begin{figure}[t]
  \centering
 \includegraphics[width=0.95\textwidth]{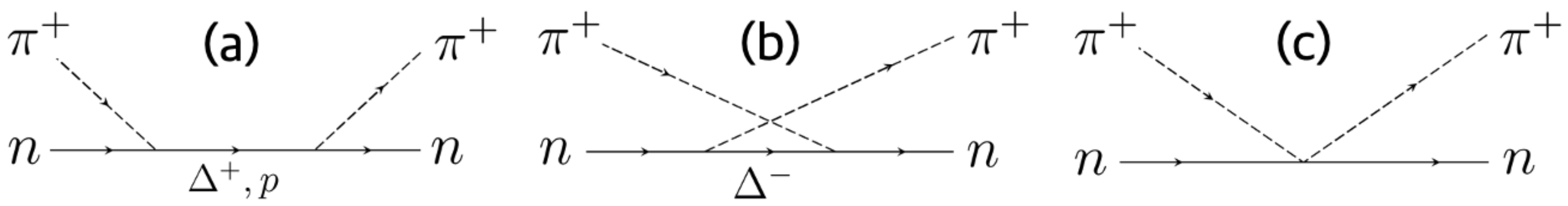}
  \caption{ \label{fig10} Feynman diagrams contribute to the elastic $\pi^+ n$ scattering for (a) the $\Delta^+$ and proton intermediate diagrams in the $s$-channel, (b) the $\Delta^-$ pole diagram in the $u$-channel, and (c) Weinberg-Tomozawa (WT) contact interaction.}
\end{figure}

At the tree level Born approximation, the Feynman diagrams of the $\pi^+ n$ scattering are depicted in Fig.~\ref{fig10}, with (a) the $\Delta^+$ and proton intermediate diagrams in the $s$-channel, (b) the $\Delta^-$ pole diagram in the $u$-channel, and (c) Weinberg-Tomozawa (WT) contact interaction. The solid and dashed lines represent the baryons (nucleon and $\Delta$) and pion, respectively. The relevant effective Lagrangians are already presented in Sec.~II. Thus, I just denote the corresponding scattering amplitudes below:
\begin{eqnarray}
\begin{split}
  \label{eq14a} 
  i\mathcal{M}_s^{\Delta^{+}} = -\frac{f_{\pi N\Delta}^2}{3M_\pi ^2}\bar{u}(p')k'_\mu G^{\mu\nu}(p+k)k_\nu u(p),\\
  i\mathcal{M}_u^{\Delta^{-}}= -\frac{f_{\pi N\Delta}^2}{M_\pi ^2}\bar{u}(p')k_\mu G^{\mu\nu}(p-k')k'_\nu u(p),
\end{split}
\end{eqnarray}
where the factor of $\frac{1}{3}$ in the $s$-channel amplitude in Eq.~\eqref{eq14a} comes from the isospin factor.

Next, the scattering amplitude for the Feynman diagram in which the proton is the intermediate particle shown in Fig.~\ref{fig10} (a) is formulated by

\begin{equation}
\label{eq14b}
    i\mathcal{M}^p_s = -i \left( \frac{2f^2 _{\pi NN}}{3M_\pi ^2} \right)\bar{u}(p')\gamma_5\slashed{k}'\frac{\slashed{p}+\slashed{k}+M_N}{(p+k)^2 -M_N^2}\gamma_5\slashed{k}u(p),
\end{equation}
where the factor of $2/3$ in Eq.~\eqref{eq14b} is the isospin factor. The scattering amplitude of the Weinberg-Tomozawa contact interaction (c) has a similar form as the $\pi^+ p$ channel.

Thus, the total amplitude of the elastic $\pi^+ n$ scattering with the phenomenological form factors~\eqref{eq10} is written as 
\begin{eqnarray}
  \label{eq14c}
  \mathcal{M}^\mathrm{total}_{\pi^+n}= i \mathcal{M}_s^{\Delta^{+}}F_s^{\Delta^{+}}
  +i \mathcal{M}_u^{\Delta^{-}}F_u^{\Delta^{-}}
  +i \mathcal{M}_s^pF_s^p
  +i \mathcal{M}_{\mathrm{WT}}F_{\mathrm{WT}}.
\end{eqnarray}

\begin{figure}[t]
  \centering
 \includegraphics[width=0.6\textwidth]{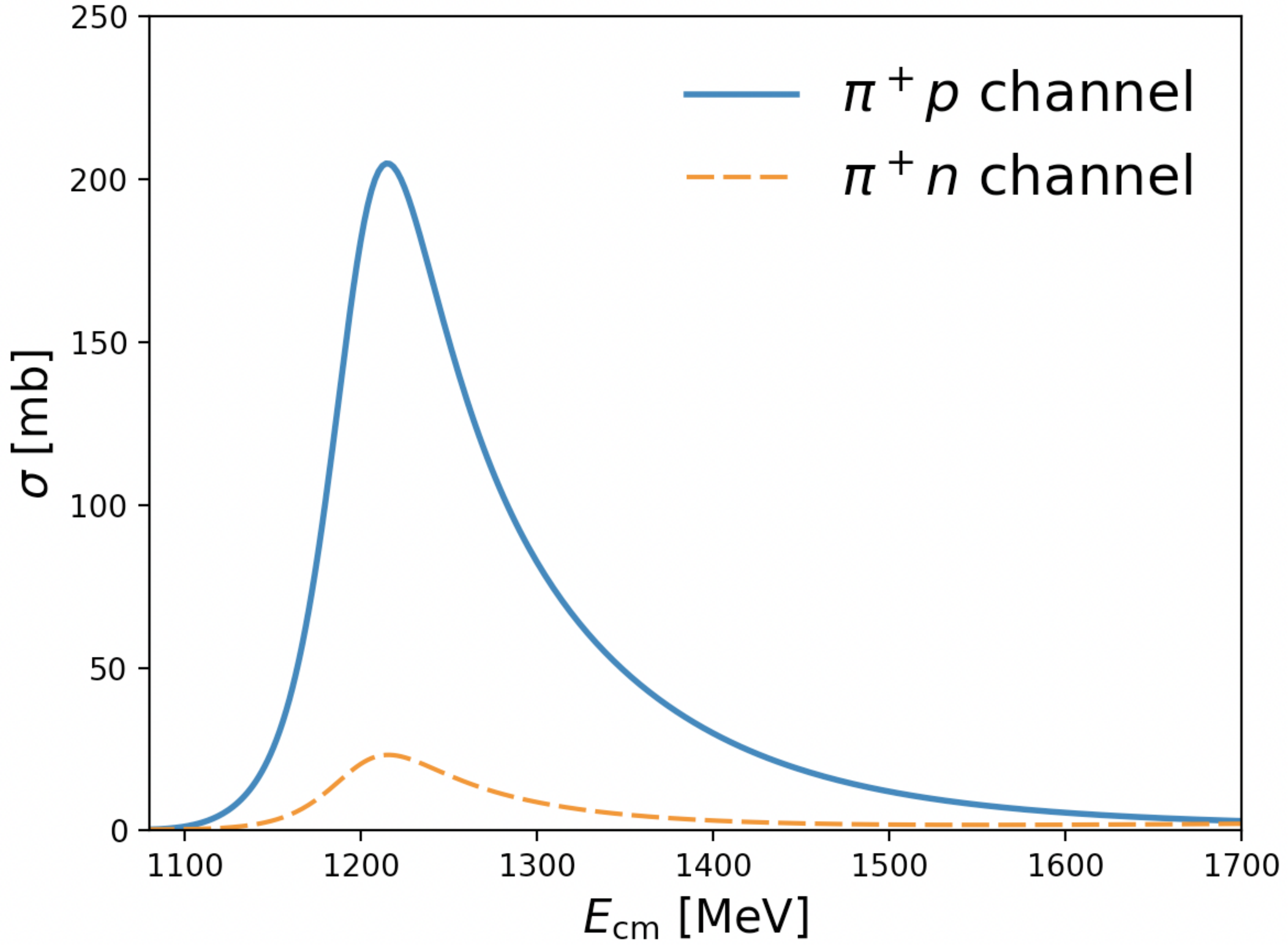}
  \caption{ \label{fig11} Total cross-section of the elastic $\pi^+p$ scattering (blue solid line) and the elastic $\pi^+n$ scattering (orange dashed line) as a function of $E_{\mathrm{cm}}$. Because of the isospin factor difference in the $s$-channel of $\Delta$ contribution, the TCS of the elastic $\pi^+ n$ scattering denotes almost 9 times smaller than the TCS of $\pi^+ p$ channel.}
\end{figure}

Using the numerical result of Eq.~\eqref{eq14c}, the total cross-section of the elastic $\pi^+ n$ scattering is calculated and represented with the TCS of the elastic $\pi^+ p$ scattering in Fig.~\ref{fig11}. As already discussed, the $s$-channel of $\Delta(1232)$ contribution is dominant in the elastic $\pi^+ N$ scattering. For this reason, due to the isospin factor difference between $\mathcal{M}^{\Delta^{++}}_s$ ($\mathcal{I}_{\pi N \Delta}$ = $1$) and $\mathcal{M}^{\Delta^+}_s$ ($\mathcal{I}_{\pi N \Delta}$ = $\frac{1}{3}$), the magnitude of the TCS of elastic $\pi^+ n$ scattering indicates that it is
almost nine times smaller than the TCS of $\pi^+p$ scattering channel.

\section{The Eikonal Glauber model}
In this section, the Eikonal Glauber model and the charge distribution function of the nucleus is described. Using the result from the elastic $\pi N$ scattering with the medium effect, I investigate the elastic $\pi A$ scattering. The Glauber model, which was first time proposed by Roy Glauber in the 1950s, was a successful phenomenological model to describe the property of the nucleus, especially the $NA$ or $AA$ interactions in the heavy-ion collision experiments~\cite{Miller:2007}.
\begin{figure}[t]
  \centering
 \includegraphics[width=0.7\textwidth]{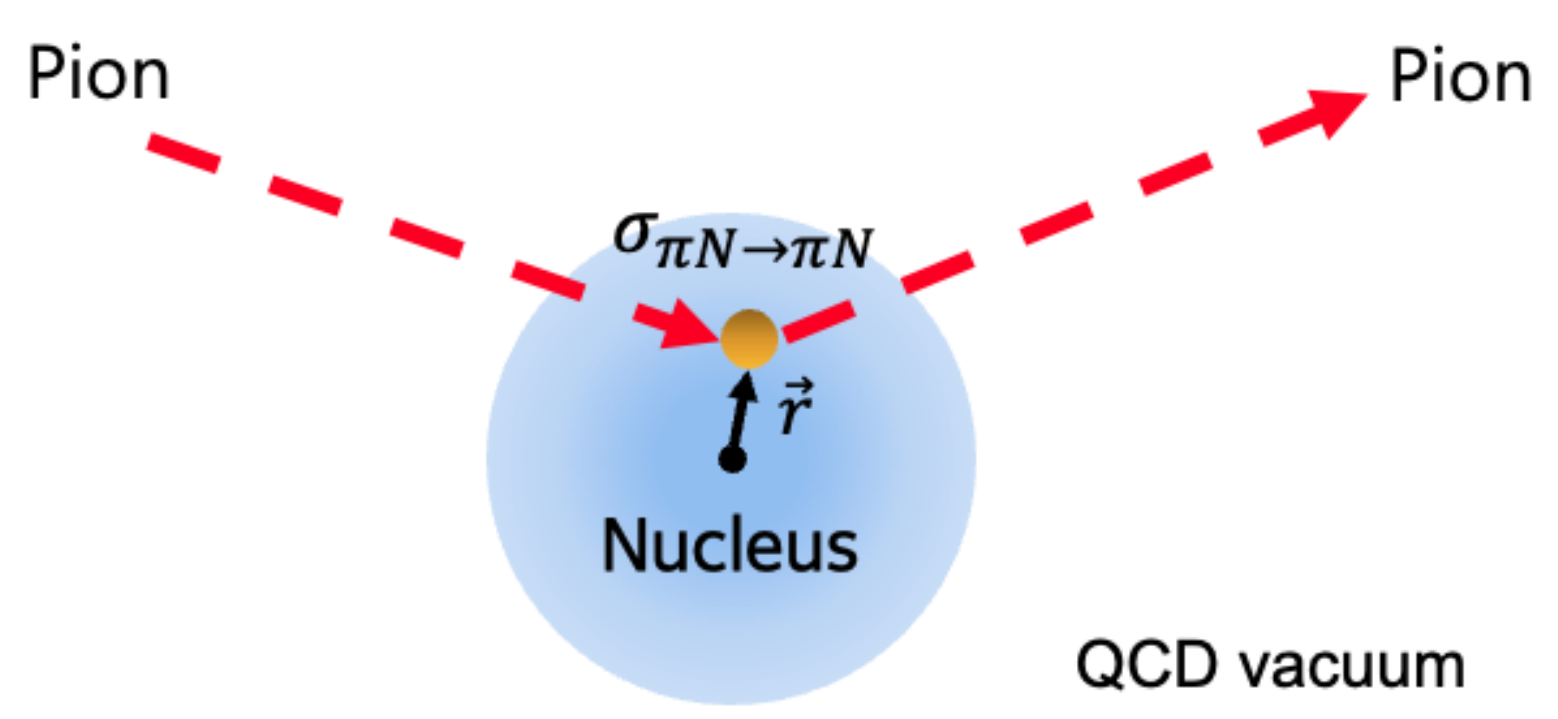}
  \caption{ \label{fig12} Schematic picture of the elastic $\pi A$ scattering with medium effect. Pion is represented by a red dashed line and $\vec{r}$ is the particle position in the nucleus.}
\end{figure}

In addition, the Glauber model explains the interaction of projectile-nucleus scattering as a multiple scattering of the elastic projectile-nucleon scattering---in this work, the total cross-section of the elastic $\pi N$ scattering, $\sigma_{\pi N\rightarrow \pi N}$ becomes the focus of this study. Here I adopt the Eikonal Glauber model, which is an updated version of the Glauber model, and contains an assumption that a projectile has a linear path, which is the so-called Eikonal approximation.

In the Glauber model, the total cross-section of the elastic $\pi A$ scattering is simply given by
\begin{align}
  \label{eqN15}
  \sigma_{\pi^+ A\rightarrow\pi^+ A}(\sqrt{s^*})&=\int d^2\boldsymbol{b}\left[1-\mathrm{exp}[-\tilde{\sigma}_{\pi^+ N\rightarrow \pi^+ N}(\sqrt{s^*},\rho^{(n,p)}_A)\cdot T^{(n,p)}_A(\boldsymbol{b})]\right], \nonumber \\T^{(n,p)}_A(\boldsymbol{b})&=\int^{\infty}_{-\infty}\rho^{(n,p)}_A(\displaystyle \sqrt{|\boldsymbol{b}|^2+z^2}) dz,
\end{align} 
where $\rho_A$ and $\boldsymbol{b}$ indicate the charge distribution of the target nucleus and the impact parameter, respectively.
$T_A$ and $z$ denote the nuclear thickness function which describes the transverse reaction probability at $\boldsymbol{b}$ and the scattering axis orthogonal to $\boldsymbol{b}$.

In Eq.~\eqref{eqN15}, as already mentioned earlier the elastic $\pi N$ scattering contains two interactions, $\pi p$, and $\pi n$. The superscript $(n,p)$ on $T^{(n,p)}_A$ and $\rho^{(n,p)}_A$ indicate $\pi n$ and $\pi p$ channel, respectively. Also, in this work, I considered $\tilde{\sigma}_{\pi^+ N\rightarrow\pi^+ N}$ which is a summation of the averaged TCS of the elastic $\pi^+N$ scattering multiplied with the neutron and proton number ratios. The total of $\tilde{\sigma}_{\pi^+ N\rightarrow\pi^+ N}$ is then given by
\begin{eqnarray}
  \label{eqN16}
  \tilde{\sigma}_{\pi^+ N\rightarrow\pi^+ N}=\sum_{N=n,p}\bar{\sigma}_{\pi^+ N\rightarrow \pi^+ N}=\left(\frac{A-Z}{A}\right)\cdot\sigma_{\pi^+ n\rightarrow \pi^+ n}+\left(\frac{Z}{A}\right)\cdot\sigma_{\pi^+ p\rightarrow \pi^+ p}.\quad\;
\end{eqnarray}

The Wood-Saxon density profile of the charge distribution function $\rho_A$ that is used in the calculation is expressed by
\begin{eqnarray}
  \label{eqN17}
  \rho^{(n,p)}_A(r)=\frac{\rho^A_0\left[1+c(r/r_A)^2\right]}{1+\mathrm{exp}[(r-r_A)/d]}\cdot\frac{(A-Z,Z)}{A},\;\;\;\;r=\sqrt{|\boldsymbol{b}|^2+z^2},
\end{eqnarray}
where $r,\;\rho^A_0,\;c,\;r_A,$ and $d$ denote the distance from the center of a nucleus, normalization constant of density, deformation parameter, average radius, and surface thickness, respectively.

\begin{table}
    \centering
    \begin{tabular}{ c | c | c | c | c | c | c }
    Nucleus & $A$ & $Z$ & $r_A$ [fm] & $d$ [fm]& $\rho^A_0$ [fm$^{-3}$]& $c$\\ \hline \hline
    He & 4 & 2 & 1.01 & 0.327 & 0.2381 & 0.445 \\ 
    C & 12 & 6 & 2.36 & 0.522 & 0.1823 & $-$0.149 \\
    Au & 197 & 79 & 6.38 & 0.535 & 0.1772 & 0 \\ 
    Pb & 208 & 82 & 6.62 & 0.549 & 0.1700 & 0
  \end{tabular}
  \caption{ Input parameters for the various nucleus in Eq.~\protect\ref{eqN17}. These parameters ($r_A, d, \mathrm{and}~c$) are obtained by fitting to experimental data~\cite{DEJAGER:1974479}, while $\rho^{A}_0$ is determined to satisfy the normalization.}
  \label{table1}
\end{table}

To satisfy the normalization of the nuclear density distribution $\rho^{(n,p)}_A$ for each nucleus, the normalization constant $\rho^A_0$ is parameterized and given by Table~\ref{table1} ~\cite{DEJAGER:1974479} and the $\rho^{(n,p)}_A$ is normalized as follows:
\begin{align}
\label{eqN18}
    A&=\displaystyle\sum_{N=n,p}\int dz d^2\boldsymbol{b}\rho^{(n,p)}_A\left(\sqrt{|\boldsymbol{b}|^2+z^2}\right)\nonumber \\ &=\displaystyle\sum_{N=n,p}2\pi\int^{\infty}_{-\infty}dz\int^\infty_0 bdb\rho^{(n,p)}_A\left(\sqrt{|\boldsymbol{b}|^2+z^2}\right).
\end{align}

\section{Nuclear medium effects in finite nuclei}

\begin{figure}[t]
  \centering
 \includegraphics[width=0.6\textwidth]{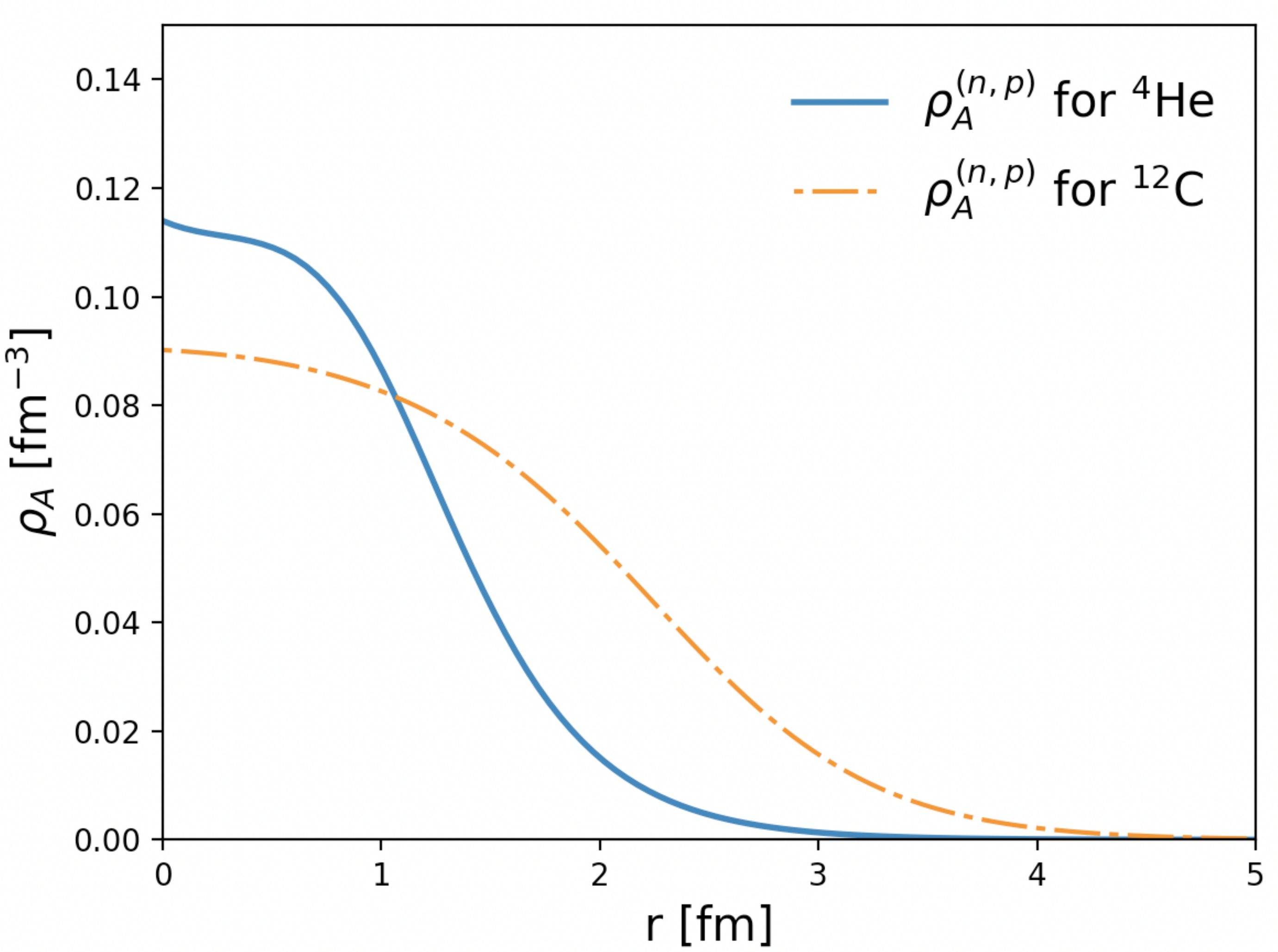}
  \caption{ \label{fig13} Nuclear density distribution $\rho^{(n,p)}_A$ as a function of $r$ for $^4\mathrm{He}$ (blue solid line), for $\displaystyle^{12}\mathrm{C}$ (orange dot-dashed line). In both nuclei cases, $\rho^n_A=\rho^p_A$ because the number of neutrons and protons is equal.}
\end{figure}

Here several properties of physical quantities for finite nuclei are analyzed, such as $\displaystyle\rho^{(n,p)}_A, M^*_B,$ and $\Gamma^*_\Delta$. Firstly the nuclear density distributions $\rho^{(n,p)}_A$ as a function of $r$ for two kinds of nuclei ($^4\mathrm{He},\displaystyle^{12}\mathrm{C}$) are depicted in Fig.~\ref{fig13}. Because Helium has a double magic number, the nuclear density distribution $\rho_A$ for $^4\mathrm{He}$ is bigger than that for $\displaystyle^{12}\mathrm{C}$ near the center of the nucleus. Also, both nuclei have the same $\rho^{(n,p)}_A$ due to the same proton and neutron numbers.

\begin{figure}[t]
  \centering
 \includegraphics[width=0.65\textwidth]{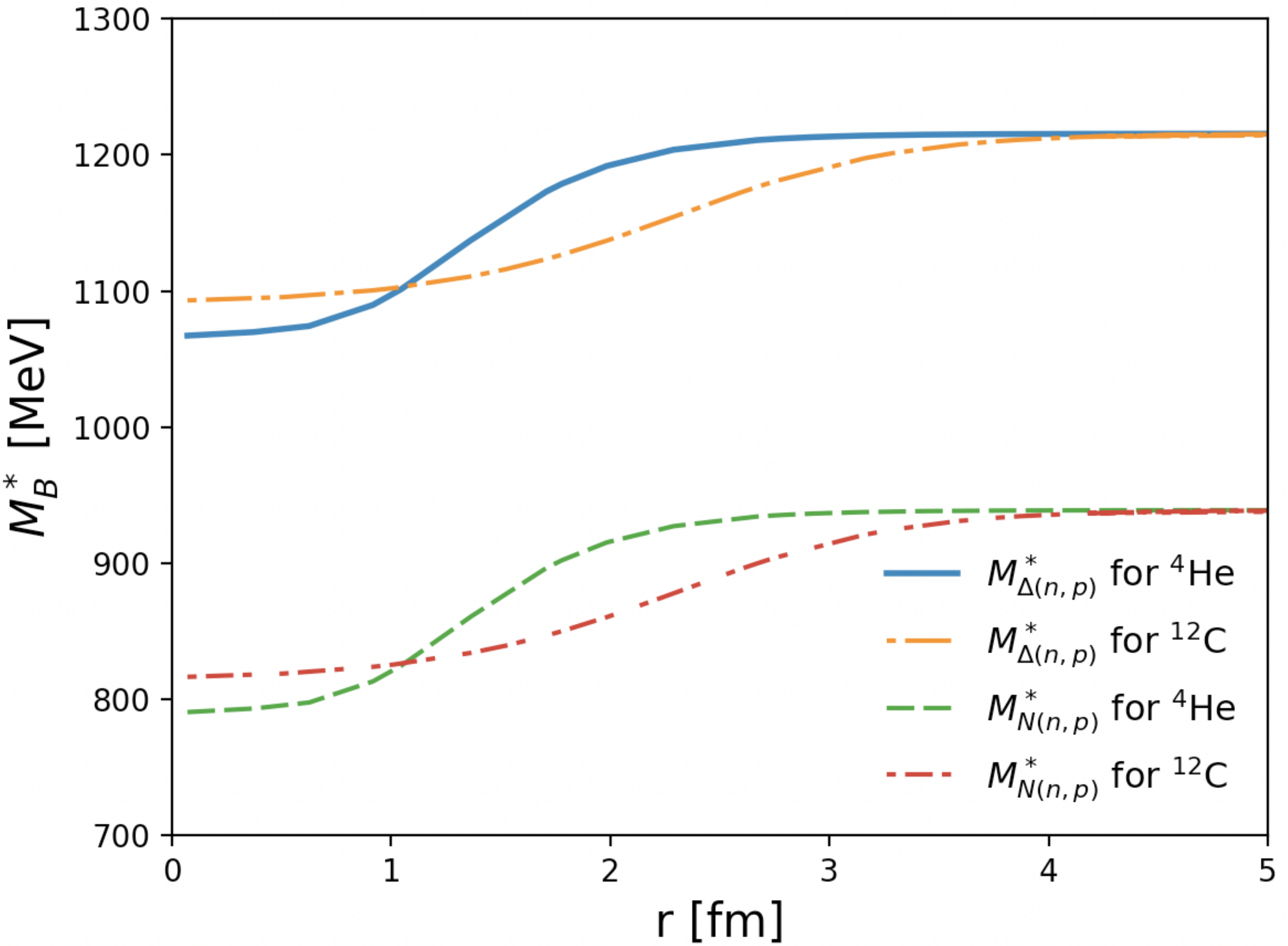}
  \caption{ \label{fig14} Effective baryon mass $M^*_B$ $(B = N,\Delta)$ as a function of $r$ for two kinds of nuclei ($^4\mathrm{He},\displaystyle^{12}\mathrm{C}$). In both two nuclei cases, $M^*_{B(n)}=M^*_{B(p)}$ because the number of neutrons and protons is equal.}
\end{figure}

In this study, I parameterized the effective baryon mass $M^*_B$ with quadratic functional form with respect to $\rho^{(n,p)}_A$, which is given by
\begin{eqnarray}
\label{eqN19}
    M^*_B=M_B+C_1\rho^{(n,p)}_A+C_2(\rho^{(n,p)}_A)^2,\; \hspace{1cm}~~B=(\Delta~ \mathrm{and}~N),
\end{eqnarray}
where $M_B$ denotes the baryon mass in a vacuum and $C_1=-1543.08\;[\mathrm{MeV\cdot fm^3}]$, $C_2=2036.14\;[\mathrm{MeV\cdot fm^3}]$ indicate the coefficients which are parameterized to define the effective baryon mass by using the QMC model.
The effective baryon masses $M^*_B$ as a function of $r$ are indicated in Fig.~\ref{fig14}. It is noticed that the effective baryon mass for $^4\mathrm{He}$ is smaller than $^{12}\mathrm{C}$ near the nuclear center because of the density distribution difference as seen in Fig.~\ref{fig13}. In Fig.~\ref{fig14}, as $r$ increases, the tendency of the distribution is reversed, and finally, effective baryon masses converge to the vacuum mass if it extends beyond a certain distance.

I then study the total cross-section of the elastic $\pi^+p$ scattering at finite baryon density, which uses a similar procedure as the TCS calculation for the elastic $\pi p$ scattering in a nuclear medium. Also similarly, here I also consider the momentum-dependent delta decay width $\Gamma^*_\Delta$ \cite{Larionov:2003av}. However, in the finite nuclei case, the decay width should be determined by the nuclear density distribution $\rho_A$, and the following form is used:
\begin{align}
\label{eqN20}
    \Gamma^*_\Delta(\sqrt{s^*},\rho_A) = \Gamma_{\mathrm{sp}} \left(\frac{\rho_A}{\rho_0}\right)&+\Gamma^0_\Delta\left[\frac{q(M^*_N,M_\pi,\sqrt{s^*})}{q(M_N,M_\pi,M_\Delta)}\right]^3  \nonumber \\
    &\times \frac{M^*_\Delta}{\sqrt{s^*}}\frac{\beta^2_0+q^2(M^*_N,M_\pi,M^*_\Delta)}{\beta^2_0+q^2(M^*_N,M_\pi,\sqrt{s^*})}.\quad
\end{align}

\begin{figure}[t]
  \centering
 \stackinset{r}{2.0cm}{t}{3.5cm}{(a)}{ \includegraphics[width=0.48\textwidth]{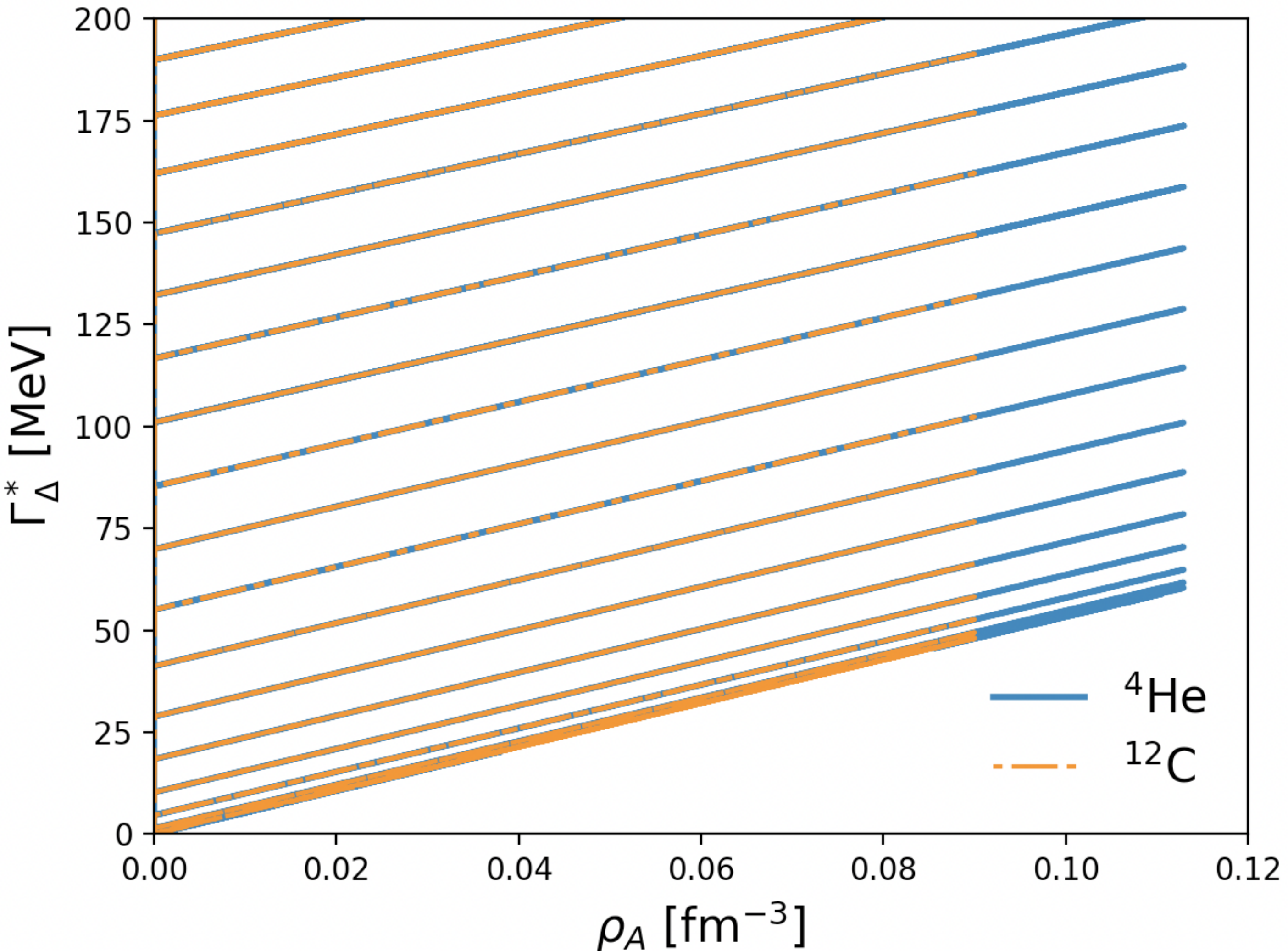}}
 \stackinset{r}{1.1cm}{t}{1.3cm}{(b)}{ \includegraphics[width=0.48\textwidth]{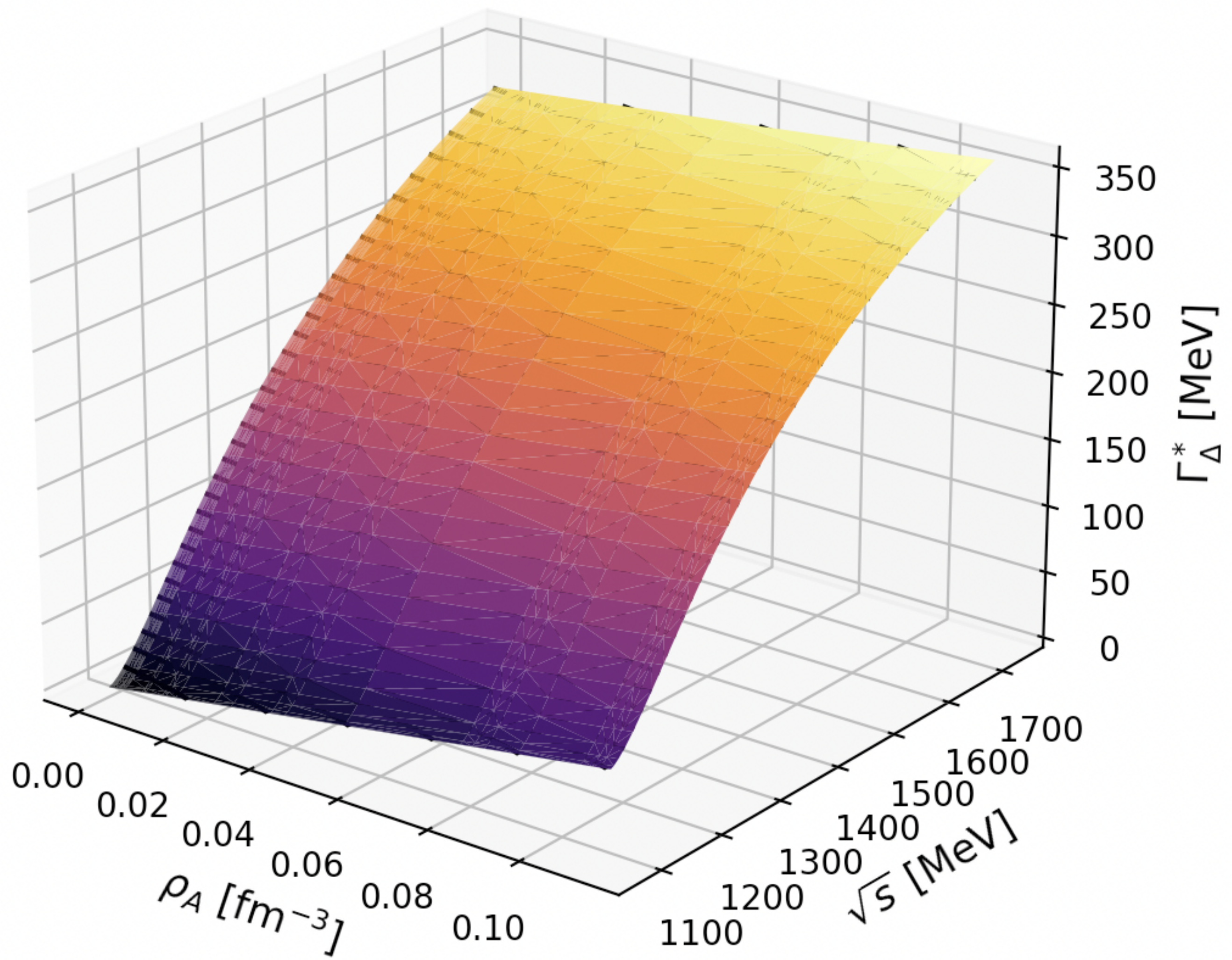}}
  \caption{ \label{fig15} (a) In-medium $\Delta$ decay width $\Gamma^*_\Delta$ as a function of $\rho_A$ for two kinds of nucleus ($^4\mathrm{He},\displaystyle^{12}\mathrm{C}$) as in Eq.~\protect\ref{eqN20}. Each line denotes the in-medium delta decay width $\Gamma^*_\Delta$ with specific energy $\sqrt{s^*}$. (b) The 3D graph of the in-medium $\Delta$ decay width $\Gamma^*_\Delta$ as a function of $\rho_A$ and $\sqrt{s^*}$.}
\end{figure}

Panel (a) of Fig.~\ref{fig15} represents the in-medium delta decay width $\Gamma^*_\Delta$ as a function of $\rho_A$ for $^4\mathrm{He}$ and $^{12}\mathrm{C}$. By the definition of $\Gamma^*_\Delta$, the density distribution of in-medium delta decay width $\Gamma^*_\Delta$ is the same for both nuclei. However, $\Gamma^*_\Delta$ of each nucleus has different values depending on their nuclear density distribution $\rho_A$. It is shown that $\Gamma^*_\Delta$ for $^4\mathrm{He}$ (blue solid line) is defined in a higher $\rho_A$ region compared to the $\Gamma^*_\Delta$ for $^{12}\mathrm{C}$ (orange dashed line) because the maximum value of $\rho_A$ for $^4\mathrm{He}$ is bigger than $^{12}\mathrm{C}$ as depicted in Fig.~\ref{fig13}. Also as seen in Fig.~\ref{fig2}, the delta decay width in the nuclear medium increases as the nuclear density distribution increases due to the same formula of $\Gamma^*_\Delta$, Eq.~(\ref{eqN12}) and ~(\ref{eqN20}). Each line describes the in-medium delta decay width corresponding to the specific energy, $\sqrt{s^*}$.

In panel (b) of Fig.~\ref{fig15}, it is shown that the 3D graph of the in-medium delta decay width $\Gamma^*_\Delta$ as a function of $\rho_A$ and $\sqrt{s^*}$, which is represented by the $\sqrt{s^*}$ dependence in $\Gamma^*_\Delta$ continuously. As previously mentioned, the behavior of $\Gamma^*_\Delta$ follows the description of Fig.~\ref{fig2} and Eq.~(\ref{eqN12}). I found that as $\sqrt{s^*}$ increases, the in-medium $\Delta$-resonance decay width $\Gamma^*_\Delta$ also increases, which is consistent with the calculation of Ref.~\cite{Cui:2020fhr}.

\begin{figure}[t]
  \centering
 \includegraphics[width=0.65\textwidth]{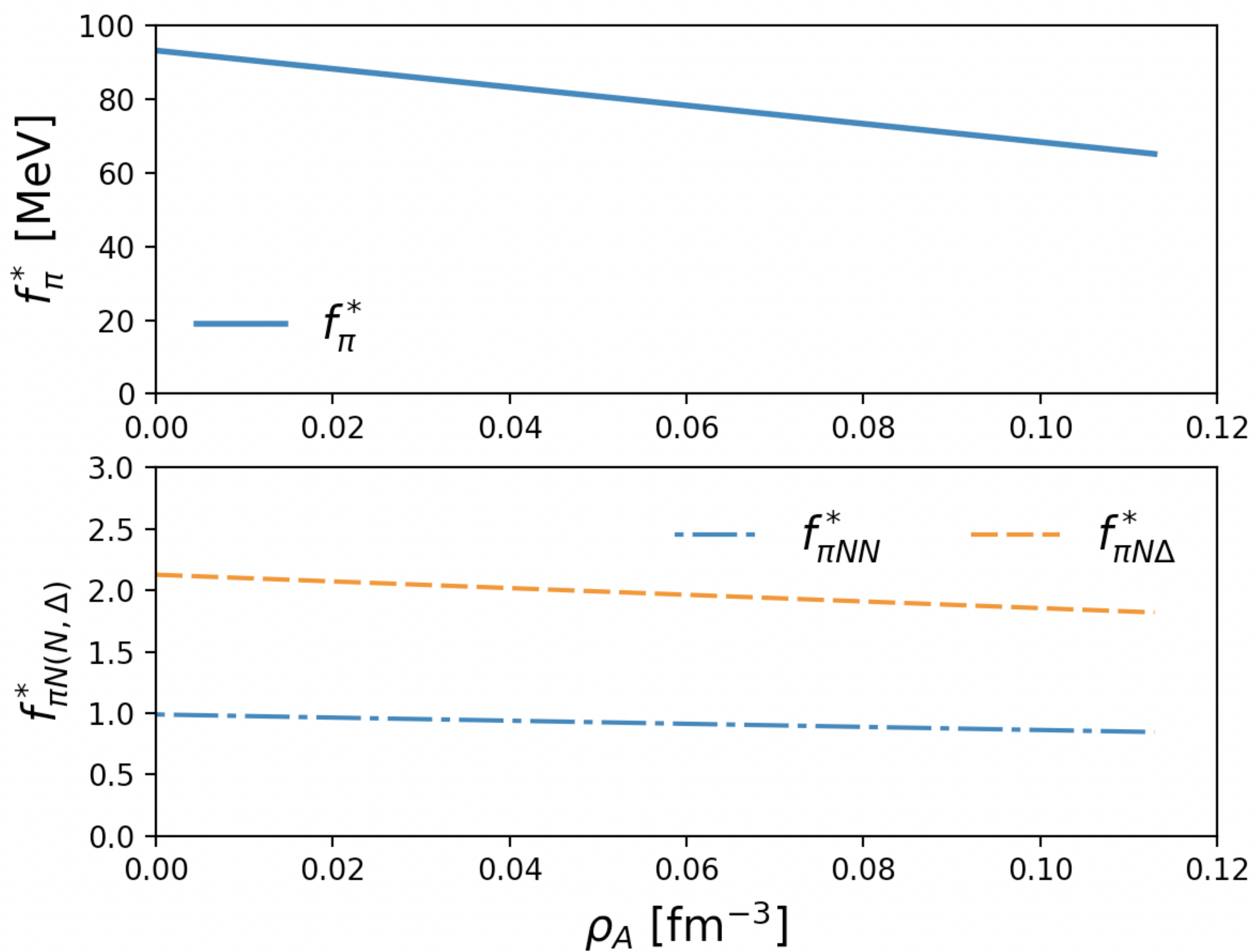}
  \caption{ \label{fig16a} Pion decay constant as a function of $\rho_A$ (upper panel) and in-medium coupling constants $f^*_{\pi NN}$ (blue dot-dashed line), $f^*_{\pi N \Delta}$ (orange dashed line) as a function of $\rho_A$ (lower panel).}
\end{figure}

Also in Sec.~III, the more realistic in-medium coupling constants are applied in this work, which do not consider in the first study. Using the Goldberger-Treiman relation (GTR), the weak axial-vector coupling constant $g_A$ is given by $g_A=\frac{g_{\pi NN}f_\pi}{M_N}$. Simply, in medium, it can be written as $g^*_A=\frac{g^*_{\pi NN}f^*_\pi}{M^*_N}$. From the combination of these vacuum and medium equations, the final expression for the in-medium coupling constants is given by
\begin{eqnarray}
\begin{split}
    \label{eqN20a}
    f^*_{\pi NN}&=\frac{g^*_A}{g_A}\frac{f_\pi}{f^*_\pi}\frac{M^*_N}{M_N}f_{\pi NN}, \\ f^*_{\pi N\Delta}&\simeq \left(\frac{f_{\pi N \Delta}}{f_{\pi NN}}\right)f^*_{\pi NN}= 2.15\times\frac{g^*_A}{g_A}\frac{f_\pi}{f^*_\pi}\frac{M^*_N}{M_N}f_{\pi NN},
\end{split}
\end{eqnarray}
where the relation $f_{\pi NN}=\sqrt{4\pi} g_{\pi NN}$ is used.
At $\rho_B$ = $\rho_0$, $g^*_A \simeq$ $0.9 g_A$ that obtained from several models~\cite{Lu:2001mf} and consistent with the empirical data of the pion atomic experiment~\cite{Kienle:2004hq}, $f^*_\pi \simeq$ $0.8 f_\pi$ taken from Ref.~\cite{Kienle:2004hq}, and $M^*_N\simeq$ $0.8 M_N$ obtained from the QMC model~\cite{Guichon:2018uew,Hutauruk:2018qku}. Therefore, the relevant in-medium coupling constants and the pion decay constant with medium effects are calculated by
\begin{eqnarray}
\begin{split}
  \label{eqN20b}
  f^*_{\pi NN}(\rho_A) &= f_{\pi NN} - 0.633\rho_A,\\
  f^*_{\pi N \Delta}(\rho_A) &= f_{\pi N\Delta} - 1.36\rho_A,\\
  f^*_\pi(\rho_A) &= f_\pi - 124.27\rho_A,
\end{split}
\end{eqnarray}
where $f_{\pi NN}$ = 0.989, $f_{\pi N \Delta}$ = 2.127, and $f_\pi = 93.2$ MeV are used, same as Sec.~II., which are given by the Nijmegen potential~\cite{Gasparyan:2003fp} and the experimental data~\cite{Janssen:1996kx}. In-medium coupling constants $f^*_{\pi NN}$, $f^*_{\pi N \Delta}$ and the pion decay constant with medium effects as a function of $\rho_A$ are provided in Fig.~\ref{fig16a}.

\section{Numerical result: elastic $\pi^+ A$ scattering with constant $\Gamma^0_\Delta$}
In this section, I present the numerical results for the elastic $\pi^+A$ scattering for $^4\mathrm{He}$ and $^{12}\mathrm{C}$ with the effective baryon mass $M^*_B$ that is calculated in the QMC model as well as the in-medium modifications of $\Delta$-resonance decay width $\Gamma^*_\Delta$. In the elastic $\pi^+ A$ scattering, the cross-section is calculated with the constant $\Gamma^0_\Delta$(=94.0 MeV). Such the $\Delta$-decay width form is also used in Ref.~\cite{Larionov:2003av}. I also consider the elastic $\pi^+ A$ scattering with $\Gamma^0_\Delta(\sqrt{s},\rho_A)$. The detail on the calculation of cross-section for the elastic $\pi^+ A$ scattering with $\Gamma^0_\Delta(\sqrt{s},\rho_A)$ will be explained in Sec.~\ref{sec:scat2}. Here I concentrate on reporting the results for the constant $\Gamma^0_\Delta$.

\begin{figure}[t]
  \centering
 \stackinset{r}{1.1cm}{t}{2.1cm}{(a)}{ \includegraphics[width=0.54\textwidth]{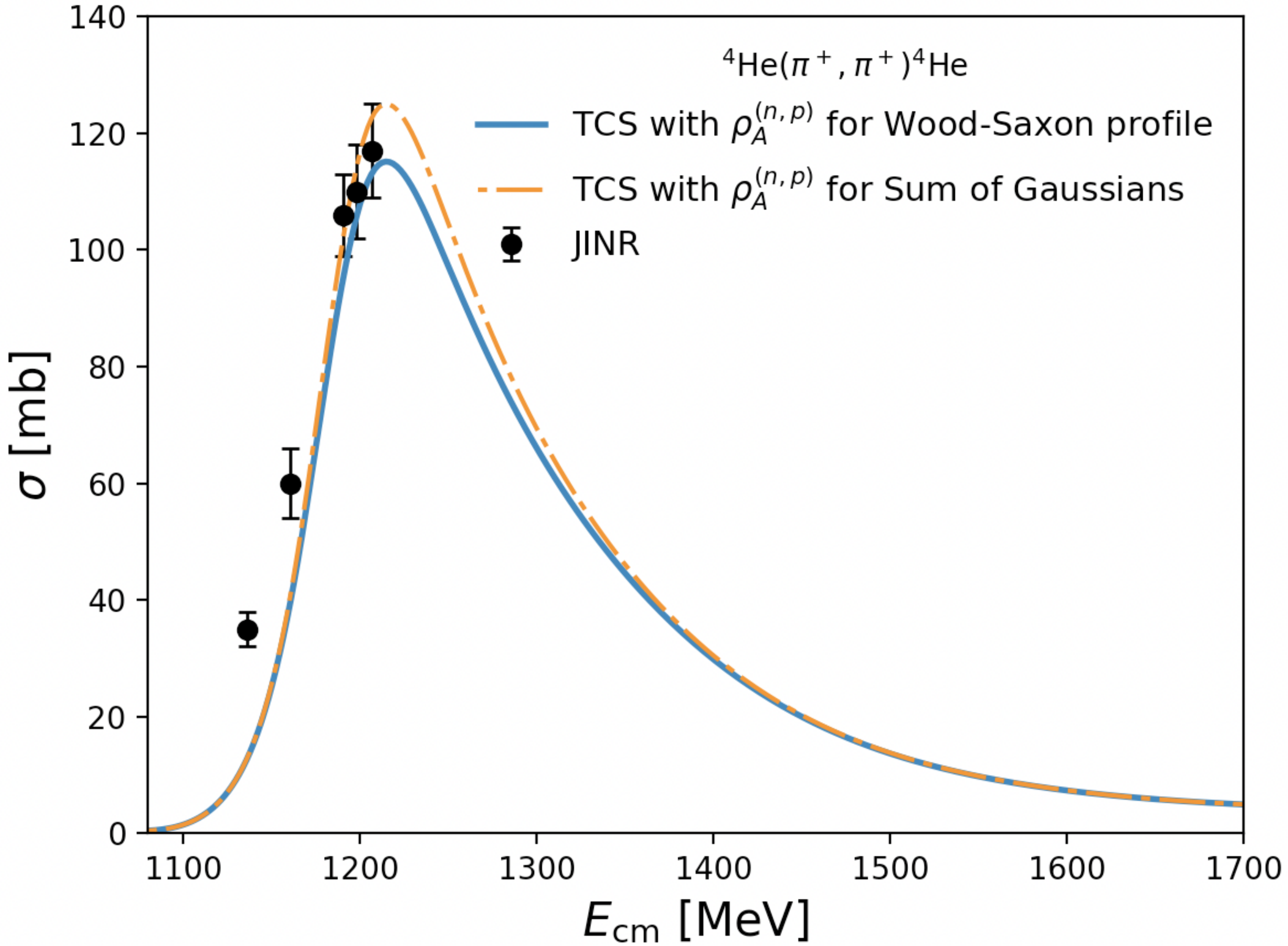}}
 \stackinset{r}{1.1cm}{t}{2.1cm}{(b)}{ \includegraphics[width=0.54\textwidth]{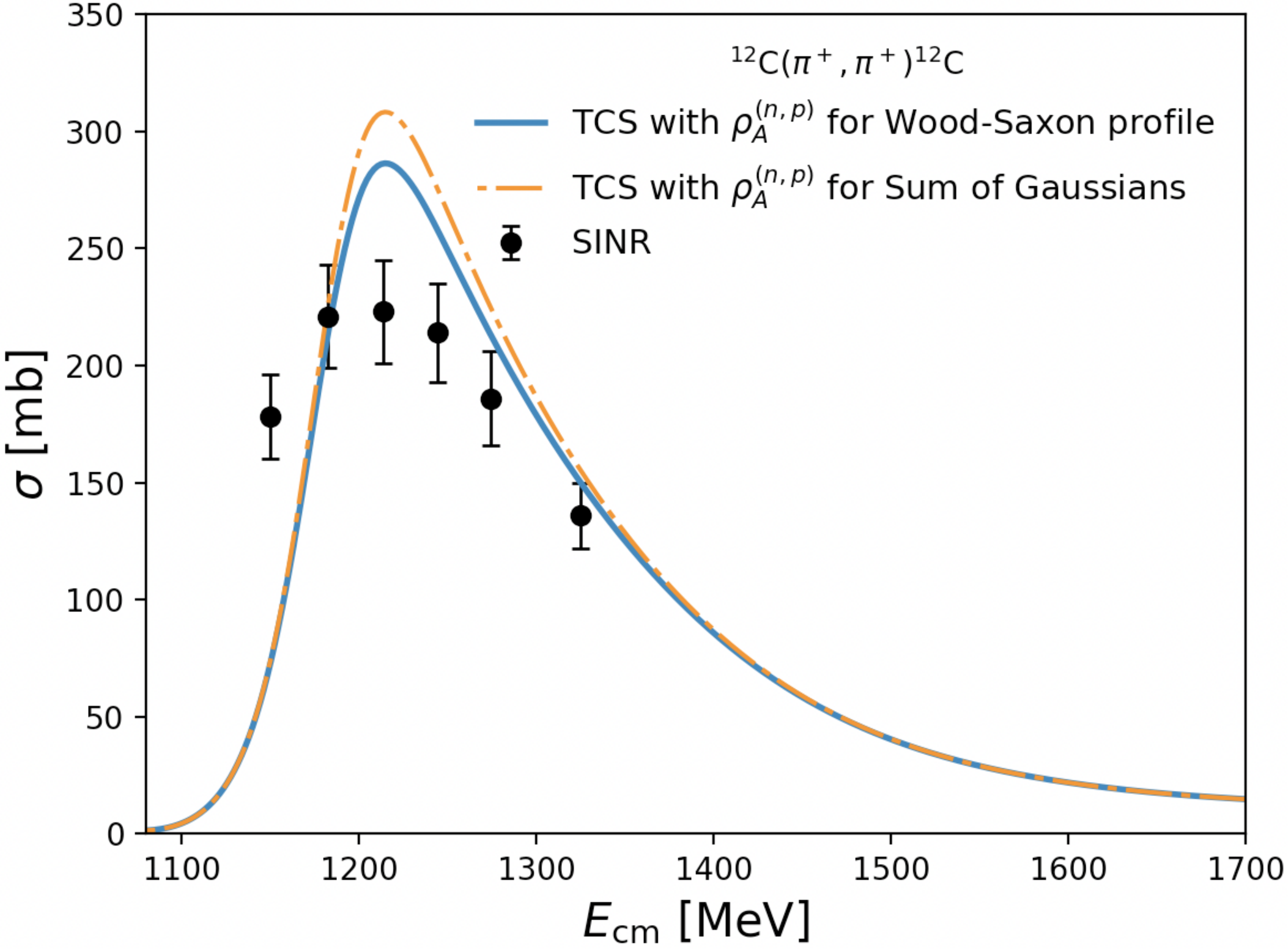}}
  \caption{ \label{fig16} (a) Total cross-section of the elastic $\pi^+$--$^4\mathrm{He}$ scattering and (b) the total cross-section of the elastic $\pi^+$--$^{12}\mathrm{C}$ with two types of $\rho^{(n,p)}_A$, the Wood-Saxon density profile and the nuclear density distribution which is given by Sum of Gaussians. The experimental data are taken from the JINR and SINR for the elastic $\pi^+ A$ scattering process.}
\end{figure}

The total cross-sections for the elastic $\pi^+A$ scattering with proton and neutron nuclear density distribution $\rho^{(n,p)}_A$ for $^4\mathrm{He}$ (upper panel) and $^{12}\mathrm{C}$ (lower panel) are depicted in Fig.~\ref{fig16}. In both figures, the blue solid line denotes the TCS with Wood-Saxon density profile $\rho_A$ as shown in Eq.~\eqref{eqN17}. The orange dot-dashed line in Fig.~\ref{fig16} represents the TCS with $\rho_A$ which is given by the sum of Gaussians (SOG). The SOG method parameterizes the charge density distribution $\rho(r)$ as a function of $r$ given through an expansion of a summation of Gaussian functions. 

With this approach, the definition of $\rho^{(n,p)}_A(r)$ is given by
\begin{align}
\label{eqN21}
    \rho^{(n,p)}_A(r)&=\sum_i A_i\left[\exp{\left(-\left\{\frac{(r+R_i)}{\gamma}\right\}^2\right)}+\exp{\left(-\left\{\frac{(r-R_i)}{\gamma}\right\}^2\right)}\right]\times\frac{(A-Z,Z)}{A}, \nonumber \\ A_i&=\frac{AQ_i}{2\pi^{3/2}\gamma^3\left[1+2\left(R_i/\gamma\right)^2\right]},
\end{align}
where $R_i$ and $Q_i$ denote respectively the position and the amplitude of the Gaussians which are fitted to the data. The values of $Q_i$ indicate the fraction of the charge in the $i$th Gaussian, which must satisfy the normalization condition $\sum_i Q_i=1$. Also, $\gamma$ represents the width of the Gaussians which is equal to the smallest width of the peaks in the nuclear radial wave functions that are calculated in the Hartree-Fock method~\cite{osti:6477756}. The values for the nuclear radial wave functions can be found in Ref.~\cite{osti:6477756}.

In panel (a) of Fig.~\ref{fig16}, the numerical results for the TCS of elastic $\pi^+$--$^4\mathrm{He}$ scattering are shown. Both TCS describe the experimental data of the elastic scattering process quite well. However, the TCS of elastic $\pi^+$--$^{12}\mathrm{C}$ scattering as shown in the panel (b) of Fig.~\ref{fig16} are rather overestimated the experimental data for the elastic scattering. This is expected because, in this work, the $NN$ correlation is not included yet. I only focus on the elastic (elastic) scattering. In addition, unlike the common calculation of the Glauber model which is mostly used in the $NN$ and the $AA$ scatterings, in this work I calculate the $\pi A$ scattering, which is a very interesting and fascinating topic because of the scarce experimental data. In finite nuclei, the contributions from such interaction (correlation)
between nucleons like ($n$-$n$), ($p$-$p$), ($n$-$p$) should be considered. For $^{12}\mathrm{C}$, which has a large number of nucleons compared to $^{4}\mathrm{He}$, such correlation effects become stronger among the nucleons inside the nucleus. This might be one of the strong reasons that our results overestimate the experimental data for the TCS of $^{12}\mathrm{C}$. 

Again, for a better understanding of the elastic $\pi^+A$ scattering processes, the more realistic descriptions of the system should be considered such as $NN$ interaction in finite nuclei, the multiple scattering contributions which are occurred in the nucleus, and so forth. Last but not least, the results of this study are expected to be relevant for the relativistic heavy-ion collision experiments as well as matter at higher density.

\section{Numerical result: elastic $\pi^+ A$ scattering with $\Gamma^{0}_\Delta(\sqrt{s^*},\rho_A)$} \label{sec:scat2}
This section contains the further investigation of the TCS of the elastic $\pi^+ A$ scattering, with $\Gamma^0_\Delta(\sqrt{s^*},\rho_A)$ which is a more realistic $\Delta$-decay width. In the $\Gamma_\Delta^*$ expression in Eq.~\eqref{eqN20}, $\Gamma^0_\Delta$ is replaced with $\Gamma^0_\Delta(\sqrt{s^*},\rho_A)$, which is calculated from the $\Delta$ decay process ($\Delta\rightarrow\pi N$)~\cite{Kim:1996ada} with density effects. It is then given by
\begin{equation}
\Gamma^0_\Delta(\sqrt{s^*},\rho_A)=\frac{\left(f^*_{\pi N \Delta}\right)^2 \mathcal{I}_{\pi N \Delta}}{12\pi M_\pi^2 M^*_\Delta}(E^*_N+M^*_N)\cdot{|\overrightarrow{p}^*_N|}^3,
\end{equation}
where $\overrightarrow{p}^*_N$ and $E^*_N$ indicate the three-momenta of the nucleon in the medium, in which the magnitude is equal to $q(M^*_N, M_\pi, M^*_\Delta)$ in Eq.~\eqref{eqN13}, and the energy of the nucleon with medium effects, represented by $E^*_N=\sqrt{{M^*_N}^2+{|\overrightarrow{p}^*_N|}^2}$, respectively. The quantity of $\mathcal{I}_{\pi N \Delta}$ denotes the isospin factor for $\pi N \Delta$ vertex, where the values of $\mathcal{I}_{\pi N \Delta}$ are equal to $1$ for the $s$-channel, and $1/3$ for the $u$-channel.

Thus, the new expression for $\Delta$-decay width can be defined by
\begin{align}
    \tilde{\Gamma}^*_\Delta(\sqrt{s^*},\rho_A) =\Gamma_{\mathrm{sp}} \left(\frac{\rho_A}{\rho_0}\right)&+\Gamma^0_\Delta(\sqrt{s^*},\rho_A)\left[\frac{q(M^*_N,M_\pi,\sqrt{s^*})}{q(M_N,M_\pi,M_\Delta)}\right]^3 \nonumber \\ &\qquad\times\frac{M^*_\Delta}{\sqrt{s^*}}\frac{\beta^2_0+q^2(M^*_N,M_\pi,M^*_\Delta)}{\beta^2_0+q^2(M^*_N,M_\pi,\sqrt{s^*})}.
\end{align}

\begin{figure}[t]
  \centering
 \stackinset{r}{1.1cm}{t}{2.1cm}{(a)}{ \includegraphics[width=0.485\textwidth]{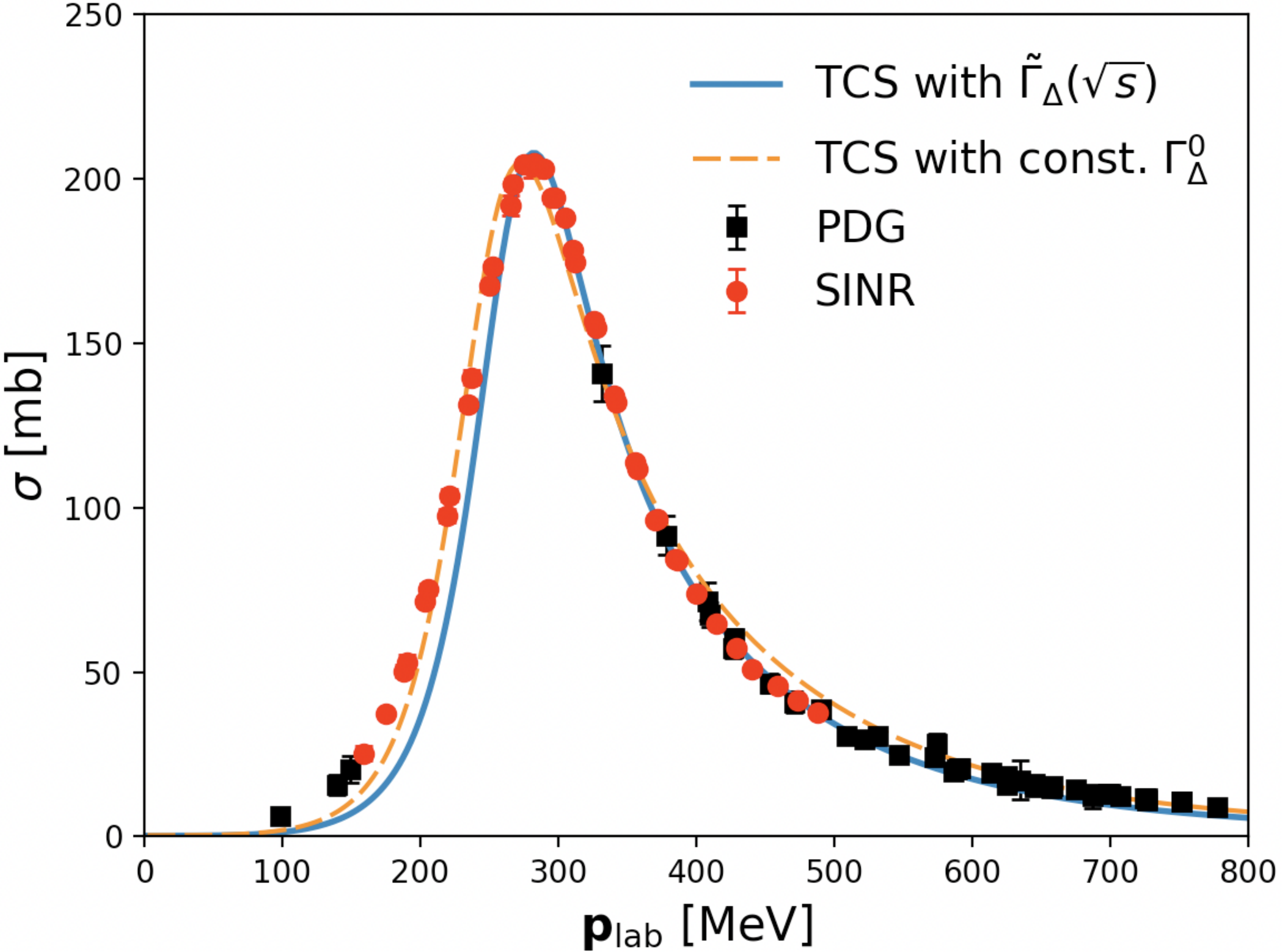}}
 \stackinset{r}{1.1cm}{t}{0.8cm}{(b)}{ \includegraphics[width=0.485\textwidth]{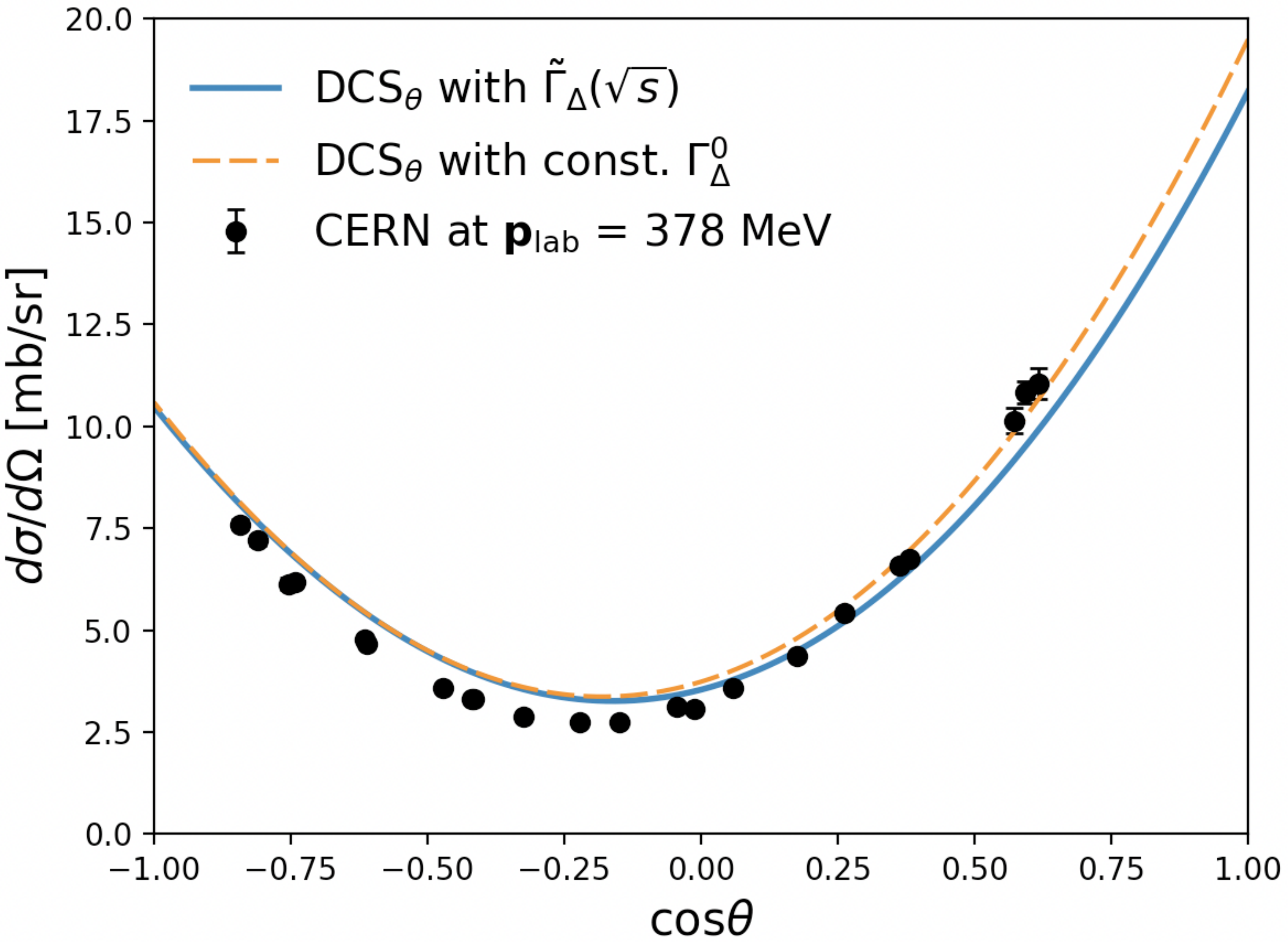}}
  \caption{ \label{fig17} (a) Total cross-section of the elastic $\pi^+ p$ scattering with two kinds form of $\Gamma_\Delta$ and (b) differential cross-section as a function of $\cos\theta$ of the elastic $\pi^+ p$ scattering with two kinds form of $\Gamma_\Delta$ at $\mathbf{p}_{\mathrm{lab}} = 378$ MeV. The blue solid lines denote results with $\tilde{\Gamma}_\Delta(\sqrt{s})$ and the orange dashed lines indicate results with const. $\Gamma^0_\Delta = 94.0$ MeV.}
\end{figure}

Here I emphasize that when I calculate a cross-section with $\tilde{\Gamma}_\Delta$ in a vacuum, the values of $M_\Delta$ and $f_{\pi N \Delta}$ are determined by reproducing the experimental data for the elastic $\pi^+ p$ scattering. So, with $\tilde{\Gamma}_\Delta$, I obtain the TCS and DCS as a function of $\cos\theta$ for the elastic $\pi^+ p$ scattering in a vacuum as shown in Fig.~\ref{fig17}.

In Fig.~\ref{fig17}, the TCS results (blue solid line) are calculated with $M_\Delta = 1222.0$ MeV, $f_{\pi N \Delta} = 1.97$, and compared to the previous results (orange dashed line) in Fig.~\ref{fig1}. Despite different values of $f_{\pi N \Delta}$, $\tilde{\Gamma}_{\Delta}$ can explain the elastic $\pi^+ p$ scattering process quite well with a more realistic $M_\Delta$ value, and also, without one free parameter, $\Gamma^0_\Delta = 94.0$ MeV.

\begin{figure}[t]
  \centering
 \includegraphics[width=0.6\textwidth]{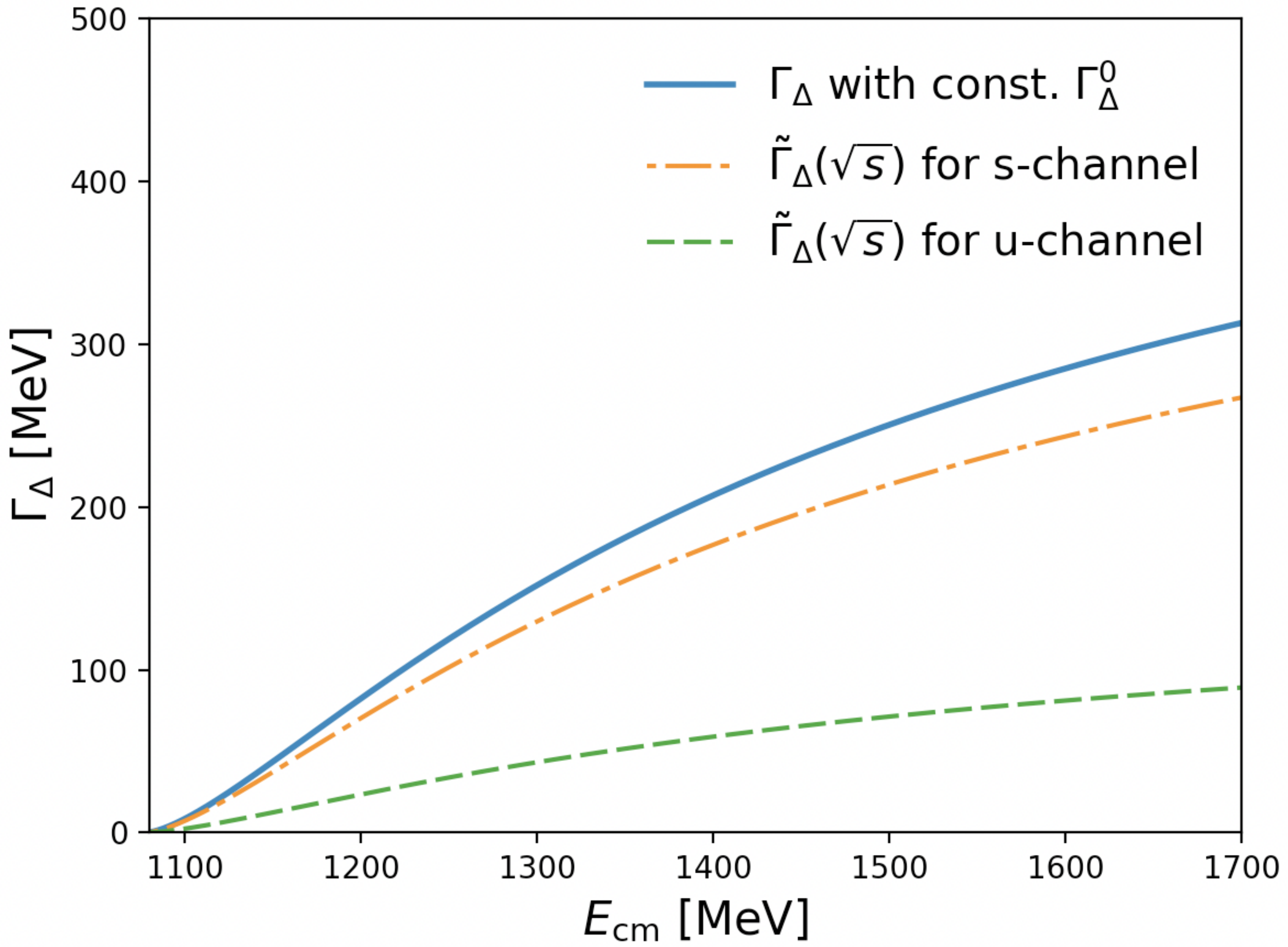}
  \caption{ \label{fig18}~Momentum dependent $\Delta$ decay width with const. $\Gamma^0_\Delta = 94.0$ MeV (blue solid line), $\tilde{\Gamma}_\Delta(\sqrt{s})$ for $s$-channel (orange dot-dashed line) and $\tilde{\Gamma}_\Delta(\sqrt{s})$ for $u$-channel (green dashed line) as a function of the energy in CM frame.}
\end{figure}

Here, it is interesting to compare the energy dependence of two different forms of $\Delta$-decay width, and to analyze the density dependence of in-medium $\Delta$ decay width. Fig.~\ref{fig18} represents two kinds of the momentum-dependent $\Delta$ decay width as a function of $E_{\mathrm{cm}}$. In Fig.~\ref{fig18}, the blue solid line describes $\Delta$ decay width with constant of $\Gamma^0_\Delta = 94.0$ MeV, and the orange dot-dashed line and green dashed line indicate $\tilde{\Gamma}_\Delta(\sqrt{s})$ for $s$-channel ($\mathcal{I}_{\pi N \Delta} = 1$) and $u$-channel ($\mathcal{I}_{\pi N \Delta} = 1/3$), respectively. Both kinds of decay width increase as the energy increases, but it is clearly shown that $\tilde{\Gamma}_\Delta$ is slowly increased with the center-mass energy compared to $\Gamma_\Delta$ with a constant of $\Gamma^0_\Delta$.

\begin{figure}[t]
  \centering
 \stackinset{r}{2.0cm}{t}{3.5cm}{(a)}{ \includegraphics[width=0.48\textwidth]{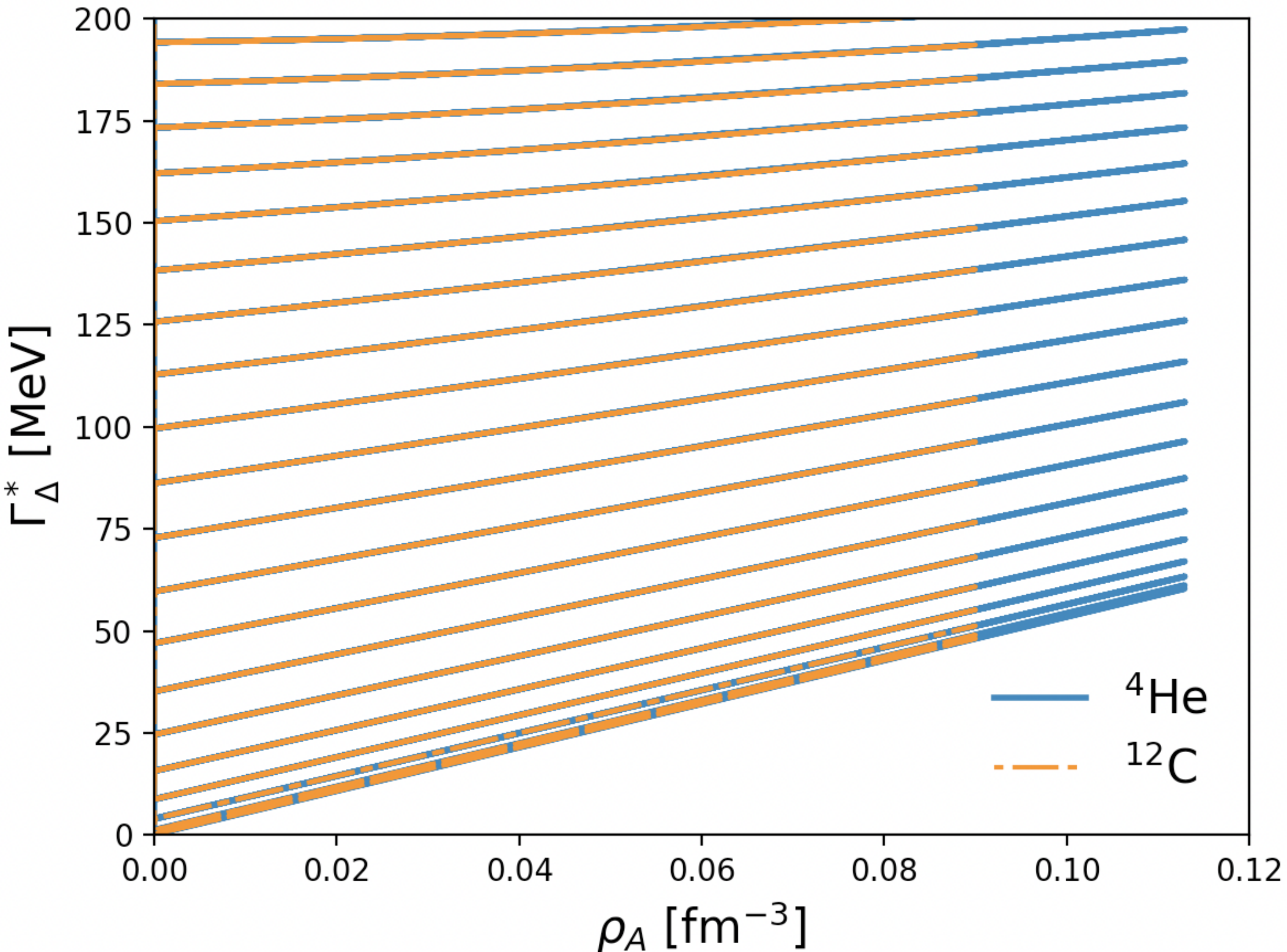}}
 \stackinset{r}{1.1cm}{t}{1.5cm}{(b)}{ \includegraphics[width=0.48\textwidth]{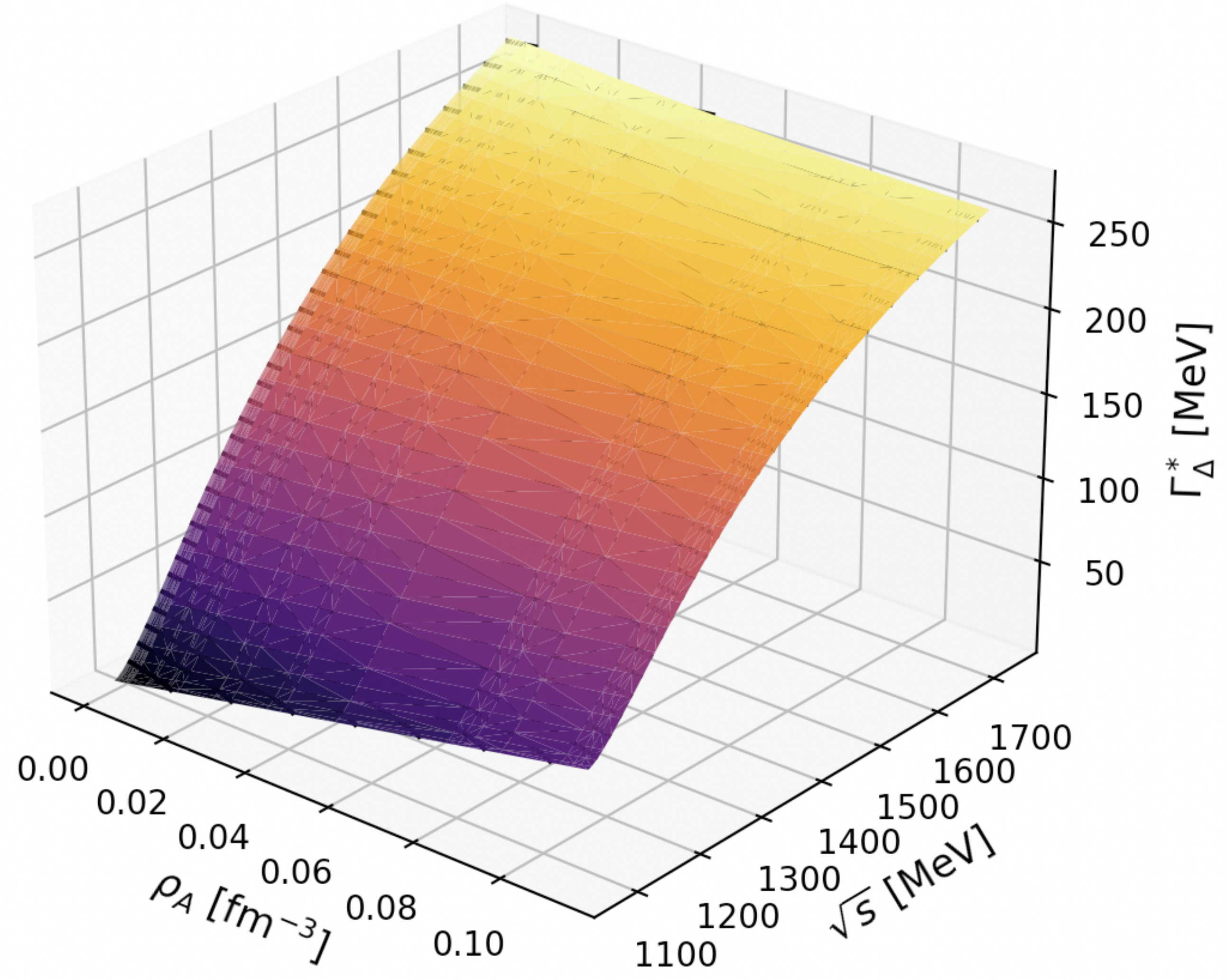}}
  \caption{ \label{fig19}~(a) In-medium $\Delta$ decay width $\tilde{\Gamma}^*_\Delta$ as a function of $\rho_A$ for two kinds of nucleus ($^4\mathrm{He},\displaystyle^{12}\mathrm{C}$). Each line denotes $\tilde{\Gamma}^*_\Delta$ with specific energy $\sqrt{s^*}$. (b) The 3D graph of  $\tilde{\Gamma}^*_\Delta$ as a function of $\rho_A$ and $\sqrt{s^*}$.}
\end{figure}

Now, the study is extended to the finite nuclei case, which is the subject of this section. By changing the coefficient in the linear equation of $f^*_{\pi N \Delta}$ to $f^*_{\pi N \Delta} = 1.97 - 1.26\rho_A$, 
I calculate the in-medium modifications of $\Delta$ decay width $\tilde{\Gamma}^*_\Delta$ at finite nuclei as a function of $\rho_A$ and $\sqrt{s^*}$, which are depicted in Fig.~\ref{fig19}.

Because the $s$-channel of $\Delta^{++}$ contribution is dominant in $\pi^+ p$ channel near the threshold energy region, $\tilde{\Gamma}^*_\Delta$ with $\mathcal{I}_{\pi N \Delta} = 1$ is indicated in Fig.~\ref{fig19}. In panel (b) of Fig.~\ref{fig19}, it is clearly shown that density dependence of $\tilde{\Gamma}^*_\Delta$ is rather similar to the previous form of $\Gamma^*_\Delta$ around $\Delta(1232)$ energy region, but, in higher energy region, the tendency about the density dependence of $\tilde{\Gamma}^*_\Delta$ is reversed. Fig.~\ref{fig18} provides a good explanation for this behavior of $\tilde{\Gamma}^*_\Delta$. In a vacuum, even though $\tilde{\Gamma}_\Delta$ increases as the energy increase, the rate of increase is smaller than the previous form of $\Gamma_\Delta$. Next, in the finite nuclei, the medium quantities like $M^*_B$, $f^*_{\pi N \Delta}$ are diminished. Such opposite tendencies (behaviors) are competing with each other in the medium, and consequently, as the energy increases, the decreasing power overcomes the increasing power. However, in this study, these properties barely appear because of the dominance of $\Delta(1232)$ resonance contribution.

\begin{figure}[t]
  \centering
 \stackinset{r}{1.1cm}{t}{2.1cm}{(a)}{ \includegraphics[width=0.55\textwidth]{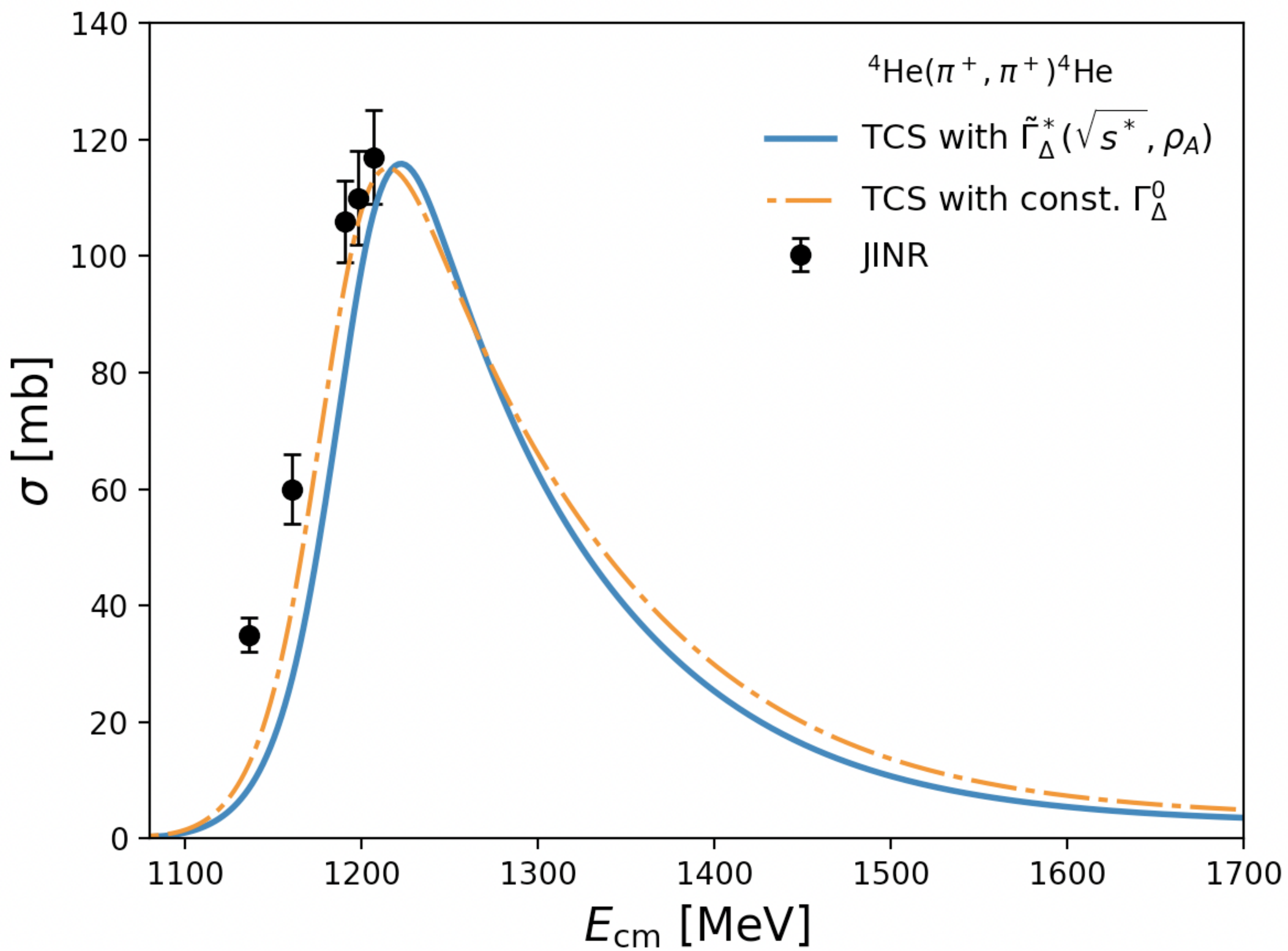}}
 \stackinset{r}{1.1cm}{t}{2.1cm}{(b)}{ \includegraphics[width=0.55\textwidth]{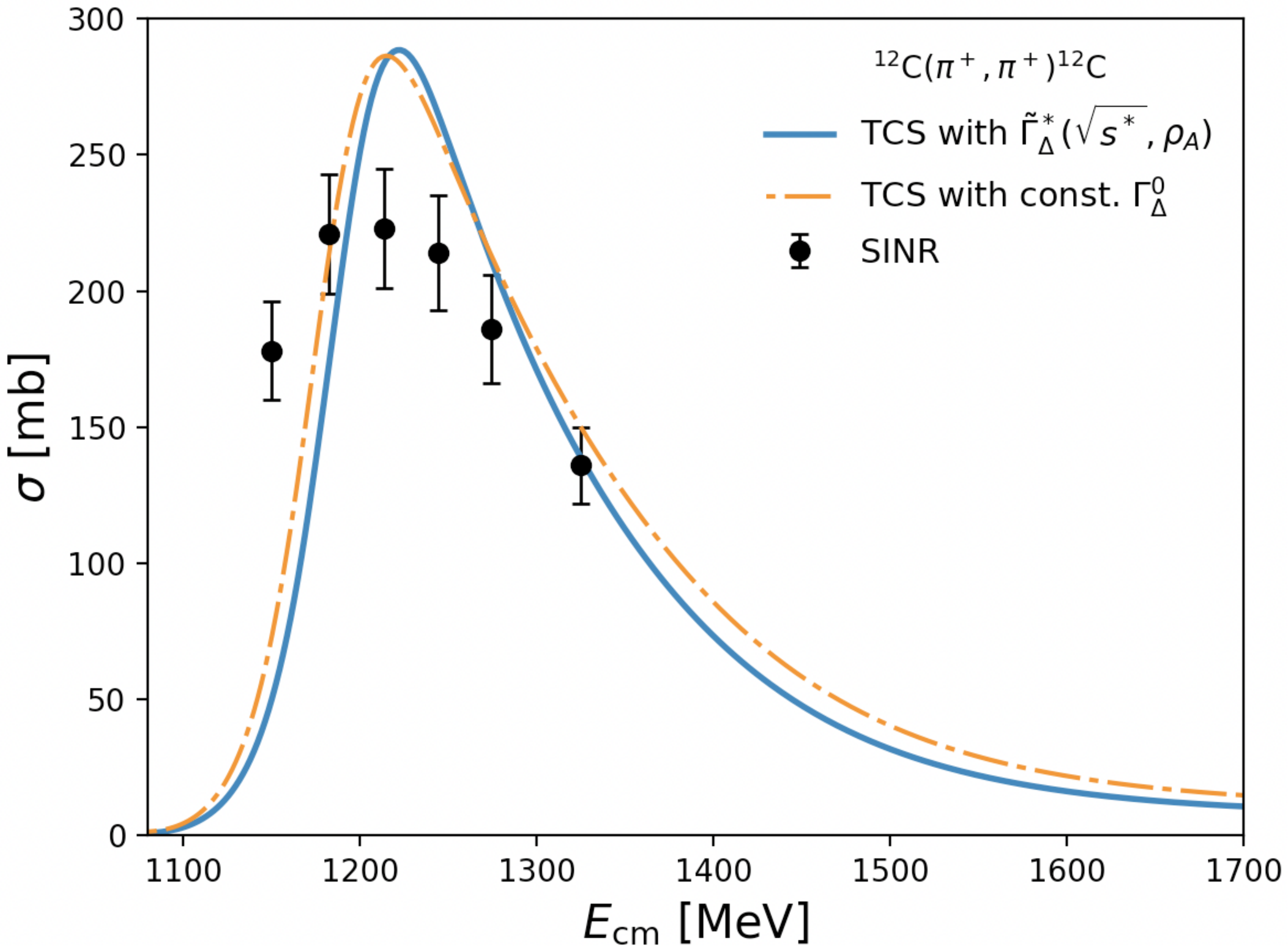}}
  \caption{ \label{fig20}\;Total cross-section of the elastic $\pi^+ A$ scattering (a) for $^4\mathrm{He}$ and (b) for $^{12}\mathrm{C}$ with $\tilde{\Gamma}^*_\Delta(\sqrt{s^*},\rho_A)$ (blue solid line), with const. $\Gamma^0_\Delta$ (orange dot-dashed line).}
\end{figure}

Results for the total cross-sections of the elastic $\pi^+ A$ scattering with considering the medium effects and $\tilde{\Gamma}^*_\Delta(\sqrt{s^*},\rho_A)$ for $^4\mathrm{He}$ and $^{12}\mathrm{C}$ are depicted in Fig.~\ref{fig20}, compared with the previous results of TCS. Similar to Fig.~\ref{fig16}, panels (a), and (b) of Fig.~\ref{fig20} denote the TCS of the elastic $\pi^+ A$ scattering with density effects for $^4\mathrm{He}$, and for $^{12}\mathrm{C}$, respectively. In both panels of Fig.~\ref{fig20}, the blue solid lines indicate the TCS of the elastic $\pi^+ A$ scattering with medium effects including $\tilde{\Gamma}^*_\Delta(\sqrt{s^*},\rho_A)$, and the orange dot-dashed lines denote the TCS of previous results in Fig.~\ref{fig16}. Because of using a different form of $\Delta$ decay width, the width of each cross-section is quite different. But, it is clearly shown that the magnitude of the TCS is almost the same, quantitatively. Results show that, despite changing $\Delta$ decay width form, the results for the TCS of the elastic $\pi^+$-$^{12}\mathrm{C}$ scattering remain unclear, compared to the experimental data. Therefore, the experiments for the pion-nucleus scattering really need to be measured in the future, and along with the experiment improvement, the theoretical study also should be further developed, to understand the complicated interactions inside the nuclei. As a reference, for interested readers, a determination of the integration limits of the elastic $\pi^+$--$^{12}\mathrm{C}$ scattering is also provided in further detail in Appendix.~C. 

\chapter{Summary and future perspectives}
As a summary, in this thesis, I have investigated the elastic $\pi N$ scattering process, which is dominated by the $\Delta(1232)$-resonance contribution, for the $I=3/2$ channel in dense medium ($\rho_B\ne0$) as well as vacuum, where the in-medium modification effects are inspired by the heavy-ion collisions (HICs) experiments. Medium modifications are considered in this study because there are pieces of evidence that the hadron properties (mass and decay width) and structure (radius and Parton distribution function (PDF)) change in the medium. This change is related to the partial restoration of the chiral symmetry. For this purpose, here I employed the effective Lagrangian approach at the tree-level Born approximation to compute the scattering amplitudes for the relevant Feynman diagrams. For the background contributions, I made use of the Weinberg-Tomozawa contact interaction, which mimics the vector meson exchange in the $t$-channel in the low-energy region, whereas 
the phenomenological form factors were taken into account by accommodating the spatial extension of the hadrons. The quark-meson coupling (QMC) model is employed to describe the medium modifications for the in-medium baryon properties, such as the baryon masses ($M^*_{N,\Delta}$) and $\Delta$ full decay width ($\Gamma^*_\Delta$). Below, I list the important findings and observations in the elastic $\pi N$ scattering:
\begin{enumerate}
\item Total and differential cross-sections in a vacuum are qualitatively well reproduced in comparison with the experimental data. Again, I confirmed that the scattering process is dominated by the $\Delta$ resonance in the $s-$channel. The angular dependence shows the strong forward and mild backward scattering enhancements in the vicinity of the $\Delta$ mass region, due to the $\Delta-$dominated $p$-wave scattering. As the energy increases, the $\Delta$-resonance and $u-$channel BKG contributions compete destructively with each other, resulting in more enhancements in the forward scattering.  
\item As for the symmetric nuclear matter (SNM), the nucleon and $\Delta$-resonance masses are decreasing at the same rate by the $\sigma$-meson exchange interaction (decreasing scalar potential) with respect to the baryon density $\rho_B$ from the QMC model. The full decay width of $\Delta$ resonance, $\Gamma^*_\Delta$ gets wider as the density and energy $E_\mathrm{cm}$ increase from the linear-density approximation. This observation indicates that it becomes hard to produce the $\Delta$ resonance in a nuclear medium (shifting the $\Delta$ production threshold).  
\item The $\Delta$-resonance invariant-mass spectrum shown in the total cross-section becomes dissolved with respect to the density, according to the change of $\pi N$ interaction in medium, and the peak position slightly moves to the higher energy regions. As for the angular-dependent differential cross-section, a similar density dependence is shown in the vicinity of the $\Delta$-resonance region, while the density dependence almost disappears beyond the region, since the BKG contributions compete and overcome the $\Delta$-resonance one. It is observed that the $t$-dependent differential cross-section also depicts a similar tendency to the angular-dependent one as well. 
\item Finally, I compute the target-recoil proton-spin asymmetry, which signals the short-range spin correlation in the medium. It is found that the asymmetry follows simple spin statistics, and the resulting curve shapes can be understood by the combination of the $S=1/2$ and $S=3/2$ spin states of the intermediate $\Delta$ resonance. In the vicinity of the $\Delta$-resonance mass region, the asymmetry is dominated by the $\Delta$ resonance and insensitive to the density changes. On the contrary, as for the regions beyond $\Delta$-resonance mass, the density dependence becomes finite and obvious, since the $\Delta$-resonance and BKG contributions start to compete. 
\end{enumerate}

Next, in the following study, I consider the elastic $\pi A$ scattering process for light nuclei ($^4\mathrm{He}$ and $^{12}\mathrm{C}$) in the $\Delta(1232)$-resonance energy region by using Eikonal Glauber model. According to the Glauber model, to get the TCS of the elastic $\pi^+ A$ scattering, firstly I consider the elastic $\pi^+ N$ scattering process at finite density $\rho_A(r)$ which is located at $r=\sqrt{\boldsymbol{b}^2+|z|^2}$. Then I integrate $\sigma_{\pi^+ N}(\sqrt{s^*},\rho_A)$ over the entire volume of the nucleus.

The hadron properties such as baryon masses $M^*_B$, $\Delta$-decay width $\Gamma^*_\Delta$ in the finite nuclei are calculated by the QMC model with the nuclear density distribution $\rho_A$. Then I calculate the in-medium coupling constants ($f^*_{\pi N \Delta}$ and $f^*_{\pi NN}$) and the pion decay constant $f^*_\pi$ at finite nuclei via GTR relations.

Results for the total cross-section of the elastic $\pi^+ A$ scattering are given for Helium-4 and Carbon-12. The TCS for $\pi^+$($ ^4\mathrm{He}$, $ ^4\mathrm{He}$)$\pi^+$ reproduces the experimental data well. But in the elastic $\pi^+$--$^{12}\mathrm{C}$ scattering, the numerical results of the TCS are a bit higher than the experimental data. Some of the reasons for this discrepancy could be attributed to not considering the multiple scattering in the finite nuclei and the interactions between the nucleons inside of nuclei. These effects will be stronger in much dense and heavier nuclei, like $^{63}\mathrm{Ca}$,  $^{208}\mathrm{Pb}$, and so forth.

From this first attempt and study, I expect that these observations will shed light on the more understanding of the $\Delta$-resonance contribution in the medium and other quantities at finite density. However, the present setup in this thesis is quite ideal and different from reality. As a further investigation, to improve these studies, the physical observations could be investigated with consideration of the more realistic description of the system, such as the asymmetric nuclear matter, multiple scattering processes, and the $NN$ interaction (correlations) in the finite nuclei, etc. Such realistic calculations and systems will be addressed to be performed for future works, as suggestions.

\chapter*{Acknowledgements}
\addcontentsline{toc}{chapter}{Acknowledgements}
Firstly, I appreciate my supervisor, Prof. S. I. Nam, who gives me meaningful bits of advice and feedback. Also, I am grateful to Dr. Parada T. P. Hutauruk, for fruitful discussions and advice. Additionally, I would like to thank Dr. S. H. Kim (Soongsil University) for answering my questions and advising me during my undergraduate study at PKNU. This work was supported by the National Research Foundation of Korea (NRF) grants funded by the Korea government (MSIT) (No. 2018R1A5A1025563).


\begin{appendices}
\chapter{Calculation of the scattering amplitude and $\Delta$ decay width}

\section{Scattering amplitudes of elastic $\pi^+ p$ scattering}
In the elastic $\pi^+ p$  scattering process, there are four Feynman diagrams at the tree-level Born approximation.
\begin{figure}[t]
\centering\includegraphics[width=0.95\textwidth]{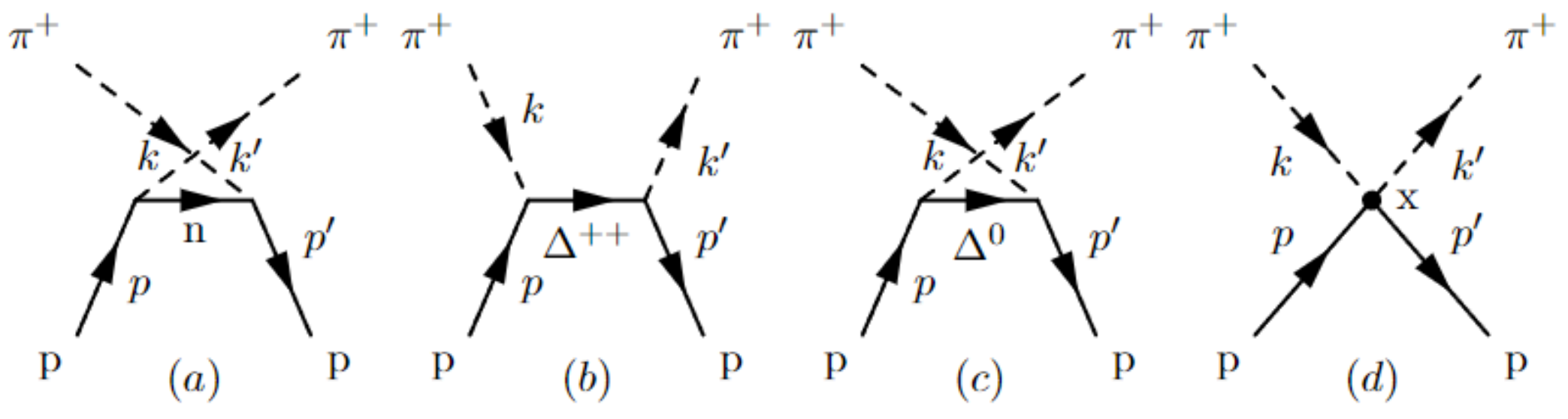}
\end{figure}
\subsection{$u$-channel of neutron}
The effective lagrangian of $\pi NN$ vertex is given by
\begin{equation}
    \mathcal{L}_{\pi NN}=-\frac{f_{\pi NN}}{m_\pi}\overline{N}\gamma_5\slashed{\partial}\pi N.
\end{equation}

Then, the scattering amplitude of ($a$) $u$-channel of neutron $\mathcal{M}^n_u$ is calculated by
\begin{align}
    \langle kp'\mid iT\mid k'p \rangle 
    &=\langle kp'\mid T\{i\int d^4 y \mathcal{L}_{\pi NN} \times i\int d^4 x \mathcal{L}_{\pi NN} \} \mid k'p \rangle \\
    &=\langle kp'\mid i\int \left[-\left(-\sqrt{\frac{2}{3}}\right)\frac{f_{\pi NN}}{m_\pi}\overline{N}\gamma_5\slashed{\partial}\pi N \right] d^4 y \nonumber \\  &\qquad\quad\;\times i\int \left[-\left(-\sqrt{\frac{2}{3}}\right)\frac{f_{\pi NN}}{m_\pi}\overline{N}\gamma_5\slashed{\partial}\pi N \right] d^4 x\mid k'p\rangle \nonumber \\
    &=-\frac{2f^2 _{\pi NN}}{3m_\pi ^2}\int d^4 y\int d^4 x\langle kp'\mid\left[\overline{N}\gamma_5\slashed{\partial}\pi N \right]\left[\overline{N}\gamma_5\slashed{\partial}\pi N \right] \mid k'p\rangle \nonumber\\
    &=-\frac{2f^2 _{\pi NN}}{3m_\pi ^2}\int d^4 y\int d^4 x\left[\overline{u}(p')e^{ip'y}\right] \gamma_5\left[i\slashed{k}e^{-iky}\right] \nonumber \\ &\qquad\quad\times\left[\int\frac{d^4 q}{(2\pi)^4}i\left(\frac{\slashed{q}+m_N}{q^2 -m_N ^2}\right)e^{iq(x-y)}\right]\gamma_5\left[-i\slashed{k}'e^{ik'x}\right]\left[u(p)e^{-ipx} \right] \nonumber\\
    &=-i\frac{2f^2 _{\pi NN}}{3m_\pi ^2}\int d^4 y\int d^4 x\left(e^{i(p'-k-q)y}\right)\left(e^{-i(p-k'-q)x}\right) \nonumber \\ &\qquad\quad\times\overline{u}(p')\gamma_5\slashed{k}\left(\frac{\slashed{q}+m_N}{q^2 -m_N^2}\right)\gamma_5\slashed{k}'u(p) \nonumber\\
    &=-i\frac{2f^2 _{\pi NN}}{3m_\pi ^2}\int\frac{d^4 q}{(2\pi)^4}(2\pi)^4\delta^4(p'-k-q)(2\pi)^4\delta^4(p-k'-q) \nonumber \\ &\qquad\quad\times\overline{u}(p')\gamma_5\slashed{k}\left(\frac{\slashed{q}+m_N}{q^2 -m_N^2}\right)\gamma_5\slashed{k}'u(p) \nonumber\\
    &=-i\frac{2f^2 _{\pi NN}}{3m_\pi ^2}(2\pi)^4\delta^4(p'-k-p+k') \nonumber \\ &\qquad\quad\times\overline{u}(p')\gamma_5\slashed{k}\frac{\slashed{p}-\slashed{k}'+m_N}{(p-k')^2 -m_N^2}\gamma_5\slashed{k}'u(p) \nonumber\\
    &=i\mathcal{M}_u^n\times(2\pi)^4\delta^4(p'+k'-p-k) 
\end{align}
\begin{align}
    \therefore\mathcal{M}_u^n=-\frac{2f^2 _{\pi NN}}{3m_\pi ^2}\overline{u}(p')\gamma_5\slashed{k}\frac{\slashed{p}-\slashed{k}'+m_N}{(p-k')^2 -m_N^2}\gamma_5\slashed{k}'u(p).
\end{align}

\subsection{$s$-channel of $\Delta^{++}$ and $u$-channel of $\Delta^0$}
The effective lagrangian of $\pi N \Delta$ vertex is given by
\begin{equation}
\label{eqA1}
\mathcal{L}_{\pi N \Delta}=\frac{f_{\pi N\Delta}}{m_\pi}\overline{\Delta}^{\mu}\boldsymbol{S}^{\dagger}\partial_\mu \boldsymbol{\pi}N + h.c.
\end{equation}

Firstly, the scattering amplitude of ($b$) $s$-channel of $\Delta^{++}$,  $\mathcal{M}^{\Delta^{++}}_s$ is calculated by
\begin{align}
    \langle k'p'\mid iT\mid kp \rangle 
    &=\langle k'p'\mid T\{i\int d^4 y \mathcal{L}_{\pi N \Delta} \times i\int d^4 x \mathcal{L}_{\pi N  \Delta} \} \mid kp \rangle \\
    &=\langle k'p'\mid i\int d^4 y\left[\frac{f_{\pi N\Delta}}{m_\pi}\left(\overline{\Delta}^{\mu}\partial_\mu \pi N+\overline{N}(\partial_\mu\pi)^\dagger\Delta^\mu\right)\right] \nonumber \\ &\qquad\quad\times i\int d^4 x\left[\frac{f_{\pi N\Delta}}{m_\pi}\left(\overline{\Delta}^{\nu}\partial_\nu \pi N+\overline{N}(\partial_\nu\pi)^\dagger\Delta^\nu\right)\right]\mid kp\rangle \nonumber\\
    &=-\frac{f_{\pi N\Delta}^2}{m_\pi ^2}\int d^4 y\int d^4 x\langle k'p'\mid\left(\overline{N}(\partial_\mu\pi)^\dagger\Delta^\mu\right)\times\left(\overline{\Delta}^{\nu}\partial_\nu \pi N\right)\mid kp\rangle \nonumber\\
    &=-\frac{f_{\pi N\Delta}^2}{m_\pi ^2}\int d^4 y\int d^4 x\left[\overline{u}(p')e^{ip'y}\right]\left[ik'_\mu e^{ik'y}\right] \nonumber \\ &\qquad\quad\times\left[\int\frac{d^4 q}{(2\pi)^4}G^{\mu\nu}(q)e^{iq(x-y)}\right]\left[-ik_\nu e^{-ikx}\right]\left[u(p)e^{-ipx}\right] \nonumber\\
    &=-\frac{f_{\pi N\Delta}^2}{m_\pi ^2}\int\frac{d^4 q}{(2\pi)^4}(2\pi)^4\delta^4(p'+k'-q)(2\pi)^4\delta^4(p+k-q) \nonumber \\ &\qquad\quad\times\overline{u}(p')k'_\mu G^{\mu\nu}(q)k_\nu u(p) \nonumber\\
    &=-\frac{f_{\pi N\Delta}^2}{m_\pi ^2}(2\pi)^4\delta^4(p'+k'-p-k)\times\overline{u}(p')k'_\mu G^{\mu\nu}(p+k)k_\nu u(p) \nonumber\\
    &=i\mathcal{M}_s^{\Delta^{++}}\times(2\pi)^4\delta^4(p'+k'-p-k)
\end{align}
\begin{align}
    \therefore \mathcal{M}_s^{\Delta^{++}}=i\frac{f_{\pi N\Delta}^2}{m_\pi ^2}\overline{u}(p')k'_\mu G^{\mu\nu}(p+k)k_\nu u(p).
\end{align}

And the scattering amplitude of ($c$) $u$-channel of $\Delta^0$,  $\mathcal{M}^{\Delta^0}_u$ is calculated by
\begin{align}
    \langle kp'\mid iT\mid k'p \rangle 
    &=\langle kp'\mid T\{i\int d^4 y \mathcal{L}_{\pi N\Delta}\times i\int d^4 x \mathcal{L}_{\pi N\Delta} \} \mid k'p \rangle \\
    &=\langle kp'\mid i\int d^4 y\left[\frac{f_{\pi N\Delta}}{\sqrt{3} m_\pi}\left(\overline{\Delta}^{\mu}\partial_\mu \pi N+\overline{N}(\partial_\mu\pi)^\dagger\Delta^\mu\right)\right] \nonumber \\ &\qquad\quad\times i\int d^4 x\left[\frac{f_{\pi N\Delta}}{\sqrt{3} m_\pi}\left(\overline{\Delta}^{\nu}\partial_\nu \pi N+\overline{N}(\partial_\nu\pi)^\dagger\Delta^\nu\right)\right]\mid k'p\rangle \nonumber\\
    &=-\frac{f_{\pi N\Delta}^2}{3m_\pi ^2}\int d^4 y\int d^4 x\langle kp'\mid\left(\overline{N}(\partial_\mu\pi)^\dagger\Delta^\mu\right)\times\left(\overline{\Delta}^{\nu}\partial_\nu \pi N\right)\mid k'p\rangle \nonumber\\
    &=-\frac{f_{\pi N\Delta}^2}{3m_\pi ^2}\int d^4 y\int d^4 x\left[\overline{u}(p')e^{ip'y}\right]\left[-ik_\mu e^{-iky}\right] \nonumber \\ &\qquad\quad\times\left[\int\frac{d^4 q}{(2\pi)^4}G^{\mu\nu}(q)e^{iq(x-y)}\right]\left[ik'_\nu e^{ik'x}\right]\left[u(p)e^{-ipx}\right] \nonumber\\
    &=-\frac{f_{\pi N\Delta}^2}{3m_\pi ^2}\int\frac{d^4 q}{(2\pi)^4}(2\pi)^4\delta^4(p'-k-q)(2\pi)^4\delta^4(p-k'-q) \nonumber \\ &\qquad\quad\times\overline{u}(p')k_\mu G^{\mu\nu}(p-k')k'_\nu u(p) \nonumber\\
    &=-\frac{f_{\pi N\Delta}^2}{3m_\pi ^2}(2\pi)^4\delta^4(p'-k-p+k')\overline{u}(p')k_\mu G^{\mu\nu}(p-k')k'_\nu u(p) \nonumber\\
    &=i\mathcal{M}_u^{\Delta^{0}}\times(2\pi)^4\delta^4(p'+k'-p-k)
\end{align}
\begin{align}
    \therefore \mathcal{M}_u^{\Delta^{0}}=i\frac{f_{\pi N\Delta}^2}{3m_\pi ^2}\overline{u}(p')k_\mu G^{\mu\nu}(p-k')k'_\nu u(p).
\end{align}
\subsection{Weinberg-Tomozawa interaction (Contact term)}
The effective lagrangian of Weinberg-Tomozawa interaction is given by
\begin{equation}
\mathcal{L}_{\mathrm{WT}}=i\frac{C_{\pi N}}{4f^2_\pi}\overline{N}\left[\pi(\slashed{\partial}\pi^\dagger)-(\slashed{\partial}\pi)\pi^\dagger\right]N.
\end{equation}

Then its scattering amplitude $\mathcal{M}_{\mathrm{WT}}$ ($d$) (contact term) is calculated by 
\begin{align}
    \langle k'p'\mid iT\mid kp \rangle 
    &=\langle k'p'\mid i\{\int d^4 x\mathcal{L}_{\mathrm{WT}}\}\mid kp\rangle \\
    &=\langle k'p'\mid i\int d^4 x\left[i\frac{C_{\pi N}}{4f^2_\pi}\overline{N}\left(\pi\slashed{\partial}\pi^\dagger-\slashed{\partial}\pi\pi^\dagger\right)N\right]\mid kp\rangle \nonumber\\
    &=-\frac{C_{\pi N}}{4f^2_\pi}\int d^4 x\left[\overline{u}(p')e^{ip'x}\right]\left[i\slashed{k}'e^{-i(k-k')x}+i\slashed{k}e^{-i(k-k')x}\right]\left[u(p)e^{-ipx}\right] \nonumber\\
    &=-i\frac{C_{\pi N}}{4f^2_\pi}\int d^4 x\left(e^{i(p'+k'-p-k)x}\right)\overline{u}(p')(\slashed{k}+\slashed{k}')u(p) \nonumber\\
    &=-i\frac{C_{\pi N}}{4f^2_\pi}(2\pi)^4\delta^4(p'+k'-p-k)\overline{u}(p')(\slashed{k}+\slashed{k}')u(p) \nonumber\\
    &=i\mathcal{M}_{\mathrm{WT}}\times(2\pi)^4\delta^4(p'+k'-p-k)
\end{align}
\begin{align}
    \therefore \mathcal{M}_{\mathrm{WT}}=-\frac{C_{\pi N}}{4f^2_\pi}\overline{u}(p')(\slashed{k}+\slashed{k}')u(p).
\end{align}

\section{$\Delta(1232)$ decay width of $\Delta\rightarrow \pi N$ process}

In this section, the calculation procedure of getting $\Delta(1232)$ decay width of $\Delta\rightarrow \pi N$ process is denoted.
\begin{align}
    d\Gamma_\Delta &=\frac{1}{2m_\Delta}\int\frac{d^3p_N}{(2\pi)^3}\int\frac{d^3p_\pi}{(2\pi)^3}\frac{1}{2E_N2E_\pi}|\mathcal{M}_{\Delta\rightarrow\pi N}|^2\\
    &\qquad\times(2\pi)^4\delta^{(3)}(\vec{p}_N+\vec{p}_\pi)\;\delta(E_N+E_\pi-m_\Delta)\nonumber \\
    &=\frac{1}{2m_\Delta}\frac{1}{2E_N2E_\pi}\int\frac{|\vec{p}_N|}{(2\pi)^2}\frac{|E_NE_\pi|d\Omega}{E_{\mathrm{CM}}}|\mathcal{M}|^2\nonumber \\ &=\frac{1}{2m_\Delta}\frac{1}{16\pi^2}\int\frac{|\vec{p}_N|}{m_\Delta}|\mathcal{M}|^2d\Omega \nonumber
\end{align}

\begin{equation}
    \therefore \Gamma_\Delta=\frac{|\vec{p}_N|}{8\pi m^2_\Delta}|\mathcal{M}_{\Delta\rightarrow \pi N}|^2.
\end{equation}

In order to get the scattering amplitude of $\Delta\rightarrow \pi N$ decay process, the effective lagrangian of $\mathcal{L}_{\pi N \Delta}$~\eqref{eqA1} is used. Specific procedures are indicated below.
\begin{align}
    \langle p_\pi p_N\mid iT \mid p_\Delta\rangle&=\langle p_\pi p_N\mid T\{i\int d^4x\mathcal{L}_{\pi N \Delta}\}\mid p_\Delta\rangle \\ &=\langle p_\pi p_N\mid i\int\frac{f_{\pi N \Delta}\sqrt{\mathcal{I}_{\pi N \Delta}}}{m_\pi}(\overline{\Delta}^\mu\partial_\mu\pi N+\overline{N}(\partial_\mu\pi)^{\dagger}\Delta^\mu)d^4x\mid p_\Delta\rangle \nonumber \\
    &=i\frac{f_{\pi N \Delta}\sqrt{\mathcal{I}_{\pi N \Delta}}}{m_\pi}\int d^4x\langle p_\pi p_N\mid \overline{N}(\partial_\mu\pi)^{\dagger}\Delta^\mu\mid p_\Delta\rangle \nonumber \\ &=i\frac{f_{\pi N \Delta}\sqrt{\mathcal{I}_{\pi N \Delta}}}{m_\pi}\int d^4x\left[\overline{u}_N(p_N)e^{i p_N\cdot x}\right]\left[i(p_\pi)_\mu e^{i p_\pi\cdot x}\right]\left[\Delta^\mu (p_\Delta)e^{-i p_\Delta\cdot x}\right] \nonumber \\ &=-\frac{f_{\pi N \Delta}\sqrt{\mathcal{I}_{\pi N \Delta}}}{m_\pi}\int e^{i(p_N+p_\pi -p_\Delta)\cdot x}d^4x\times\overline{u}_N(p_N)(p_\pi)_\mu\Delta^\mu (p_\Delta) \nonumber \\ &=-\frac{f_{\pi N \Delta}\sqrt{\mathcal{I}_{\pi N \Delta}}}{m_\pi}(2\pi)^4\delta^{(4)}(p_N+p_\pi -p_\Delta)\overline{u}_N(p_N)(p_\pi)_\mu\Delta^\mu (p_\Delta) \nonumber \\
    &=i\mathcal{M}_{\Delta\rightarrow\pi N}\times(2\pi)^4\delta^{(4)}(p_N+p_\pi -p_\Delta).
\end{align}
\begin{equation}
    \therefore i\mathcal{M}(p_\Delta\rightarrow p_\pi p_N)=-\frac{f_{\pi N \Delta}\sqrt{\mathcal{I}_{\pi N \Delta}}}{m_\pi}\overline{u}_N(p_N)(p_\pi)_\mu\Delta^\mu (p_\Delta).
\end{equation}

And, as a final calculation, the spin sum of the scattering amplitude $\mathcal{M}_{\Delta\rightarrow \pi N}$ is given by 
\begin{align}
    |\mathcal{M}_{\Delta\rightarrow\pi N}|^2&=\frac{1}{4}\sum_{\mathrm{Spins}}|\mathcal{M}(p_\Delta\rightarrow p_\pi p_N)|^2 \\ &=\frac{f_{\pi N \Delta}^2\mathcal{I}_{\pi N \Delta}}{4m_\pi^2}\mathrm{Tr}\left[\left\{\overline{u}_N(p_N)(p_\pi)_\mu \Delta^\mu(p_\Delta)\right\}\left\{\overline{\Delta}^\nu(p_\Delta)(p_\pi)_\nu u_N(p_N)\right\}\right]  \nonumber \\
    &=\frac{f_{\pi N \Delta}^2\mathcal{I}_{\pi N \Delta}}{4m_\pi^2}\mathrm{Tr}\left[u_N(p_N)\overline{u}_N(p_N)(p_\pi)_\mu \Delta^\mu(p_\Delta)\overline{\Delta}^\nu(p_\Delta)(p_\pi)_\nu \right] \nonumber \\
    &=\frac{f_{\pi N \Delta}^2\mathcal{I}_{\pi N \Delta}}{4m_\pi^2}\mathrm{Tr}\left[(\slashed{p}_N+m_N)(p_\pi)_\mu G^{\mu\nu}(p_\Delta)(p_\pi)_\nu \right] \nonumber \\
    &=\frac{f_{\pi N \Delta}^2\mathcal{I}_{\pi N \Delta}}{4m_\pi^2}\mathrm{Tr}[(\slashed{p}_N+m_N)(\gamma^0 m_\Delta+m_\Delta) \nonumber \\ &\qquad\times\left\{-(p_\pi)^2+\frac{1}{3}(\slashed{p}_\pi)^2+\frac{2}{3}\frac{(m_\Delta E_\pi)^2}{m_\Delta^2}-\frac{1}{3}\frac{m_\Delta E_\pi \slashed{p}_\pi-m_\Delta E_\pi \slashed{p}_\pi}{m_\Delta}\right\} ] \nonumber \\
    &=\frac{f_{\pi N \Delta}^2\mathcal{I}_{\pi N \Delta} m_{\Delta}}{4m_\pi^2}\mathrm{Tr}\left[(\slashed{p}_N+m_N)(\gamma^0 +1)\left[-(p_\pi)^2+\frac{1}{3}(\slashed{p}_\pi)^2+\frac{2}{3}E_\pi^2\right]\right] \nonumber \\
    &=\frac{f_{\pi N \Delta}^2\mathcal{I}_{\pi N \Delta} m_\Delta}{m_\pi^2}(E_N+m_N)\left[-\frac{2}{3}m_\pi^2+\frac{2}{3}(m_\pi^2+{\vec{p}_\pi}^2)\right], \nonumber
\end{align}
where $G^{\mu\nu}(p_\Delta)$ is represented by
\begin{equation}
    G^{\mu\nu}(p_\Delta)=(\slashed{p}_\Delta+m_\Delta)\left[-g^{\mu\nu}+\frac{1}{3}\gamma^\mu\gamma^\nu+\frac{2}{3}\frac{p_\Delta^\mu p_\Delta^\nu}{m_\Delta^2}-\frac{1}{3}\frac{p_\Delta^\mu\gamma^\nu-p_\Delta^\nu\gamma^\mu}{m_\Delta}\right].
\end{equation}

Especially, when we calculate the spin sum of the scattering amplitude, the trace part should be calculated with care. For example, $\mathrm{Tr}[\slashed{p}_N \gamma^0 (\slashed{p}_\pi)^2]$ is calculated by
\begin{align}
    \mathrm{Tr}[\slashed{p}_N \gamma^0 (\slashed{p}_\pi)^2]&=\mathrm{Tr}[\gamma^\mu(p_N)_\mu\cdot\gamma^0\cdot\gamma^\nu(p_\pi)_\nu\cdot\gamma^\lambda(p_\pi)_\lambda] \\ &=4\{g^{\mu 0}g^{\nu\lambda}-g^{\mu\nu}g^{0\lambda}+g^{\mu\lambda}g^{0\nu}\}(p_N)_\mu(p_\pi)_\nu(p_\pi)_\lambda \nonumber \\ &=4\{E_N p^2_\pi-(p_N\cdot p_\pi)E_\pi+(p_N\cdot p_\pi)E_\pi\} \nonumber \\ &=4E_N p^2_\pi =4E_N(E^2_\pi-(\vec{p}_\pi)^2)=4E_N m^2_\pi, \nonumber
\end{align}
where the relation $\mathrm{Tr}(\gamma^\mu\gamma^\nu\gamma^\lambda\gamma^\tau)=4(g^{\mu\nu}g^{\lambda\tau}-g^{\mu\lambda} g^{\nu\tau}+g^{\mu\tau}g^{\nu\lambda})$ is used. the proof for this expression is given by the anti-commutation relation between the gamma matrices, $\{\gamma^\mu,\gamma^\nu \}=2g^{\mu\nu}$.

Finally we can get the $\Delta$ decay width of $\Delta\rightarrow\pi N$ decay, which is given by
\begin{equation}
    \therefore \Gamma_\Delta=\frac{|\vec{p}_N|}{8\pi m^2_\Delta}\times\frac{1}{4}\sum_{\mathrm{Spins}}|\mathcal{M}_{\Delta\rightarrow \pi N}|^2=\frac{f_{\pi N \Delta}^2\mathcal{I}_{\pi N \Delta}}{12\pi m_\pi^2 m_\Delta}(E_N+m_N)|\vec{p}_N|^3,
\end{equation}
where $\mathcal{I}_{\pi N \Delta}$ denotes the square of the isospin factor for a $\pi N \Delta$ vertex, which are given 1 for the $s$-channel of $\Delta^{++}$ and 1/3 for the $u$-channel of $\Delta^0$ in this study. Also the magnitude of the three-momenta for the pion $|\vec{p}_\pi|$ is equal to the absolute value of the three-momenta of nucleon $|\vec{p}_N|$, thus $|\vec{p}_\pi|^2=|\vec{p}_N|^2$.
\chapter{The Gaussian quadrature}

To calculate the elastic $\pi$-nucleus scattering with the Eikonal Glauber model using a programming language(e.g. Fortran), we need to consider how to compute the integration of equations. Fortunately, there is a well-known method called the Gaussian quadrature. But firstly we should know about a  quadrature rule, which means an approximation of the continuous integration of a function as a summation of weighted value multiplied to function values at specified points in the integration interval region.

The Gaussian quadrature is a quadrature rule, which is proposed to calculate an exact result for polynomials of degree $2n-1$ or less by a proper value of the nodes $x_i$ and weights $w_i$ for $i$ = 1, $\dots$, $n$. Due to the integration interval, various methods for the Gaussian quadrature are used with specific orthogonal polynomials, such as Legendre polynomials, Hermite polynomials, Laguerre polynomials, and so on. 

In this study, the integration intervals within the equation for TCS of the elastic $\pi^+ A$ scattering process using the Glauber model are given ($-\infty,\infty$) for $z$, [0,$\infty$) for $\boldsymbol{b}$. So the Gauss-Hermite quadrature for integration of $z$ and the Gauss-Laguerre quadrature for integration of $\boldsymbol{b}$ were used. Therefore the brief explanation of these two quadrature methods is denoted below.
\section{The Gauss-Hermite quadrature}
The Gauss-Hermite quadrature is simply used to integrate within ($-\infty,\infty$) interval, using the Hermite polynomials $H_n(x)$. The formula is given by
\begin{equation}
    \int^\infty_{-\infty}f(x)dx=\int^\infty_{-\infty}e^{-x^2}\left[e^{x^2}f(x)\right]dx\simeq\sum^n_{k=1}w(x_k)\left[e^{x^2_k}f(x_k)\right]+R_n(x),
\end{equation}
where $x_k$ is the $k$-th zero of $H_n(x)$, and $w(x_k)$, $R_n(x)$ are represented by
\begin{equation}
    w(x_k)=\frac{2^{n-1}n!\sqrt{\pi}}{n^2[H_{n-1}(x_k)]^2},\quad R_n(x)=\frac{n!\sqrt{\pi}}{2^n(2n)!}f^{(2n)}(x).
\end{equation}
\section{The Gauss-Laguerre quadrature}
The Gauss-Laguerre quadrature is used to integrate within [0,$\infty$) interval, using the Laguerre Polynomials $L_n(x)$. The Gauss-Laguerre formula is given by
\begin{equation}
    \int^\infty_0 f(x)dx=\int^\infty_0 e^{-x}\left[e^x f(x)\right]dx\simeq\sum^n_{k=1}w(x_k)\left[e^x f(x_k)\right]+R_n(x),
\end{equation}
where $x_k$ is the $k$-th zero of $L_n(x)$, and $w(x_k)$, $R_n(x)$ are represented by
\begin{equation}
    w(x_k)=\frac{x_k}{(n+1)^2[L_{n+1}(x_k)]^2},\quad R_n(x)=\frac{(n!)^2}{(2n)!}f^{(2n)}(x).
\end{equation}

As a reference, there is useful website to calculate weight $w(x_k)$ corresponding to the number of $n$, named `efunda (engineering fundamentals)'~\cite{efunda}. So if you don't have proper library to calculate Hermite or Laguerre polynomials in the programming language what you use, this site will be helpful. 

\chapter{A check for the integration limits of the elastic $\pi^+ (^{12}\mathrm{C},$ $^{12}\mathrm{C})\pi^+$ scattering}

\begin{figure}[t]
  \centering
 \includegraphics[width=0.7\textwidth]{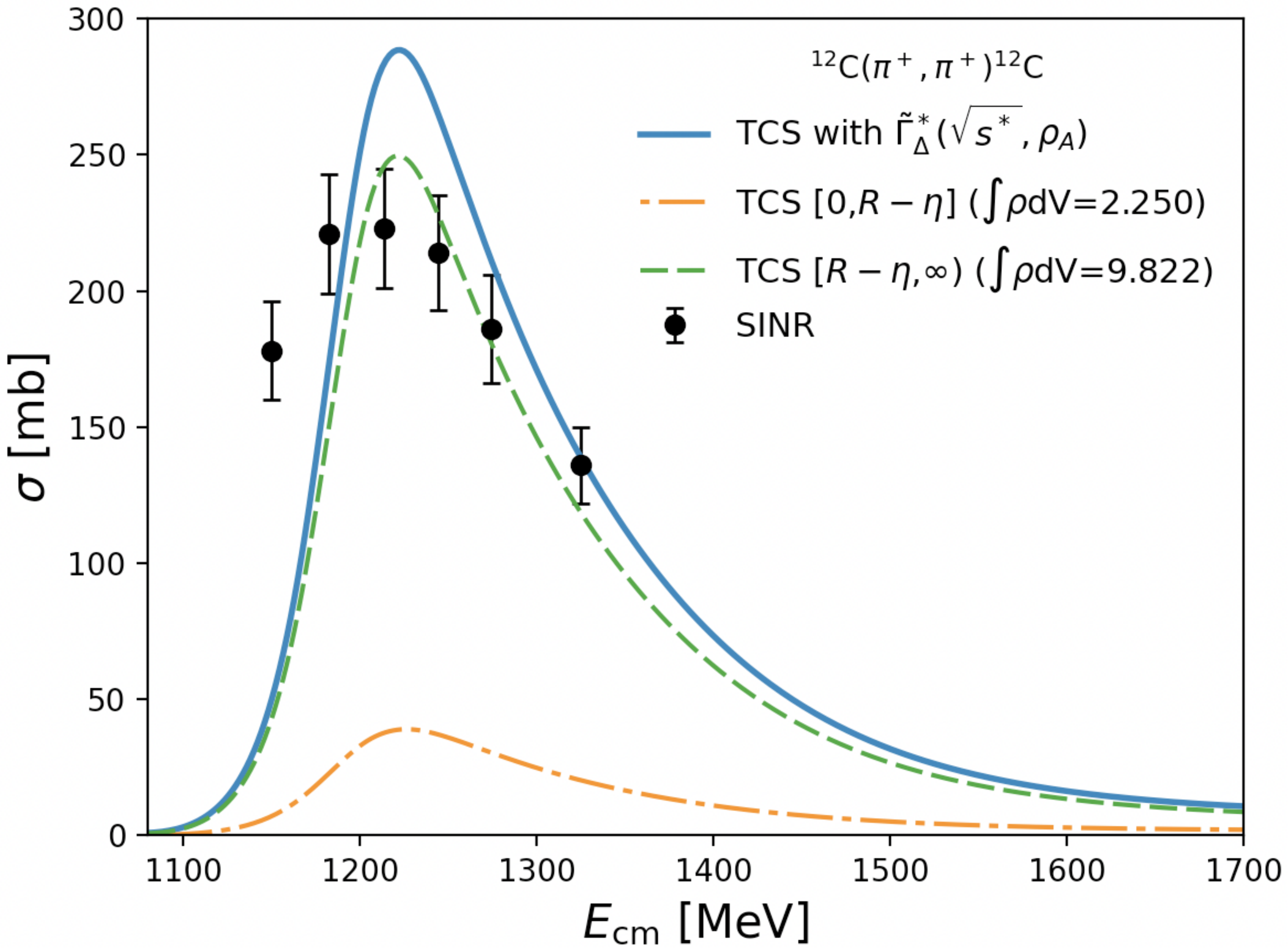}
  \caption{Total cross-section of the elastic $\pi^+$--$^{12}\mathrm{C}$ scattering within the interval [0,$\infty$) (blue solid line), [0,$R-\eta$] (orange dot-dashed line), and [$R-\eta$,$\infty$) (green dashed line) for $r$. \label{fig21}}
\end{figure}

It is known that the elastic $\pi^+ A$ scatterings happen around the surface of the nucleus. Thus, as an additional investigation, I try to reconstruct the elastic $\pi^+ A$ scattering for $^{12}\mathrm{C}$, by limiting the integration interval from [0,$\infty$) to [$R-\eta$,$\infty$) for radius $r$ of nuclei within the equation of the Eikonal Glauber model. It is quite factitious deciding the length of $\eta$, so I varied $\eta$ until the numerical result of TCS of the elastic scattering for $^{12}\mathrm{C}$ fit the experimental data of the elastic $\pi^+$--$^{12}\mathrm{C}$ scattering. Then, the numerical results within different integration intervals are depicted in Fig.~\ref{fig21}.

In Fig.~\ref{fig21}, the total cross-section of the elastic $\pi^+$--$^{12}\mathrm{C}$ scattering within the interval [0,$\infty$) (blue solid line), [0,$R-\eta$] (orange dot-dashed line), and [$R-\eta$,$\infty$) (green dashed line) for $r$, individually. In this calculation, the total cross-sections are given with $R-\eta = 1.5$ fm. The green dashed line can handle the explanation of the experimental data, just except for the threshold energy region. This approach seems quite reasonable analysis, but the critical problem is remained, that the normalization of the nuclear density distribution $\rho_A$ is not satisfied as denoted in the legend of Fig.~\ref{fig21}.

So to explain the elastic $\pi A$ scattering for heavier nuclei than $^4\mathrm{He}$ by using the Glauber model with hadron and quark level constructions, additional investigations should be considered.

\end{appendices}

\end{document}